\documentclass{article}

\usepackage[english]{babel}

\usepackage[letterpaper,top=2cm,bottom=2cm,left=3cm,right=3cm,marginparwidth=1.75cm]{geometry}

\usepackage{amsmath}
\usepackage{graphicx}
\usepackage[colorlinks=true, allcolors=black]{hyperref}

\usepackage{times}
\usepackage{dsfont,bm,color}
\usepackage{amsfonts}
\usepackage{natbib}
\usepackage{subcaption} 
\usepackage{float}
\usepackage{booktabs}
\usepackage{xurl}
\usepackage{changepage}
\usepackage{multirow,adjustbox}
\usepackage{algorithm}
\usepackage{algpseudocode}
\allowdisplaybreaks

\newtheorem{example}{Example}
\newtheorem{definition}{Definition}
\newtheorem{proposition}{Proposition}
\newtheorem{theorem}{Theorem}
\newtheorem{remark}{Remark}
\newtheorem{assumption}{Assumption}
\newtheorem{lemma}{Lemma}

\DeclareMathOperator*{\argmax}{\arg\!\max}

\newenvironment{keywords}{%
  \par\medskip\noindent
  \small
  \textbf{Keywords:}\ \ignorespaces
}{%
  \par\medskip
}

\title{DEEPEAST technique to enhance power in two-sample tests via the same-attraction function}

\author{Yiting Chen \\ 
Department of Computer Science, Mathematics, Physics and Statistics, \\ University of British Columbia, Kelowna, Canada V1V 1V7
\and Min Gao \\ 
School of Big Data and Statistics, 
Anhui University, Hefei, PR China 230601;\\ Department of Computer Science, Mathematics, Physics and Statistics, \\ University of British Columbia, Kelowna, Canada V1V 1V7
\and Wei Lin \\
Department of Statistics and Actuarial Science, Simon Fraser University, \\ Burnaby, BC, Canada V5A 1S6
\and Andrew Jirasek \\
Department of Computer Science, Mathematics, Physics and Statistics, \\ University of British Columbia, Kelowna, Canada V1V 1V7
\and Kirsty Milligan \\
Department of Computer Science, Mathematics, Physics and Statistics, \\ University of British Columbia, Kelowna, Canada V1V 1V7
\and Xiaoping Shi\\
Department of Computer Science, Mathematics, Physics and Statistics, \\ University of British Columbia, Kelowna, Canada V1V 1V7
}
\date{August 2024}

\begin{document}
\maketitle

\begin{abstract}
Data depth has emerged as an invaluable nonparametric measure for the ranking of multivariate samples. The main contribution of depth-based two-sample comparisons is the introduction of the $Q$ statistic \citep{liu1993quality}, a quality index. Unlike traditional methods, data depth does not require the assumption of normal distributions and adheres to four fundamental properties: affine invariance, maximality at the center, monotonicity relative to the deepest point, and vanishing at infinity \citep{liu1993quality, Serfling2000}. Many existing two-sample homogeneity tests, which assess mean and/or scale changes in distributions often suffer from low statistical power or indeterminate asymptotic distributions. To overcome these challenges, we introduced a DEEPEAST (\underline{de}pth-\underline{e}x\underline{p}lored  sam\underline{e}-\underline{a}ttraction \underline{s}ample-to-sample cen\underline{t}ral-outward ranking) technique for improving statistical power in two-sample tests via the same-attraction function. We proposed two novel and powerful depth-based test statistics: the sum test statistic and the product test statistic, which are rooted in $Q$ statistics,  share a `common attractor' and are applicable across all depth functions.
We further proved the asymptotic distribution of these statistics for various depth functions, in addition to the minimum statistics valid for all depth functions. To assess the performance of power gain, we apply three depth functions: Mahalanobis depth \citep{liu1993quality}, Spatial depth \citep{Brown58, Gower74}, and Projection depth \citep{Liu92}. All of these functions are implemented in the R package \textit{ddalpha}. Through two-sample simulations, we have demonstrated that our sum and product statistics exhibit superior power performance, utilizing a strategic block permutation algorithm and compare favourably with popular methods in literature. Our tests are further validated through analysis on Raman spectral data, acquired from cellular and tissue samples, highlighting the effectiveness of the proposed tests highlighting  the effective 
discrimination between health and cancerous samples. 
\end{abstract}


\section{Introduction}
Prostate cancer is one of the most common diseases in Canada and the second most common cancer in the world.  Its potential to be fatal underscores the critical importance of early diagnosis \citep{Berenguer et al. (2023)}. Additionally, a significant challenge in optimizing treatment protocols is the lack of consideration for individual patient radiosensitivity when prescribing radiation doses. Consequently, there is a pressing need to develop methods for monitoring radiation response in individuals undergoing radiation therapy. Various techniques have been explored for this purpose \citep{Jiang2008, Devpura2014}. In recent years, Raman spectroscopy (RS) has been investigated as a potential augmentative tool for biochemical analysis of tumour response \citep{Devpura2014,Auner2018,Corsetti2018,Picot2022,Sigle2023}. RS provides detailed `fingerprint' biochemical information on various biomolecules (e.g., protein, lipid, DNA, etc.) through a vibrational inelastic light scattering process \citep{Movasaghi 2007}. Recent studies have indicated that RS can offer predictive capabilities regarding tumour proliferation status \citep{Milligan2022, Fuentes2023}. Moreover, when RS is combined with group and basis-restricted non-negative matrix factorization along with random forest strategies, this enhanced technique can yield valuable ranked information about the biochemical dynamics within irradiated tumours \citep{Milligan2021}.

Despite the potential of Raman Spectroscopy (RS) in cancer diagnosis, several systematic issues in data processing need to be addressed. These include data interference and subjective determination errors \citep{Brewer 2023}. Challenges such as baseline variability between sample acquisitions are prevalent. Notably, approximately 10\% of Raman spectra suffer interference from cosmic rays, leading to spikes and potential false peaks in the spectra. Furthermore, the analysis of most Raman spectra relies on manual evaluation, resulting in subjective determination errors due to the lack of a uniform and efficient automated method.

Our research aims to statistically ascertain the significance of the peak at 1524 $cm^{-1}$ in Raman spectra acquired on prostate biopsy samples, which is an indicator of prostate cancer \citep{Movasaghi 2007}. Identifying the presence of this peak accurately is crucial for initial screening and efficient analysis of spectral data acquired on patient samples.
The comparison of different shapes of spectral data can be formulated as a homogeneity test for two multivariate samples with distributions $F$ and $G$, i.e. testing 
\begin{equation}\label{T1}
    H_0: F=G \text{ vs } H_1: F \ne G.
\end{equation}

Some existing two-sample tests have clear drawbacks, such as the need for strong assumptions or reduced statistical power due to increased variance from inefficient pairwise comparisons, which hinders their ability to differentiate between distributions effectively. For instance, parametric methods like Multivariate Analysis of Variance (MANOVA) \citep{Fisher1936} require an assumption of normality, which can be a significant limitation. 
In contrast, one-dimensional non-parametric tests, such as the Cramér test \citep{Anderson1962}, Wilcoxon Rank-Sum test \citep{Wilcoxon1992}, and Energy Distance test \citep{szekely2013}, bypass the need for normal data assumption. These tests are more adaptable to various data distributions. However, extending these approaches to the multivariate scenario presents challenges. For example, the multivariate extension of Cramér's test, as proposed by \citep{kim2020}, introduces point-to-point (PtP) distances to evaluate the distance between pairs of points. While innovative, this approach can inadvertently increase variance in pairwise comparisons, potentially reducing the test's power, especially in scenarios with finite dimensions. An alternative is the extension of Wilcoxon's Rank-Sum test \citep{Liu2022}, which aligns with the idea of PtP distance. 

However, a promising development in this field is the concept of data depth,  $D(x; F)$, which measures the centrality of a point $x$ in a distribution $F(x)$ within a $d$-dimensional space. This approach maps $x$ from $R^d$ to the interval $[0,1]$, providing a point-to-sample (PtS) central-outward ranking. 
Despite this advancement, the Depth-based Rank Statistic (DbR) \citep{Small2011}, which is based on the PtS central-outward ranking, still incorporates further PtP comparisons through univariate Kruskal-Wallis type tests. This addition can increase variance and subsequently decrease the test's power, as evidenced in simulations discussed in Section 3.

Innovative and powerful two-sample tests are essential for detecting varied shapes of spectra. In this regard, we explore the depth-based quality index $Q(F,G)$ \citep{liu1993quality}, measuring the relative ``outlyingness'' of distribution $F$ in comparison to distribution $G$, which is defined as $$Q(F,G)=P\{D (X;F)\leq D(Y;F)|X\sim F, Y\sim G\},$$ where $F$ is the reference distribution. When the two distributions, $F$ and $G$, are unknown, the empirical distributions of $F_m$ and $G_n$ can be employed, assuming sample sizes of $X$ and $Y$ are $m$ and $n$, respectively. The $Q$ index can be estimated by the $Q$ statistic  $$ Q(F_m, G_n)= \frac{1}{nm} \sum_{i=1}^m\sum_{j=1}^n I(D(x_i,F_ m) \leq D(y_j,F_m)), $$ where $I(\cdot)$ is an indicator function that takes 1 if true and 0 otherwise. 
The detailed description of the properties of data depth and the $Q$ index by \cite{liu1993quality} highlights the importance of using $Q$ statistics in homogeneity tests.

The $Q(F_m, G_n)$ statistic serves as a sample-to-sample (StS) central-outward ranking, averaging the depth-based PtS central-outward ranking. The method provides a more accurate comparison than the PtP distance, as it encompasses broader information about the distribution rather than focusing on single points. In addition to its enhanced power, the depth-based StS central-outward ranking benefits from the free distribution assumption and adheres to four fundamental properties of data depth: affine invariance, centroid maximality, monotonicity about the deepest point, and vanishing at infinity \citep{Serfling2000}, which allows for the acquisition of valuable ranked information on the biochemical dynamics inside the RS.

While the depth-based StS central-outward ranking effectively handles multivariate data comparisons, it may lose certain information, such as  data direction, by the one-dimensional projection of the data depth. This loss can be critical in achieving higher statistical power. Addressing this, a recent advancement by \cite{Shi2023} suggests preserving power by considering the maximum of two $Q$ statistics $(Q(F_m, G_n)-\frac{1}{2})^2$ and $(Q(G_n, F_m)-\frac{1}{2})^2$. This insight has led us to explore a new approach to pairwise StS central-outward ranking, derived from $Q$ statistics. Our method maintains the direction of the StS central-outward ranking by analyzing the sign of the partial derivatives of the $Q$ statistics. $Q$ statistics sharing the same sign indicate a unified direction of change under the alternative hypothesis. By considering making a combination of two $Q$ statistics with the property of ``same-attraction'', where the two $Q$ statistics have the same limit under the null hypothesis and approach to the same value under the alternative hypothesis, we can enhance the power of a test based on the derived $Q$ statistic.
In physics, an attractor refers to a set of numerical values toward which a system tends to evolve, regardless of its starting conditions. This can present the long-term behavior of the system. For example, consider all objects near a black hole; they are attracted in the same direction. We apply this concept of attractors, with the ``same attractor” referring to the convergence of distributions over time toward a specific limit. Specifically, the statistic $Q(F_m, G_n)$ is attracted to $\frac{1}{2}$ under $H_0$, while under $H_{\alpha}$, it is attracted to the limit of 0 or 1. Here, the term ``attraction” denotes the direction of movement.
While considering the maximum of $Q(F_m, G_n)$ and $Q(G_n, F_m)$ is one such combination, it is not the most efficient one because their partial derivatives with respect to each $Q$ statistic are not strictly positive or negative, and may be zero. Our two new combinations, $Q(F_m, G_n)+Q(G_n, F_m)$ and $Q(F_m, G_n)\times Q(G_n,F_m)$, promise a greater improvement in power since their partial derivatives have the same sign and are almost never zero; for more details, see Section 2.1.
We have named our proposed technique DEEPEAST, short for \underline{de}pth-\underline{e}x\underline{p}lored  sam\underline{e}-\underline{a}ttraction \underline{S}tS cen\underline{t}ral-outward ranking.

The structure of our paper is as follows: Section 2 introduces the concepts of same attractive $Q$ statistics. We also present a strategic block permutation algorithm accompanied with theoretical justification. We showed the general form of asymptotic distribution of the same attractive $Q$ statistics applicable across all depth functions, as well as a specific one-dimensional case in Euclidean depth, which is related to the Craig distribution \citep{Craig}. Inspired by \cite{Gao24}, our proof utilizes a second-order approximation to the $Q$ statistics, Hoeffding decomposition \citep{Ustat} and Cox-Reid expansion method \citep{Cox}, contrasting with existing methods that yield asymptotically normal distributions through first-order approximations \citep{zuo2006limiting}. In addition, our proof differs significantly from that in \cite{Gnettner2024}, which establishes an asymptotic chi-squared distribution for $Q(F_m, G_n)+Q(G_n, F_m)-1$ using the Halfspace depth \cite{tukey} in one-dimensional Euclidean space, whereas we obtain the Craig distribution for one-dimensional Euclidean depth. Furthermore, we justified that the power can be obtained under the alternative hypothesis.
In Section 3, we employ the strategic block permutation algorithm for various depth functions, conducting numerous simulations and offering a detailed comparison with other popular multivariate data methods in the literature. Section 4 applies the DEEPEAST technique to compare samples across differentiated spectra. Finally, Section 5 summarizes our findings and discusses potential future work.

\section{DEEPEAST technique}

The use of $Q$ statistics as StS central-outward ranking leads to a natural question: How can we ensure functional consistency across all $Q$ statistics to enhance statistical power? This section is dedicated to address this question. Let us consider the scenario where we aim to combine $L$ $Q$ statistics, denoted as $Q_1,\dots, Q_L$. We present the combined function as $\mathcal{G}(Q_1, \dots, Q_L)$ .
To optimally gauge similarity within the same distribution and dissimilarity across different distributions, the combined function $\mathcal{G}$ should ideally satisfy two properties: selfsame and coordinate, which are crucial for ensuring both the efficacy and reliability of the function in different statistical context. We have formalized and detailed these properties in Definition \ref{def1}, providing a framework for evaluating and applying the combined $Q$ statistics function in practical scenarios.

\subsection{Definitions and properties}

 \begin{definition}[Same-attraction function] 
Assume the following properties for $Q_1,\dots, Q_L$:\\
\label{def1}
\vspace{-3mm}
\begin{itemize}
\item[(i)] P1. \textit{Selfsame:} $Q_1, \dots, Q_L$ share the asymptotic ``same'' null distribution.
\item[(ii)] P2. \textit{Coordinate:} The partial derivative $\frac{\partial \mathcal{G}(Q_1, \dots, Q_L)}{\partial Q_\ell}$ is non-negative ($\geq 0$) or non-positive ($\leq 0$) almost surely for all $\ell=1, \dots, L$ under the alternative hypothesis.
\end{itemize}
\end{definition}

Note that a same-attraction function $\mathcal{G}(Q_1, \dots, Q_L)$ is strictly same-attraction if the inequalities in P2 are strict almost surely, meaning:
 $$\frac{\partial \mathcal{G}(Q_1, \dots, Q_L)}{\partial Q_\ell} > 0\quad \text{or}~<0.$$

It can be shown that a collection of same-attraction functions  $\mathcal{G}_s(Q_1, \dots, Q_L)$, $1\leq s\leq S$ is closed under countable additions. This means that the sum of a number of same-attraction functions remains to be a same-attraction function. However, it is important to note that this closure may not apply to subtraction.

As indicated above, the definition of a strictly same-attraction function imposes a constraint on the sign of the partial derivatives. They must be consistently positive or negative and cannot equal zero. This characteristic potentially makes a strictly same-attraction function more effective than a non-strict same-attraction function in most cases, as zeros do not maintain the directionality of the StS central-outward ranking. Nevertheless, there are exceptions to this generalization, as illustrated in Example \ref{example-weighted}. Therefore, a more strong criterion is needed to determine the benefit in power of one same-attraction function to another. This criterion is elaborated in Proposition \ref{theorem-powerful}.
\vspace{2mm}

\begin{proposition}
\label{theorem-powerful}
Consider two same-attraction functions, $\mathcal{G}_1(Q_1, \dots, Q_L)$ and $\mathcal{G}_2(Q_1, \dots, Q_L)$. For a specified type I error probability $\alpha$, we defined two decision rules: $\mathcal{G}_1(Q_1, \dots, Q_L)>c_{\alpha,1}$ and $\mathcal{G}_2(Q_1, \dots, Q_L)>c_{\alpha,2}$, such that  $P_{H_0}[\mathcal{G}_r(Q_1, \dots, Q_L)>c_{\alpha,r}] = \alpha$ for $r=1,2.$ If the inequality $\frac{\mathcal{G}_1(Q_1, \dots, Q_L)}{\mathcal{G}_2(Q_1, \dots, Q_L)}\geq\frac{c_{\alpha,1}}{c_{\alpha,2}}$ holds under that alternative hypothesis $H_1$, then $\mathcal{G}_1(Q_1, \dots, Q_L)$ is more powerful than $\mathcal{G}_2(Q_1, \dots, Q_L)$.
\end{proposition}
\vspace{2mm}

The proof of Proposition \ref{theorem-powerful} hinges on the fact that $P_{H_1}[\mathcal{G}_1(Q_1, \dots, Q_L)>c_{\alpha,1}]\geq P_{H_1}[\mathcal{G}_2(Q_1, \dots, Q_L)>c_{\alpha,2}]$. If the conditions hold  asymptotically, then the more powerful test is also in terms of asymptotics.

Among a family of same-attraction functions, the optimal same-attraction function can be found through the following property. 
Consider $G^*$  a set of all possible combinations of function $\mathcal{G}(Q_1, \dots, Q_L)$, the most powerful test statistics $G^0 $ can be selected according to taking the maximum of equations below: 
$$ \argmax_{G \in G^*} \frac{G(Q_1, \dots, Q_L)}{c_{\alpha, G}} =G^0,$$ 
where the $c_{\alpha, G}$ are defined as $P_{H_0}[G>c_{\alpha, G}]=\alpha$ with type I error probability $\alpha$ under null hypothesis.

This criterion provides a framework for comparing the efficacy of various same-attraction functions. The following examples demonstrate its application in evaluating the power of different types of same-attraction functions.
\vspace{2mm}
 
\begin{example}[Maximum statistic \citep{Shi2023}]
\label{example-maximum}
Consider the maximum statistic
\begin{equation}\label{Msta}
   M_{m,n}=\max(Q_1,Q_2),
\end{equation}
where 
$Q_1=\left[\frac{1}{12}(\frac{1}{m}+\frac{1}{n})\right]^{-1} (Q( {F}_{m}, {G}_{n})-\frac{1}{2} )^2$
and 
$Q_2=\left[\frac{1}{12}(\frac{1}{m}+\frac{1}{n})\right]^{-1} (Q( {G}_{n}, {F}_{m})-\frac{1}{2} )^2.$
Both $Q_1$ and $Q_2$ are selfsame (P1), as they follow the same asymptotic null chi-squared distribution \citep{zuo2006limiting}. The coordinate (P2) is also met under $H_1$ since for $r=1,2$,
\begin{equation*}
\frac{\partial M_{m,n} }{\partial Q_r} =
    \begin{cases}
      1 & \text{if }  M_{m,n}=Q_r\\
      0 & \text{otherwise }. \\
    \end{cases}       
\end{equation*}
It is worth noting that $M_{m,n}$ is non-differentiable at $Q_1=Q_2$ with zero probability almost surely. Thus, by Definition \ref{def1}, $M_{m,n}$ qualifies as a same-attraction function.

The maximum statistic $M_{m,n}$ is more powerful than either $Q_1$ or $Q_2$, which can be further verified by the Proposition \ref{theorem-powerful}. Let $\mathcal{G}_1=M_{m,n}$, $\mathcal{G}_2=Q_1$ or $Q_2$. We observed that $\mathcal{G}_1\geq \mathcal{G}_2$. Moreover, both $M_{m,n}$ and $\mathcal{G}_2$ converge in distribution to $\chi^2_1$ under $H_0$, leading $c_{\alpha,1}=c_{\alpha,2}$, where $P(\chi^2_1>c_{\alpha,1})=\alpha$. Therefore, $\frac{\mathcal{G}_1}{\mathcal{G}_2} \geq \frac{c_{\alpha,1}}{c_{\alpha,2}} =1$, indicating that $\mathcal{G}_1$ is asymptotically more powerful than $\mathcal{G}_2$.
\end{example}

\begin{example}[Weighted average statistic \citep{Shi2023}]
\label{example-weighted}
The weighted average statistic, $W_{m,n} (w_1, w_2)$, is defined as
\begin{equation}\label{wst}
W_{m,n} (w_1, w_2)= w_1Q_1+w_2Q_2,
\end{equation}
where $w_1, w_2 > 0$, $w_1+w_2=1$, and $Q_1, Q_2$ are defined in \eqref{Msta}.

Contrasting with the maximum statistic in Example \ref{example-maximum}, for $r=1,2$ we observe:
\begin{equation*}
\frac{\partial W_{m,n} (w_1, w_2) }{\partial Q_r}=w_r>0.
\end{equation*}
Thus, $W_{m,n} (w_1, w_2)$ qualifies as a strictly same-attraction function.

Despite being strictly same-attraction, $W_{m,n} (w_1, w_2)$ is asymptotically less powerful than $M_{m,n}$. This is due to the fact that $M_{m,n}\geq W_{m,n} (w_1, w_2)$ and that $W_{m,n} (w_1, w_2)$ converges in distribution to $\chi^2_1$.
\end{example}

\vspace{2mm}
\begin{example}[Minimum statistic \citep{Chen23}]
The minimum statistic, $M_{m,n}^*$, is defined as
\begin{equation}\label{mist}
M_{m,n}^*=-\min(Q_1, Q_2),
\end{equation}
where 
$Q_1=\left[\frac{1}{12}(\frac{1}{m}+\frac{1}{n})\right]^{-1/2} (Q( {F}_{m}, {G}_{n})-\frac{1}{2} )$ and 
$Q_2=\left[\frac{1}{12}(\frac{1}{m}+\frac{1}{n})\right]^{-1/2} (Q( {G}_{n}, {F}_{m})-\frac{1}{2} ).$

Both $Q_1$ and $Q_2$ fulfill the selfsame property (P1).
The coordinate (P2) is also met under the alternative hypothesis $H_1$, as for $r=1,2$,
\begin{equation*}
\frac{\partial M_{m,n}^* }{\partial Q_r} =
    \begin{cases}
      -1   & \text{if }  \min( Q_1, Q_2)=Q_1\\
      0 & \text{otherwise }. \\
    \end{cases}       
\end{equation*}

Thus, $M_{m,n}^*$ is classified as a same-attraction. Moreover,   $M_{m,n}^*$ and $M_{m,n}$ have the same asymptotic power. This equivalence is demonstrated by $M_{m,n}^*$ converging in distribution to $|\mathcal{N}(0,1)|$ and $(M_{m,n}^*)^2$ converging to $\chi^2_1$. Detailed explanations and proofs can be found in Appendices \ref{min-proof}, \ref{max=min}.

\end{example}

\vspace{2mm}
\begin{example}[Sum statistic \citep{Chen23,Gnettner2024}]
The sum statistic $S_{m,n}$ was firstly proposed by \cite{Chen23} and later studied by \cite{Gnettner2024}  and is defined as
\begin{equation}\label{sum-def}
S_{m,n}=-\frac{mn}{m+n} (Q(F_m, G_n)+Q(G_n, F_m)-1),
\end{equation}
where $F_m$ and $G_n$ represent the empirical distributions of $F$ and $G$, respectively.

Both $Q( {F}_{m}, {G}_{n})$ and $Q( {G}_{n}, {F}_{m})$ are selfsame. The partial derivatives of $S_{m,n}$ with respective to both $Q(F_m, G_n)$ and $Q(G_n, F_m)$ are $ -\frac{mn}{m+n}$ which are less than zero. Hence, $S_{m,n}$ qualifies as a strictly same-attraction function. 

We note that the rate of convergence of $S_{m,n}$ is very different from that of $Q_1$ or $Q_2$.
Determining the asymptotic distribution of $S_{m,n}$ poses a challenge. \cite{Gnettner2024} derived an asymptotic approximation of $\chi^2_1$ for the Halfspace depth in one-dimensional Euclidean space. Consequently, based on a specific case of one-dimensional Euclidean depth \citep{liu1993quality,Kosiorowski and Zawadzki 2017}, we set $c_{\alpha, 1} =1.6566$, where $P(S_{m,n} > c_{\alpha, 1}) \rightarrow0.05$. 
This leads to the conclusion that the sum statistic $S_{m,n}$ is asymptotically more powerful than the maximum statistic $M_{m,n}$ in certain scenarios. For instance, consider the condition: 
  \begin{equation}\label{CS}
      Q(F_m, G_n)+Q(G_n, F_m) \leq q^+,\quad (0\leq q^+<1).
  \end{equation} 
This condition often applies in cases of mean shifts. Then,  
\begin{align*}
    \frac{S_{m,n}}{\sqrt{M_{m,n}}}&= \frac{-(\frac{1}{m}+\frac{1}{n})^{-\frac{1}{2}} (Q(F_m, G_n)+Q(G_n, F_m)-1)}{\sqrt{12} \max(|Q(F_m, G_n)-\frac{1}{2}|, |Q(G_n, F_m)-\frac{1}{2}| ) } \\
    & \geq \frac{2 (1-q^+) (\frac{1}{m}+\frac{1}{n})^{-\frac{1}{2}} }{\sqrt{12}},  \\
    & \geq \frac{1.6566}{\sqrt{3.84}}, \text{ as } (\frac{1}{m}+\frac{1}{n})^{-\frac{1}{2}} \rightarrow \infty,
\end{align*}
suggesting that $S_{m,n}$ is asymptotically more powerful than $M_{m,n}$ under the specified condition in \eqref{CS}.  Using an alternative method based on the Hoeffding decomposition \citep{Ustat} and the Cox-Reid expansion \citep{Cox}, we derive a general form of asymptotic distribution capable for all depth functions.
For more details, see Section 2.3.
\end{example}

\vspace{2mm}

\begin{example}[Product statistic \citep{Chen23}]
The product statistic $P_{m,n}$ is defined as 
\begin{equation}\label{prod-def} 
P_{m,n}=-\frac{mn}{m+n} \Big(Q(F_m, G_n) Q(G_n, F_m)-\frac{1}{4} \Big)
\end{equation}

  Consider the partial derivatives of $P_{m,n}$, we have$$\frac{\partial P_{m,n}}{\partial Q(F_m, G_n)}=-\frac{mn}{m+n} Q(G_n, F_m) <0$$ and $$\frac{\partial P_{m,n}}{\partial Q(G_n, F_m)}=-\frac{mn}{m+n} Q(F_m, G_n) <0$$ almost surely as  $Q( {F}_{m}, {G}_{n})$ and $Q( {G}_{n}, {F}_{m})$ are almost surely positive. Hence, $P_{m,n}$ is strictly same-attraction.

The asymptotic distribution of $P_{m,n}$ for univariate Euclidean depth can be obtained in a manner similar to that of $S_{m,n}$ ; detailed explanations are provided in Section 2.3. Setting $\alpha=0.05$ and $c_{\alpha, 1} =0.9384$, we find $P(P_{m,n}>c_{\alpha, 1}) \rightarrow 0.05$. Additionally, under the condition specified in \eqref{CS}, the inequality $Q(F_m, G_n)\times Q(G_n, F_m)\leq (q^+)^2/4<1/4$ holds. Consequently, we derive 
 \begin{align*}
    \frac{P_{m,n}}{\sqrt{M_{m,n}}}&=\frac{-(Q(F_m, G_n) Q(G_n, F_m)-\frac{1}{4}) (\frac{1}{m}+\frac{1}{n})^{-\frac{1}{2}} }{\sqrt{12} \max(|Q(F_m, G_n)-\frac{1}{2}|, |Q(G_n, F_m)-\frac{1}{2}| ) } \\
    & > \frac{2(1/4-(q^+)^2/4)  (\frac{1}{m}+\frac{1}{n})^{-\frac{1}{2}} }{ \sqrt{12} }  \\
   & \geq \frac{0.9384}{\sqrt{3.84}}, \text{ as } (\frac{1}{m}+\frac{1}{n})^{-\frac{1}{2}} \rightarrow \infty,
\end{align*}
indicating that $P_{m,n}$ is asymptotically more powerful than $M_{m,n}$ under condition in \eqref{CS}. It is also noteworthy that $S_{m,n}$ and $P_{m,n}$ are comparable, as they have similar asymptotic distributions.
\end{example}

Moreover, we could visualize the rejection region through figures. 
The Figure \ref{RR} illustrates the rejection region of $Q(F_m,G_n)$, $Q(G_n, F_m)$, $M_{m,n}$, $S_{m,n}$, and $P_{m,n}$, respectively under univariate Euclidean depth.
We conducted 1000 simulations with $m=n=100$. 
Purple dots are null $Q$ statistics under $F=G=\mathcal{N}(0,1)$; blue triangles are alternative $Q$ statistics under  $F=\mathcal{N}(0,1)$ and $G=\mathcal{N}(0.8,1.2)$.  The rejection regions are shaded with colored lines. For a larger detailed version of Figure \ref{RR}, see Appendix \ref{fig1-appendix}.

\begin{figure}
\centering
\includegraphics[width=0.195\linewidth]{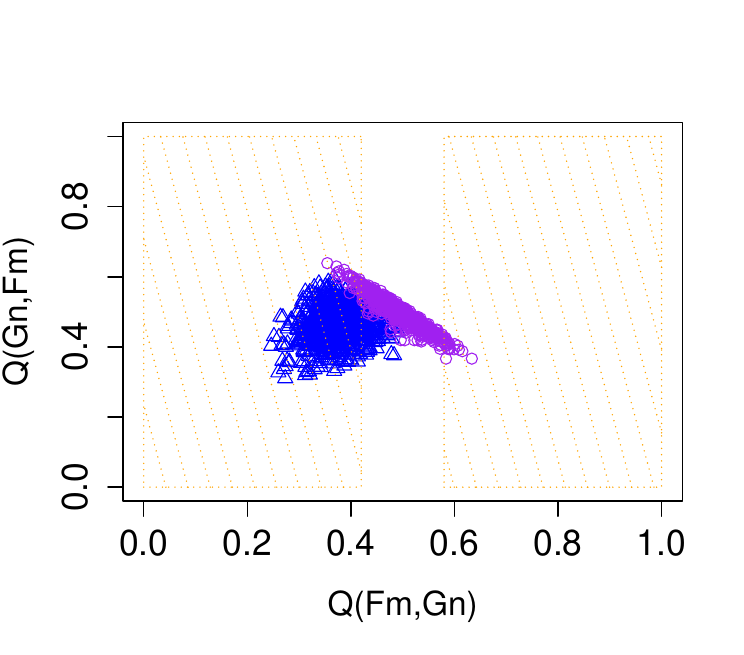}
\includegraphics[width=0.195\linewidth]{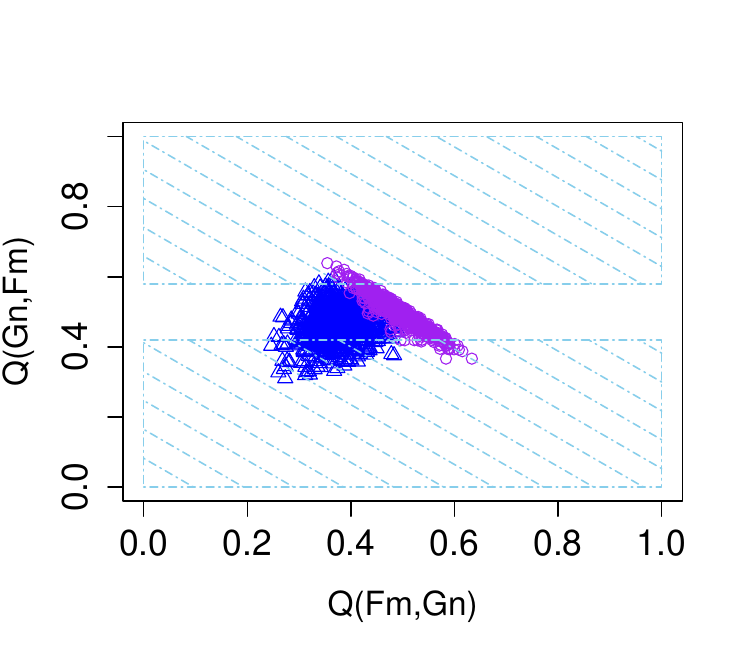}
\includegraphics[width=0.195\linewidth]{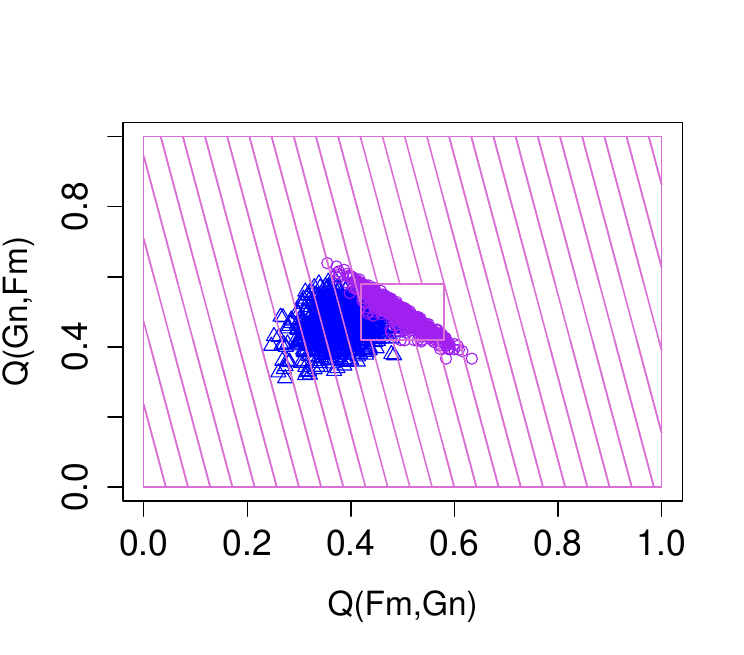}
\includegraphics[width=0.195\linewidth]{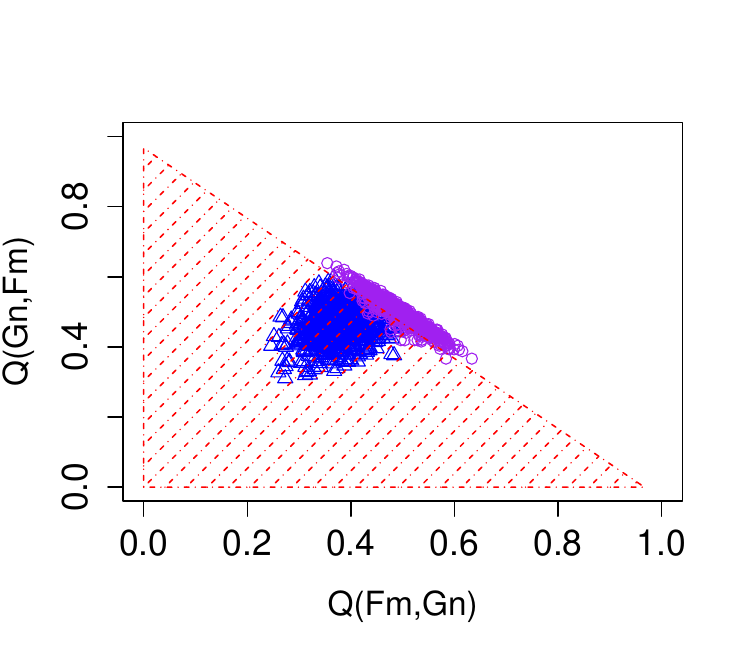}
\includegraphics[width=0.195\linewidth]{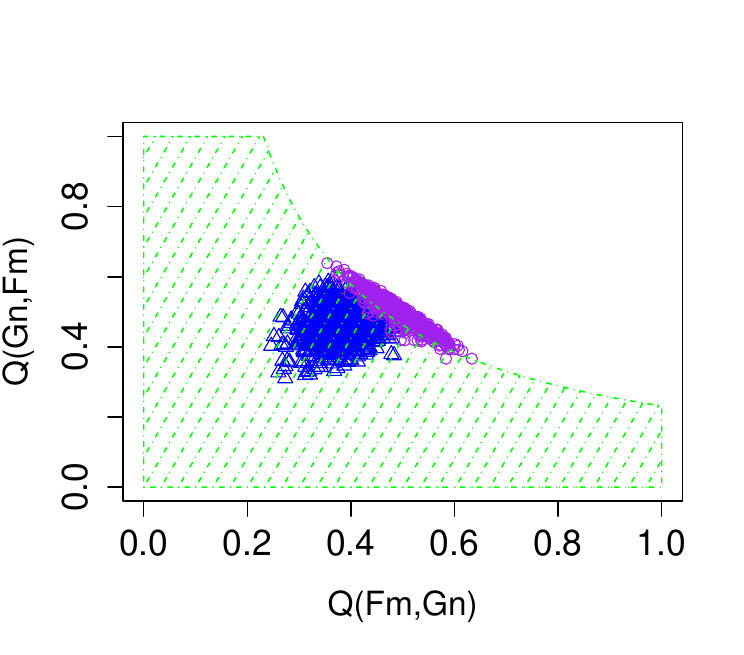}
\caption{Plots of $Q(F_m,G_n)$ and $Q(G_n, F_m)$, respectively for the univariate Euclidean depth under the null hypothesis $F=G=\mathcal{N}(0,1)$ (purple dots) and the alternative hypothesis $F=\mathcal{N}(0,1)$ and $G=\mathcal{N}(0.8,1.2)$ (blue triangles) corresponding to each shaded rejection region based on $Q(F_m,G_n)$, $Q(G_n, F_m)$, $M_{m,n}$, $S_{m,n}$, and $P_{m,n}$ from the left panel to the right panel, respectively.}
\label{RR}
\end{figure}

To determine if a test is good, we need to control for both types of errors. In other words, the shaded area should not include the null $Q$ statistics (purple) for reducing  the type I error, but should include almost all of the alternative $Q$ statistics (blue) for decreasing the type II error or increasing the power.
As shown from Figure \ref{RR}, it turns out that $S_{m,n}$ and $P_{m,n}$ have almost the same performance and they all outperform others except  $M_{m,n}$, but there is a type I error control problem for $M_{m,n}$.

  In the subsequent section, we introduce a permutation algorithm, which is non-parametric  and  can  be applied to other depth measures such as Mahalanobis depth \citep{liu1993quality}, Halfspace depth \citep{zuo2006limiting},  Spatial depth \citep{Brown58, Gower74}, and Projection depth \citep{Liu92}.

\subsection{Strategic Block Permutation Algorithm}

Acknowledging the permutation tests are inherently time-intensive,  we employ the Strategic Block Permutation Algorithm \citep{Welch 1990,Chung 2016}. Initially, the raw test statistic $T$ such as $S_{m,n}$ and $P_{m,n}$ is calculated from the empirical distributions $F_m$ and $G_n$.
We divide all samples in $F_m$ into $b_1$ blocks of size $s$, i.e., $b_1=\frac{m}{s}$, and all samples in $G_n$ into $b_2$ blocks, i.e., $b_2=\frac{n}{s}$. Assume that the total number of blocks for all samples $x_1, \dots, x_m, y_1, \dots, y_n$ is $N=b_1+b_2$. Combining all $N$ sample blocks together denotes all sample blocks as $Z=(Z_1, \dots, Z_{b_1}, Z_{b_1+1}, \dots, Z_{b_1+b_2} )$, where the first $b_1$ sample blocks come from $F_m$ and the second $b_2$ sample blocks come from $G_n$. Then, by randomizing all the blocks, there is a total of $N!$ permutations, denoting the set of all permutations as $(\pi(1),\dots,\pi(N))$. After randomizing all blocks, we have $\tilde Z=(Z_{\pi(1)},\dots, Z_{\pi(N)})$. Considering the first $b_1$ blocks as $\tilde F_m$ and the next $b_2$ blocks as the $\tilde G_n$, we derive the new test statistic $T^*$ from the $\tilde F_m$ and $\tilde G_n$. By calculating all $T^*$ values, in a one-sided test, the $p$-value is calculated based on the proportion of all $T^*$ for which $T^* >T$.

The pseudo-code for implementing the Strategic Block Permutation Algorithm to compute the $p$-value is as follows:

\begin{algorithm}
\caption{Pseudo-code for the Strategic Block Permutation Algorithm}
\begin{algorithmic}
\State Determine $S^0_{m,n}$ and $P^0_{m,n}$ test statistic values from $F_m$ and $G_n$. 
\State  Set block size $s$ and number of repetitions $\mathcal{C}$. 
\State Calculate the total number of blocks $B=(m+n)/s$. 
\For{$i=1$ to $\mathcal{C}$}
    \State Permute all $B$ blocks, 
    \State Produce  $\tilde F_m$ from the first $b_1$ permuted blocks, 
    \State Produce  $ \tilde G_n$ from remaining blocks, 
    \State  Calculate new $P_{m,n}^*$, $S_{m,n}^*$ values from  $\tilde F_m$ and $\tilde G_n$.
\EndFor
\State Let  $p$-value$_S$=$\sum[I(S_{m,n}^* >  S^0_{m,n} )]/\mathcal{C}$.
\State  Let $p$-value$_P$=$\sum[I(P_{m,n}^* >  P^0_{m,n})]/\mathcal{C}$. 
\State Output $p$-value$_S$ and $p$-value$_P$.
\end{algorithmic}
\end{algorithm}

At a predetermined significance level of $\alpha$, for the Sum statistic, we reject the null hypothesis if $p$-value$_S<\alpha$.
As with the Product statistic, we reject the null hypothesis if $p$-value$_P<\alpha$.

In the subsequent section, we will show the asymptotic distributions of $S_{m,n}$ and $P_{m,n}$ for general multidimensional depth functions.

\subsection{Asymptotic distribution}
 
 For convenience, we denote 
\begin{align}
I(x,y,F)&=I\Big(D(x;F)\leq D(y;F)\Big), \label{I1}\\
I(x,y,F_{m},G_{n})&= I(x,y,F_m) -I(x,y,G_n),\label{I2}\\
 \rho_{s}(y;F_{m},F)&=E_{x}\Big[\Big(D(y;F_{m})-D(x;F_{m})\Big)^{s}\Big|D(x;F)=D(y;F)\Big], s=1,2,3.\label{I3} 
 \end{align}
Inspired by \cite{Gao24}, we rely on the Hoeffding decomposition and the Cox-Reid expansion to study the asymptotic distributions of $S_{m,n}$ and $P_{m,n}$. 
First, we apply the following Hoffding decomposition as follows:
\begin{align}
-\frac{m+n}{mn}S_{mn}=Q(F_{m},G_{n})+Q(G_{n},F_{m})-1=M_{mn1}+M_{mn2}+R_{mn},\label{zo}  
\end{align}
where
$$M_{mn1}=\int\int I(x,y,F_{m},G_{n})dF_{m}(x)dG(y),$$
$$M_{mn2}=\int\int I(x,y,F_{m},G_{n})dF(x)dG_{n}(y),$$
$$R_{mn}=-\frac{m+n}{mn}S_{mn}-M_{mn1}-M_{mn2}.$$
Here $M_{mn1}$ and $M_{mn2}$ are the two main terms, which are conditionally independent given some conditions related to $F_{m}$ and $G_{n}$, and $R_{mn}$ is the higher order term. Meanwhile, we will also apply the Cox-Reid expansion to the conditional probabilities to obtain exact expansions of $M_{mn1}$ and $M_{mn2}$, where
\begin{eqnarray}
\int &&I(x,y,F_{m})dF(x)=F_{D(X;F)}\Big(D(Y;F)\Big)+f_{D(X;F)}\Big(D(Y;F)\Big)\rho_{1}(Y;F_{m},F)\nonumber\\
&&+\frac{\partial}{\partial(D(Y;F))}\Big[f_{D(x;F)}\Big(D(Y;F)\Big)\rho_{2}(Y;F_{m},F)\Big]+O_{p}(m^{-\frac{3}{2}}).  \label{z1}
\end{eqnarray}
Note in particular that we consider the third-order expansion different from the second-order expansion in \cite{Gao24}. Similar to the $P_{mn}$, we observe that 
\begin{eqnarray}
&&-\frac{m+n}{mn}P_{mn}=Q(F_m, G_n)\times Q(G_n, F_m)-\frac{1}{4}\nonumber\\
&=&\Big(Q(F_m, G_n)-\frac{1}{2}\Big)\times \Big(Q(G_n, F_m)-\frac{1}{2}\Big)+\frac{1}{2}\Big(Q(F_m, G_n)+Q(G_n, F_m)-1\Big).\label{z2}
\end{eqnarray}

For the expansion of $-\frac{m+n}{mn}P_{mn}$ in \eqref{z2}, we use Lemma \ref{lem 6} in Appendix E to expand the first term, and for the second term, we use  Appendix F to expand $-\frac{m+n}{mn}S_{m,n}$  in \eqref{zo}.

To construct the complete proof, we list assumptions as follows:
\begin{assumption} \label{ass 1}
$E\Big[\Big(\sup\limits_{x\in R^{d}}|D(x;F_{m})-D(x;F)|\Big)^{\alpha}\Big]=O(m^{-\alpha/2})$. \end{assumption}
\begin{assumption} \label{ass 2}
$E(\Sigma_{i}p_{ix}(F_{m})p_{iy}(F_{m}))=o(\Delta_{m})$  (where $\Delta_{m}\rightarrow 0$) if there exists $c_{i}$ such that $p_{ix}(F_{m})>0$ and
$p_{iy}(F_{m})>0$ for $p_{iZ}(F_{m})=:P(D(Z;F_{m})=c_{i}|F_{m}), i=1,2,\ldots.$
\end{assumption}

\begin{assumption} \label{ass 3}
For $i\neq k$, $j\neq l$, let $\rho_{1}(x_{i};F_{m},F)\perp\rho_{1}(x_{k};F_{m},F)|\Lambda_{m}$ and $\rho_{1}(y_{j};F_{m},F)\perp\rho_{1}(y_{l};F_{m},F)|\Lambda_{m}$, where $\perp$ denotes independence. Assume that
\begin{align}
E\Big[f_{D(y;F)}\Big(D(x;F)\Big)\rho_{1}(x;F_{m},F)\Big|\Lambda_{m}\Big]&=0,\label{r0}\\
E\Big[f_{D(x;F)}\Big(D(y;F)\Big)\rho_{1}(y;F_{m},F)\Big|
\Lambda_{m}\Big]&=0,\label{r1}\\
\frac{1}{n}\sum_{j=1}^{n}\frac{\partial}{\partial(D(y_{j};F))}\Big[f_{D(x;F)}\Big(D(y_{j};F)\Big)\rho_{2}(y_{j};F_{m},F)\Big]&=O_{P}(m^{-3/2}),\label{r2}\\ 
\frac{1}{m}\sum_{i=1}^{m}\frac{\partial}{\partial(D(x_{i};F))}\Big[f_{D(y;F)}\Big(D(x_{i};F)\Big)\rho_{2}(x_{i};F_{m},F)\Big]&=O_{P}(m^{-3/2}),\label{rr2}\\
\sup\limits_{\zeta}\Big\{\Big[\frac{\partial^{2}}{\partial^{2} D(y;F)}\Big(f_{D(x;F)}\Big(D(y;F)\Big)\rho_{3}(y;F_{m},F)\Big)\Big]\Big|_{D(y;F)=\zeta}\Big\}&=O_{P}(m^{-3/2}),\label{r3}\\ 
\sup\limits_{\zeta}\Big\{\Big[\frac{\partial^{2}}{\partial^{2} D(x;F)}\Big(f_{D(y;F)}\Big(D(x;F)\Big)\rho_{3}(x;F_{m},F)\Big)\Big]\Big|_{D(x;F)=\zeta}\Big\}&=O_{P}(m^{-3/2}),\label{rr3} 
\end{align}
where $F_{D(x;F)}(\cdot)$ be the distribution function of $D(x;F)$, $f_{D(x;F)}(\cdot)$ be the derivative of $F_{D(x;F)}(\cdot)$, and $\Lambda_{m}$ is the condition related to $F_{m}$,  which may vary for different depth functions.
\end{assumption}

\begin{assumption} \label{ass 4}
For $i\neq k$ and $j\neq l$, suppose that $E[I(x_{i},y_{j},F_{m},G_{n})|\Lambda_{mn}]=0$ and $I(x_{i},y_{j},F_{m},G_{n})\perp I(x_{k},y_{l},F_{m},G_{n})|\Lambda_{mn}$, where $\Lambda_{mn}$ is the condition related to $F_{m}$ and $G_n$, and it may vary for different depth functions. 
\end{assumption}
\begin{assumption} \label{ass 5} 
Under alternative hypothesis, $F\neq G$, such that $||\bm{\theta}(F)-\bm{\theta}(G)||\neq0$, with $\bm{\theta}(F)$ and $\bm{\theta}(G)$ are the parameters of the distributions $F$ and $G$, respectively, we further suppose $-[Q(F_m,G_n)+Q(G_n,F_m)-1]$ and $-[Q(F_m,G_n)Q(G_n,F_m)-1/4]$ can be approximated by $q(||\bm{\theta}(F)-\bm{\theta}(G)||)+o_p(1)$, where $q(x)$ is a monotonically increasing function of $x$.
\end{assumption}

\begin{remark} \label{rem 1}
Assumptions  \ref{ass 1} and \ref{ass 2} origin from \cite{zuo2006limiting}, but we require $\alpha=4$ for the third-order expansion.   
Assumption \ref{ass 3} relates to the Cox-Reid expansion, while Assumption \ref{ass 4} relates to the Hoeffding decomposition; see more details in \cite{Gao24}. 
Assumptions  \ref{ass 1}-\ref{ass 4} apply to a wide range of depths, such as Euclidean depth, Mahalanobis  depth, Halfspace depth, Projection depth, and Spatial depth.  
See Appendix \ref{validcation-assumption1-4} for these verifications. 
\end{remark}

\begin{remark} \label{rem 3}
Assumption \ref{ass 5} ensures that the power can be obtained under the alternative hypothesis.  
For the verification of Assumption \ref{ass 5}, see Appendix \ref{validation-assumption5}.

\end{remark}

\begin{theorem}\label{the 1}
 Under the null hypothesis $F=G$ and ${m}/{n}\rightarrow c>0$ ($c$ is a constant), then under Assumptions  \ref{ass 1}-\ref{ass 4}, it has
\begin{eqnarray}
S_{m, n}
&=&-\frac{m n}{m+n}\Big\{-\frac{1}{m}\sum_{i=1}^{m}f_{D(y;F)}\Big(D(x_{i};F)\Big)\Big(\rho_{1}(x_{i};F_{m},F)-\rho_{1}(x_{i};G_{n},F)\Big)\nonumber\\
&&+\frac{1}{n}\sum_{j=1}^{n}f_{D(x;F)}\Big(D(y_{j};F)\Big)\Big(\rho_{1}(y_{j};F_{m},F)-\rho_{1}(y_{j};G_{n},F)\Big)\Big\}\nonumber\\
&&+o\Big(\frac{m n}{m+n}\Delta_{m}\Big)+O_{P}(m^{-1/2}),\label{s1}
\end{eqnarray}
and
\begin{eqnarray}
P_{m, n}&=&\frac{m n}{m+n}\Big\{\Big[\frac{1}{n}\sum\limits_{j=1}^{n}\Big(F_{D(x;F)}\Big(D(y_{j};F)\Big)-\frac{1}{2}\Big)+\frac{1}{m}\sum\limits_{i=1}^{m}\Big(\frac{1}{2}-F_{D(y;F)}\Big(D(x_{i};F)\Big)\Big)\Big]^2\Big\}\nonumber\\
&&-\frac{mn}{2(m+n)}\Big\{-\frac{1}{m}\sum_{i=1}^{m}f_{D(y;F)}\Big(D(x_{i};F)\Big)\Big(\rho_{1}(x_{i};F_{m},F)-\rho_{1}(x_{i};G_{n},F)\Big)\nonumber\\
&&+\frac{1}{n}\sum_{j=1}^{n}f_{D(x;F)}\Big(D(y_{j};F)\Big)\Big(\rho_{1}(y_{j};F_{m},F)-\rho_{1}(y_{j};G_{n},F)\Big)\Big\}\nonumber\\
&&+o\Big(\frac{m n}{m+n}\Delta_{m}\Big)+O_{P}(m^{-1/2}). \label{s2}
\end{eqnarray}
\end{theorem}

The proof of Theorem \ref{the 1} is presented in Appendix \ref{proof-thm1}, as well as the relevant lemmas in Appendix \ref{lemma-proof}.

\begin{remark}  \label{theorem1}
  Under the null hypothesis $F=G$, the asymptotic distribution of $S_{m,n}$ and $P_{m,n}$ in one-dimensional Euclidean depth \citep{liu1993quality,Kosiorowski and Zawadzki 2017} follows the related Craig distribution \citep{Craig}: 
\begin{eqnarray}
S_{m,n} \rightarrow   -Z_1 Z_2, \label{s3} 
\end{eqnarray}
\begin{eqnarray}
P_{m,n} \rightarrow  Z_3^2-\frac{1}{2} Z_1 Z_2,\label{s4}   
\end{eqnarray}
where $Z_1 \sim \mathcal{N}(0,1)$,  $Z_2 \sim \mathcal{N}(0,\frac{2}{\sqrt{3} \pi})$,   $Z_3 \sim \mathcal{N}(0,\frac{1}{12})$,  $Cov(Z_1, Z_2)=-\frac{1}{\pi}$, and $Z_3$ is independent of $Z_1$ and $Z_2$. 

The proof of Remark \ref{theorem1} can be found in Appendix \ref{euclidean-proof}. Moreover, in the Appendix \ref{remark3-density}, we compare the approximated density and the simulated density. It turns out that the approximation is very accurate.
\end{remark}

For the power of permutation test,  we have the following Theorem \ref{theorem2}.

\begin{theorem} \label{theorem2}
Assume Assumption \ref{ass 5} holds, under the alternative hypothesis, the power of the permuted Sum or Product tests at the significance level $\alpha$ approaches 1 as both the block size $s$ and the number of repetitions $\mathcal{C}$ in the Strategic Block Permutation Algorithm go to infinity. 
\end{theorem}

The proof of Theorem 2 is presented in Appendix \ref{thm2-proof}.

\section{Simulation studies: two-sample test}

We consider power comparisons between the Sum statistic $S_{m,n}$ \eqref{prod-def} and the Product statistic $P_{m,n}$ \eqref{sum-def} alongside of other existing methods: $M_{m,n}$ \eqref{Msta}, $M_{m,n}^*$ \eqref{mist}, Depth-Based Rank Statistics (DbR) \citep{Small2011}, and Modified Depth-Based Rank Statistics (BDbR) \citep{Barale&Shirke} within the context of multivariate data, focusing on multivariate distributions and employing various depth functions. There are various possible depth functions, here we present simulations on three popular depth functions: Mahalanobis depth, spatial depth and projection depth. Under the null hypothesis, we assume $F=G=N(\bm{0},I_{2\times2})$, where $N(\bm{0},I_{2\times2})$ denotes the bivariate normal distribution with a mean vector $\bm{0}$ and a two-by-two identity covariance matrix $I_{2\times 2}$. Our objective is to assess the power of different test statistics under three distinct scenarios: changes in scale, mean, and both scale and mean. 

We set the significance level at $\alpha=0.05$. For $M_{m,n}$, $M_{m,n}^*$, DbR and BDbR, critical values were determined as the upper 95\% quantiles from 1000 replications under the null hypothesis ($F =G= \mathcal{N}(0,1)$). Power was then calculated as the proportion of instances across 1000 repetitions where the test statistic exceeded these critical values. For permutation tests based on $P_{m,n}$ and $S_{m,n}$ using the Strategic Block Permutation algorithm, we set the threshold for the $p$-value to be the lower 5\% quantile of the simulated 1000 $p$-values under the null hypotheses, with the number of repetitions $\mathcal{C}=200$ and block size $s=25$. The power of the permutation test is then calculated using the proportion of times in 1000 repetitions that the statistic is less than the lower 5\% quantile.

\begin{figure}[H]
\begin{center}
\includegraphics[width=0.75\textwidth]{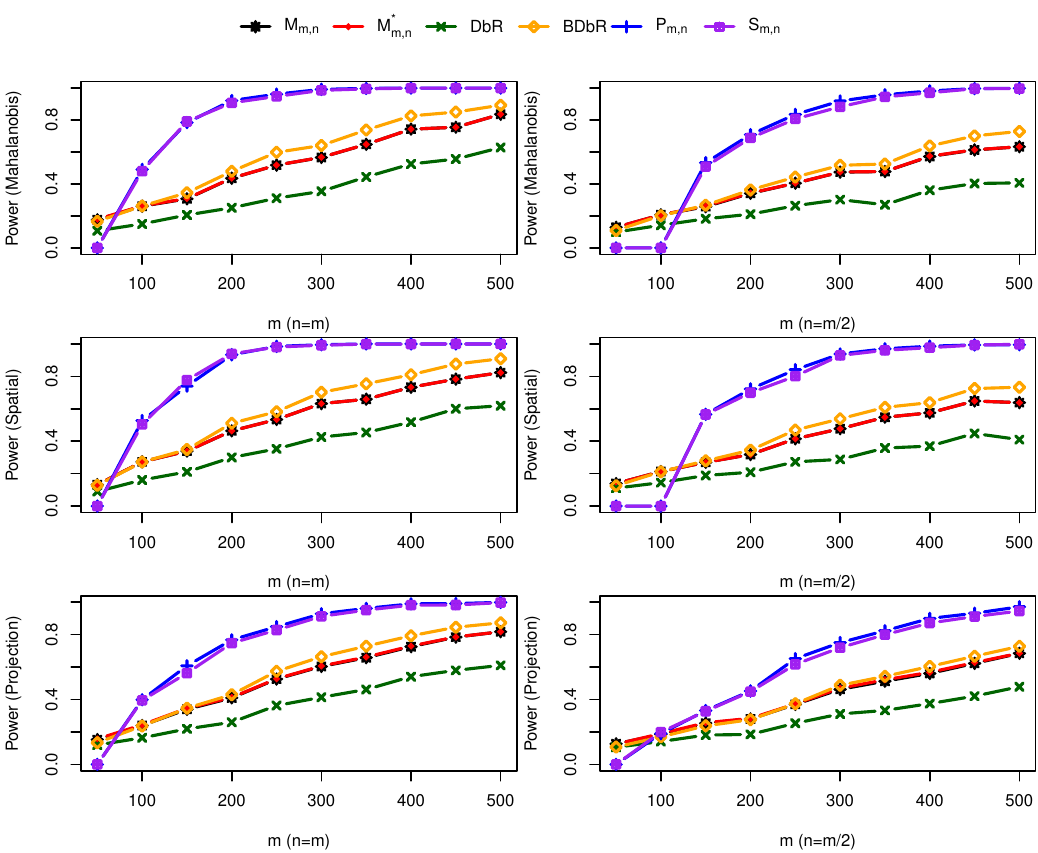}
\caption{Power comparison under alternative hypothesis $F=N(\bm{0},I_{2\times2})$ against $G=N(\bm{0},I_{2\times2}+0.5\tilde{I}_{2\times2})$ for $m$= 50, 100, $\ldots$, 500 and $n=m$ (1st column) or $n=m/2$ (2nd column) for Mahalanobis depth (Row 1), Spatial depth (Row 2), and Projection depth (Row 3). } 
\label{fig:power1}
\end{center}
\end{figure}

For the case of scale change, we consider the distributions $F=N(\textbf{0},I_{2\times2})$ and $G=N(\textbf{0},I_{2\times2}+0.5\tilde{I}_{2\times2})$, where $\tilde{I}_{2\times2 }=((0,1)^\top, (1,0)^\top)$. Power comparisons, depicted in Figure \ref{fig:power1}, are based on Mahalanobis depth, spatial depth and projection depth, for each scenario. It is evident from the Figure \ref{fig:power1} that all three depth functions exhibit similar trends in the power of the various test statistics under examination. Notably, $P_{m,n}$ and $S_{m,n}$ are comparable and outperform all other tested statistics across all depth functions, attributed to their efficacy in detecting scale changes. Additionally, $M_{m,n}$, $M_{m,n}^*$, and BDbR test statistics show similar levels of performance.

\begin{figure}[H]
\begin{center}
\includegraphics[width=0.75\textwidth]{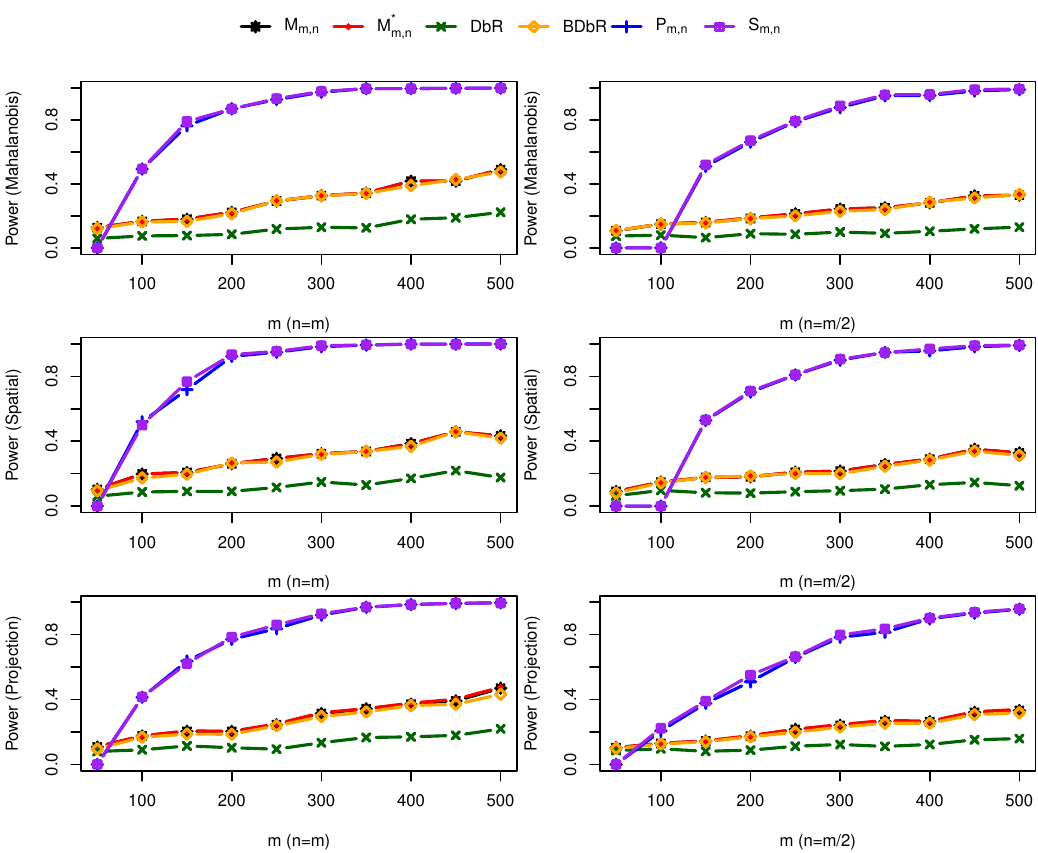}
\caption{Power comparison under alternative hypothesis $F=N( \bm{0},I_{2\times2})$ against $G=N((0.3,0.3)^\top, I_{2\times2})$ for $m= 50, 100, \ldots, 500$ and $n=m$ (1st column) or $n=m/2$ (2nd column) for Mahalanobis depth (Row 1), Spatial depth (Row 2), and Projection depth (Row 3).}
\label{fig:power2}
\end{center}
\end{figure}

In assessing the performance of a change in mean on test statistic power, we consider two specific distributions: $F=N(\textbf{0},I_{2\times2})$ and $G=N((0.3,0.3)^\top, I_{2\times2})$. Power comparisons for this setting are illustrated in Figure \ref{fig:power2}. The results reveal that $P_{m,n}$ and $S_{m,n}$ statistics not only demonstrate comparable performance to each other but also consistently outperform the other tested statistics.

\begin{figure} 
\begin{center}
\includegraphics[width=0.75\textwidth]{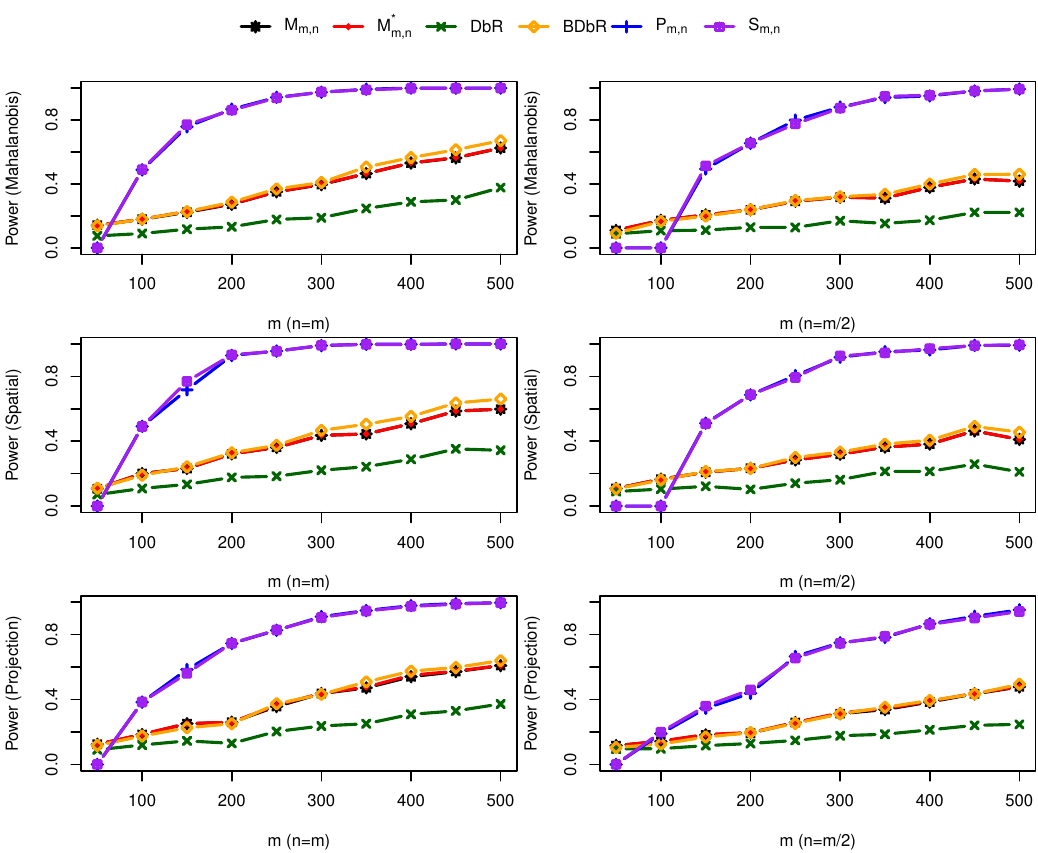}
\caption{Power comparison under alternative hypothesis $F=N( \bm{0},I_{2\times2})$ against $G=N((0.2,0.2)^\top, I_{2\times2}+0.4\tilde{I}_{2\times2})$ for $m= 50, 100, \ldots, 500$ and $n=m$ (1st column) or $n=m/2$ (2nd column) for Mahalanobis depth (Row 1), Spatial depth (Row 2), and Projection depth (Row 3).}
\label{fig:power3}
\end{center}
\end{figure}

In the scenario where both mean and scale change, we consider the distributions $F=N(\textbf{0},I_{2\times2})$ and $G=N((0.2,0.2)^\top, I_{2\times2}+0.4\tilde{I}_{2\times2})$. The power comparisons, as shown in Figure \ref{fig:power3}, further validate the superior performance of the $P_{m,n}$ and $S_{m,n}$ statistics in this multivariate setting.

In conclusion, both $P_{m,n}$ and $S_{m,n}$ demonstrate comparable and notably high efficacy across various multivariate depth functions. These statistics have proven to be promising tools, particularly in scenarios involving changes in mean, scale, or both, within multivariate distributions. Their consistent performance across different scenarios highlights their potential as versatile and robust choices for statistical testing in multivariate analyses.

\section{Raman spectrum}

\begin{figure} 
\centering
    \includegraphics[width=\linewidth]{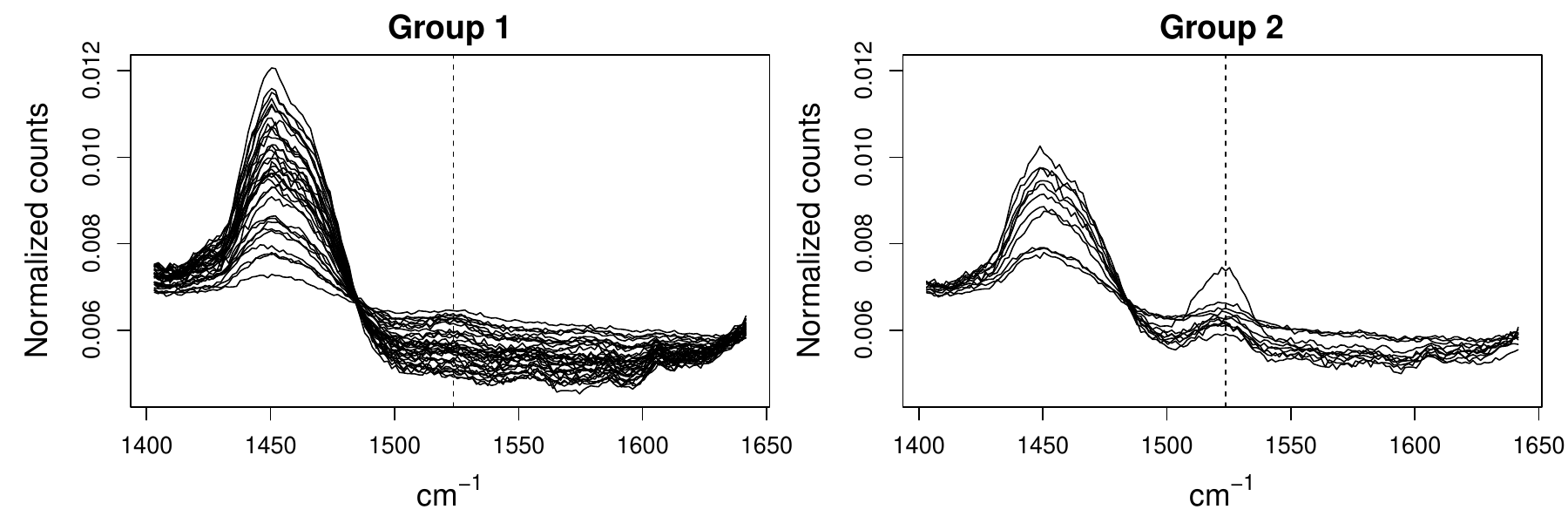}
\caption{Plots of normalized intensity vs wavenumber of all the spectra in Group 1 (left) and Group 2 (right) with a possible peak at 1524 $cm^{-1}$.
}
\label{fig:spectral_peak}
\end{figure}

As introduced earlier in the Introduction section, the classification of different shapes in a range of spectra is crucial in the context of prostate cancer. 
Our dataset comprised 48 spectra, 46 of which were usable (two were discarded due to detector saturation), each containing 1019 wavenumbers.
The dataset includes a column of wavenumber values, ranging from 147 $cm^{-1}$ to 1870 $cm^{-1}$, alongside a corresponding column of intensity values measured in counts. Our primary focus is on the spectral shapes that exhibit a peak at 1524 $cm^{-1}$, a wavenumber indicative of carotenoids presence, which is a crucial biomarker for the diagnosis and prognosis of prostate cancer \citep{Movasaghi 2007}. 
For the analysis, we concentrated on the wavenumber range from 1503 $cm^{-1}$ to 1544 $cm^{-1}$, centered at 1524 $cm^{-1}$. This range encompasses 27 wavenumber values, referred to as the data dimension $d$. Our approach involved a two-stage classification process. In the first stage, we employed a linear regression model to fit the quadratic values to the spectral intensities, using the R-squared value as a measure of fit. A spectrum with a peak should resemble a quadratic curve, as illustrated in Figure \ref{fig:spectral_peak}. Spectra were then categorized into two groups based on an R-squared threshold of 0.5. Specifically, the quadratic function used was $(x-x_0)^2/27^2$, where $x$ ranges from 1 to 27 and $x_0$ is the central point at 14. The R-squared value determines the fit of this quadratic model to the index values of $x$, with a threshold of 0.5 employed to distinguish between two possible peak types. Spectra with R-squared values below 0.5 were classified into Group 1, while those with higher values were categorized into Group 2. In total, 35 spectra were assigned to Group 1 and 11 spectra to Group 2.

In the second stage of our analysis, aimed at more accurately distinguishing between the two spectral shapes, we conducted a two-sample test considering various dimensions. Initially, based on the classifications from the first stage, we focused on the central 27 points around 1524 $cm^{-1}$, denoted as $27M$. To assess the impact of smaller dimensions on test power, we also considered 15 points in the middle ($15M$), 5 points to the left of the center ($5L$), and 5 points to the right of the center ($5R$). We consider different dimensions to examine their effect as collecting more counts in raw data asks for additional costs and requires a larger budget.

For each dimension, we computed the $p$-values for $S_{m,n}$ and $P_{m,n}$  with a block size $s=2$ and repetition number $\mathcal{C}=1000$ , and compared these with  $M_{m,n}$, $M_{m,n}^*$, DbR, and BDbR as shown in the Table \ref{p-value}. It is important to note that in larger dimensions such $27M$ and $15M$, Mahalanobis and Spatial depths were not applicable due to the non-invertibility of the sample covariance matrix, and thus are omitted from these comparisons.

Additionally, we investigated the effects of logarithmic transformation on the depth measures. To distinguish between the original and log-transformed depths, we added an ``O" suffix for the original depths and an ``L" suffix for log-transformed depths in our notation. For example, ``MO" signifies Mahalanobis depth applied to the original counts, while ``ML" refers to Mahalanobis depth on log-transformed counts. Using significance level $\alpha=0.1$ and comparing the resulting $p$-values, we observed that $P_{m,n}$, $S_{m,n}$, and BDbR consistently ranked among the top three, with projected depths showing greater dominance in higher dimensions. Conversely, Mahalanobis depths were more influential in lower dimensions. We also noted that logarithmic transformations tended to slightly improve consistency in inference.

\begin{table} 
\centering
\caption{$p$-values of $P_{m,n}$, $S_{m,n}$, $M_{m,n}$, $M_{m,n}^*$, DbR, BDbR under different depth, $d$, and data transformation of counts.}
\begin{adjustbox}{width=1\textwidth}
  \begin{tabular}{lcccccccccccccccc}
    \toprule
   $d$&
      \multicolumn{2}{c}{27M} &
      \multicolumn{2}{c}{15M} &
      \multicolumn{6}{c}{5L}&\multicolumn{6}{c}{5R} \\ 
      \cmidrule(r){2-3} \cmidrule(r){4-5} \cmidrule(r){6-11} \cmidrule(r){12-17}
      Depth & PO & PL  & PO & PL & MO & ML & SO & SL & PO & PL & MO & ML & SO & SL & PO & PL \\
      $P_{m,n}$& 0.006 & 0  & 0.004 & 0.010 & 0.079 & 0.087 & 0.086 &{\bf 0.129} & 0.054 &{\bf 0.293} & 0.051 & 0.051 & 0.037 & 0.057 & 0.045 & {\bf 0.221}\\

$S_{m,n}$& 0.006 & 0 & 0.004 & 0.006 & 0.018 & 0.086 & 0.023 & {\bf 0.109} &0.028 & {\bf 0.240} & 0.051 & {\bf 0.246} & 0.070 & {\bf 0.345} & 0.027 & {\bf 0.224}\\

$M_{m,n}$& 0.008& 0.045  & 0.021 & 0.031 &{\bf 0.569} &{\bf 0.145} &{\bf 0.558} &{\bf 0.149} &{\bf 0.298} &{\bf 0.767} &{\bf 0.199} & 0.029 &{\bf 0.101} & 0.034 &{\bf 0.417} & 0.071\\

$M_{m,n}^*$& 0.008 & 0.045 & 0.02 & 0.031 &{\bf 0.569} &{\bf 0.145} &{\bf 0.558} &{\bf 0.149} &{\bf 0.294} &{\bf 0.759} &{\bf 0.199} & 0.029 &{\bf 0.101} & 0.034 &{\bf 0.411} & 0.07 \\

DbR& 0.006 & 0.004 & 0.036 & {\bf 0.185} &{\bf 0.384} &{\bf 0.148} &{\bf 0.325} &{\bf 0.133} &{\bf 0.263} &{\bf 0.68} &{\bf 0.244} & 0.044 &{\bf 0.145} & 0.067 &{\bf 0.256} & 0.096\\

BDbR& 0.004 & 0.013 & 0.027 & 0.04 &{\bf 0.613} &{\bf 0.149} &{\bf 0.595} &{\bf 0.137} &{\bf 0.23} &{\bf 0.541} &{\bf 0.163} & 0.03 & 0.077 & 0.038 &{\bf 0.145} & 0.015 \\\bottomrule 
  \end{tabular}
  \end{adjustbox}
  \label{p-value}
\end{table}

\begin{figure} 
\centering
\begin{minipage}{.49\linewidth}
    \includegraphics[width=\linewidth]{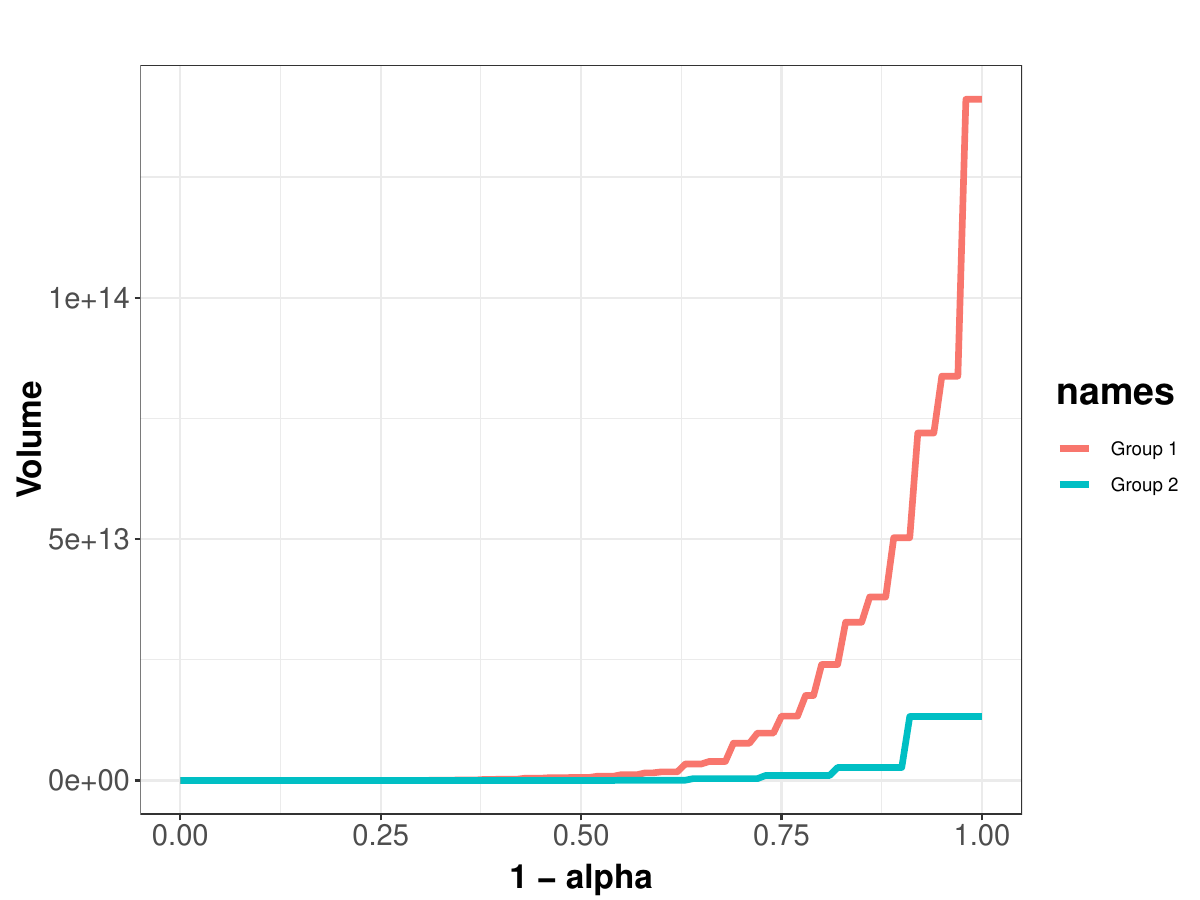}
    \label{fig:scale-maha}
\end{minipage} 
\begin{minipage}{.49\linewidth}
    \includegraphics[width=\linewidth]{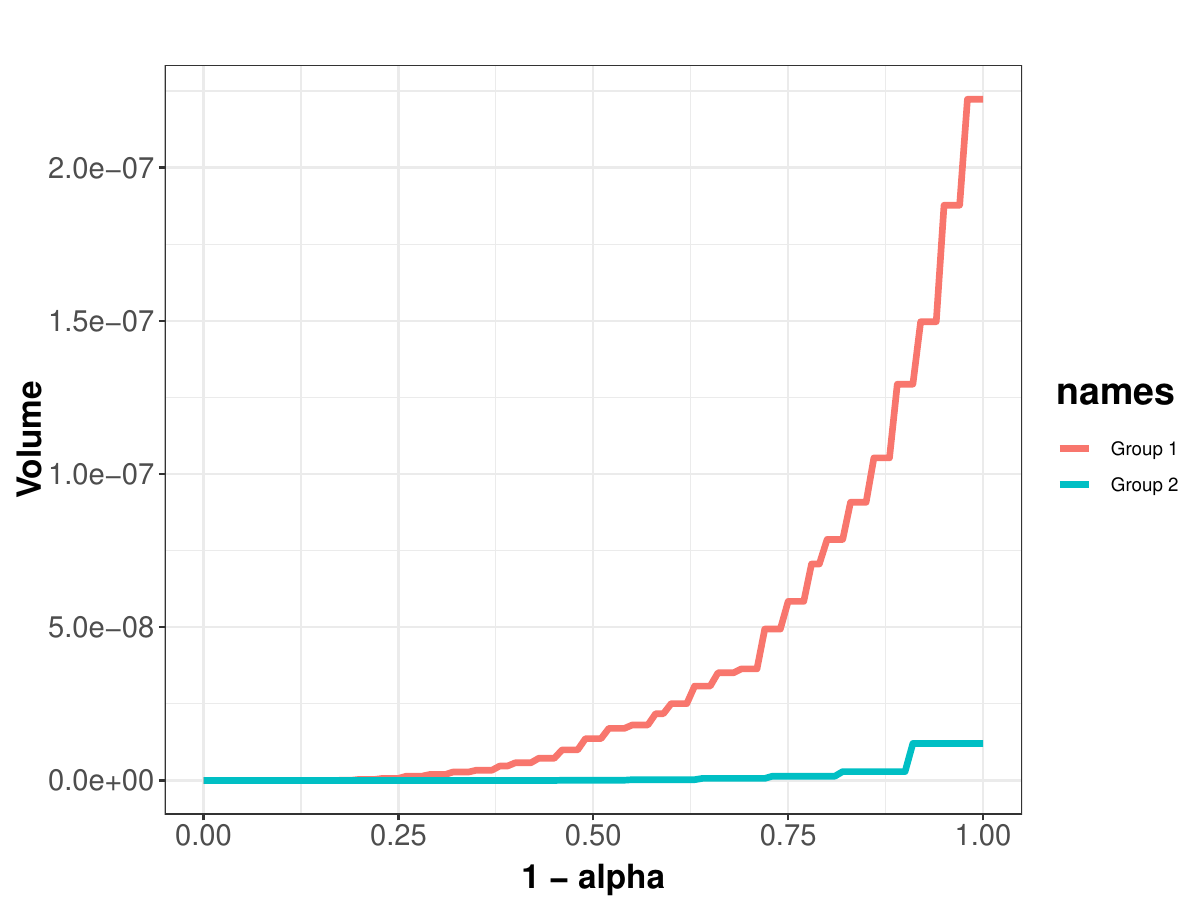}
    \label{fig:scale-maha-log}
\end{minipage}
\caption{Scale curves for  $5L$ derived from Mahalanobis depth to the original intensity values (left) and log-transformed intensity values (right).}
 \label{fig:spectra}
\end{figure}

In addition to the methods previously discussed, we employed the concept of a scale curve, introduced by \citep{Liu99}, to compare the dispersion or scale of two samples. The scale curve quantifies the volume of the $\alpha$-trimmed region of distribution $F$, denoted as $D_\alpha(F)$, which is defined as 
$$D_\alpha (F)=\left\{x\in\mathds{R}^d: D(x; F) \geq \alpha \right \}.$$
Consequently, we plotted the volume of this convex region $V(\alpha; F_m)$ against the $1-\alpha$ scale. Figure \ref{fig:spectra} displays the scale curves derived from the Mahalanobis depth, illustrating both the raw (left) and log-transformed (right) intensity values for the $5L$ spectral dimension. This analysis further validates the differences between the two samples.

Additionally, adjusting the R-squared threshold in the initial spectral classification step results in varying compositions of spectra in each group, as detailed in Table \ref{R^2}. It is noteworthy that the spectra in Group 2 exhibit more complexity compared to those in Group 1, as depicted in Figure \ref{fig:spectral_peak}. For instance, setting the R-squared threshold at 0.4 yields small 
$p$-values for all test statistics (as seen in Table \ref{p-value-0.4}), suggesting a significant difference between the two groups. This significance is attributed to the reclassification of some spectra from Group 1 into Group 2, leading to a smaller yet still discernible difference. Conversely, when the threshold is increased to 0.6, most 
$p$-values become larger (refer to Table \ref{p-value-0.6}), indicating that the differences between the groups are not statistically significant. This shift results from some spectra originally in Group 2 being categorized into Group 1, further narrowing the differences.

\begin{table}
\centering
\caption{ Classification   of 46 spectra into 2 groups with different R-squared threshold values (1: Group 1; and 2: Group 2)}
\begin{adjustbox}{width=1\textwidth}
\begin{tabular}{ |p{0.5cm}|p{0.2cm}|p{0.2cm}|p{0.2cm}|p{0.2cm}|p{0.2cm}|p{0.2cm}|p{0.2cm}|p{0.2cm}|p{0.2cm}|p{0.2cm}|p{0.2cm}|p{0.2cm}|p{0.2cm}|p{0.2cm}|p{0.2cm}|p{0.2cm}|p{0.2cm}|p{0.2cm}|p{0.2cm}|p{0.2cm}|p{0.2cm}|p{0.2cm}|p{0.2cm}|} 
\hline 
 $R^2$ & 1 & 2 & 3 & 4 & 5 & 6 & 7 & 8 & 9 & 10 & 11 & 12 & 13 & 14 & 15 & 16 & 17 & 18 & 19 & 20 & 21 & 22 & 23  \\
\hline 
 0.3 & 1 & 1 & 1 & 1 & 1 & 1 & 1 & 1 & 2 & 2 & 1 & 1 & 1 & 1 & 1 & 1 & 2 & 1 & 1 & 1 & 1 & 1 & 2 \\ 
\hline 
 0.4 & 1 & 1 & 1 & 1 & 1 & 1 & 1 & 1 & 2 & 2 & 1 & 1 & 1 & 1 & 1 & 1 & 2 & 1 & 1 & 1 & 1 & 1 & 2 \\ 
\hline 
 0.5 & 1 & 1 & 1 & 1 & 1 & 1 & 1 & 1 & 1 & 2 & 1 & 1 & 1 & 1 & 1 & 1 & 1 & 1 & 1 & 1 & 1 & 1 & 1 \\
\hline 
 0.6 & 1 & 1 & 1 & 1 & 1 & 1 & 1 & 1 & 1 & 2 & 1 & 1 & 1 & 1 & 1 & 1 & 1 & 1 & 1 & 1 & 1 & 1 & 1 \\
\hline 
 0.7 & 1 & 1 & 1 & 1 & 1 & 1 & 1 & 1 & 1 & 2 & 1 & 1 & 1 & 1 & 1 & 1 & 1 & 1 & 1 & 1 & 1 & 1 & 1 \\
\hline 
\end{tabular}
\end{adjustbox}
\vspace{4mm}
\begin{adjustbox}{width=1\textwidth}
\begin{tabular}{ |p{0.5cm}|p{0.2cm}|p{0.2cm}|p{0.2cm}|p{0.2cm}|p{0.2cm}|p{0.2cm}|p{0.2cm}|p{0.2cm}|p{0.2cm}|p{0.2cm}|p{0.2cm}|p{0.2cm}|p{0.2cm}|p{0.2cm}|p{0.2cm}|p{0.2cm}|p{0.2cm}|p{0.2cm}|p{0.2cm}|p{0.2cm}|p{0.2cm}|p{0.2cm}|p{0.2cm}|} 
\hline 
 $R^2$ & 24 & 25 & 26 & 27 & 28 & 29 & 30 & 31 & 32 & 33 & 34 & 35 & 36 & 37 & 38 & 39 & 40 & 41 & 42 & 43 & 44 & 45 & 46 \\
\hline 
 0.3 & 2 & 2 & 2 & 1 & 1 & 1 & 1 & 2 & 2 & 2 & 1 & 2 & 2 & 2 & 1 & 2 & 2 & 1 & 2 & 2 & 2 & 2 & 2 \\ 
\hline 
 0.4 & 2 & 2 & 2 & 1 & 1 & 1 & 1 & 2 & 2 & 1 & 1 & 2 & 2 & 1 & 1 & 2 & 2 & 1 & 1 & 2 & 2 & 2 & 2 \\ 
\hline 
 0.5 & 2 & 2 & 2 & 1 & 1 & 1 & 1 & 2 & 2 & 1 & 1 & 1 & 1 & 1 & 1 & 2 & 2 & 1 & 1 & 1 & 2 & 2 & 2 \\
\hline 
 0.6 & 1 & 2 & 2 & 1 & 1 & 1 & 1 & 2 & 1 & 1 & 1 & 1 & 1 & 1 & 1 & 2 & 2 & 1 & 1 & 1 & 2 & 1 & 2 \\
\hline 
 0.7 & 1 & 1 & 1 & 1 & 1 & 1 & 1 & 2 & 1 & 1 & 1 & 1 & 1 & 1 & 1 & 1 & 1 & 1 & 1 & 1 & 2 & 1 & 2 \\
\hline 
\end{tabular}
\end{adjustbox}
\label{R^2}
\end{table}

\begin{table}
\centering
\caption{$p$-values of $P_{m,n}$, $S_{m,n}$, $M_{m,n}$, $M_{m,n}^*$, DbR, BDbR under different depth, $d$, and data transformation of counts with R-squared threshold 0.4.}
\begin{adjustbox}{width=1\textwidth}
  \begin{tabular}{lcccccccccccccccc}
    \toprule
   $d$&
      \multicolumn{2}{c}{27M} &
      \multicolumn{2}{c}{15M} &
      \multicolumn{6}{c}{5L}&\multicolumn{6}{c}{5R} \\ 
      \cmidrule(r){2-3} \cmidrule(r){4-5} \cmidrule(r){6-11} \cmidrule(r){12-17}
Depth & PO & PL  & PO & PL & MO & ML & SO & SL & PO & PL & MO & ML & SO & SL & PO & PL \\

$P_{m,n}$& 0 & 0 & 0 & 0.001 & 0.01 & 0.001 & 0.009 & 0.007 & 0.001 & 0.004 & 0.008 & 0.002 & 0.01 & 0 & 0.003 & 0.006 \\

$S_{m,n}$& 0 & 0 & 0 & 0.001 & 0.012 & 0 & 0.005 & 0.005 & 0 & 0.002 & 0.003 & 0 & 0.004 & 0.002 & 0.004 & 0.003 \\

$M_{m,n}$& 0 & 0 & 0.001 & 0 & 0.052 & 0.022 & 0.064 & 0.037 & 0 & 0 & {\bf 0.106} & 0 & {\bf 0.19} & 0 & 0.005 & 0.001\\

$M_{m,n}^*$& 0 & 0 & 0.001 & 0 & 0.052 & 0.022 & 0.064 & 0.037 & 0 & 0 & {\bf 0.106} & 0 & {\bf 0.19} & 0 & 0.005 & 0.001\\

DbR& 0.003 & 0.001 & 0.002 & 0.001 & 0.084 & 0.004 & 0.082 & 0.002 & 0.002 & 0 & 0.098 & 0 & 0.089 & 0 & 0.012 & 0.002\\

BDbR& 0 & 0 & 0 & 0.001 & 0.053 & 0.025 & 0.066 & 0.034 & 0 & 0.001 & {\bf 0.116} & 0 & {\bf 0.199} & 0.001 & 0.009 & 0.006 \\
\bottomrule 
  \end{tabular}
  \end{adjustbox}
  \label{p-value-0.4}
\end{table}

\begin{table}
\centering
\caption{$p$-values of $P_{m,n}$, $S_{m,n}$, $M_{m,n}$, $M_{m,n}^*$, DbR, BDbR under different depth, $d$, and data transformation of counts with R-squared threshold 0.6.}
\begin{adjustbox}{width=1\textwidth}
  \begin{tabular}{lcccccccccccccccc}
    \toprule
   $d$&
      \multicolumn{2}{c}{27M} &
      \multicolumn{2}{c}{15M} &
      \multicolumn{6}{c}{5L}&\multicolumn{6}{c}{5R} \\ 
      \cmidrule(r){2-3} \cmidrule(r){4-5} \cmidrule(r){6-11} \cmidrule(r){12-17}
Depth & PO & PL  & PO & PL & MO & ML & SO & SL & PO & PL & MO & ML & SO & SL & PO & PL \\

$P_{m,n}$& 0.009 & 0.001 & 0.017 & {\bf 0.106} & {\bf 0.533} &{\bf 0.286} &{\bf 0.469} &{\bf 0.218} & 0.060 &{\bf 0.525} & 0.050 & {\bf 0.304} & 0.048 &{\bf 0.372} &{\bf 0.292} &{\bf 0.193 } \\

$S_{m,n}$& 0.005 & 0 & 0.003 &{\bf 0.124} &{\bf 0.311} &{\bf 0.171} &{\bf 0.246} &{\bf 0.136} & 0.035 &{\bf 0.421} &{\bf 0.211} & {\bf 0.732} &{\bf 0.322} &{\bf 0.814} &{\bf 0.351} &{\bf 0.273} \\

$M_{m,n}$& 0.055 &{\bf 0.11} &{\bf 0.167} & 0.054 &{\bf 0.783} &{\bf 0.462} &{\bf 0.842} &{\bf 0.456} &{\bf 0.754} &{\bf 0.397} & 0.081 &{\bf 0.217} &{\bf 0.104} &{\bf 0.29} &{\bf 0.429} & 0.009 \\

$M_{m,n}^*$& 0.055 &{\bf 0.11} &{\bf 0.167} & 0.054 &{\bf 0.783} &{\bf 0.462} &{\bf 0.842} &{\bf 0.456} &{\bf 0.754} &{\bf 0.397} & 0.081 &{\bf 0.217} &{\bf 0.104} &{\bf 0.29} &{\bf 0.427} & 0.009 \\

DbR& 0.014 & 0.004 & 0.094 &{\bf 0.332} &{\bf 0.809 }&{\bf 0.456} &{\bf 0.859} &{\bf 0.433} &{\bf 0.292} & 0.071 &{\bf 0.137} & {\bf 0.25} &{\bf 0.208} &{\bf 0.352} &{\bf 0.606} & 0.014 \\

BDbR& 0.013 & 0.002 & 0.013 & 0.061 &{\bf 0.77}&{\bf 0.477} &{\bf 0.847} &{\bf 0.459} &{\bf 0.28} &{\bf 0.69} & 0.082 &{\bf 0.207} &{\bf 0.104} &{\bf 0.277 }&{\bf 0.351} & 0.008 \\
\bottomrule 
  \end{tabular}
  \end{adjustbox}
  \label{p-value-0.6}
\end{table}

The comprehensive analysis conducted on various spectral shapes firmly establishes the efficacy of our proposed statistical method. This method not only effectively differentiates between distinct types of tumor samples but also demonstrates superior performance compared to other competing methodologies. The ability to discern subtle variations in spectral data is crucial in the context of tumor sample classification, and our approach has proven to be a robust tool in this regard.

Through the use of scale curves, depth functions, and strategic permutation testing, our method offers a reasonable and precise means of detecting and categorizing spectral differences. This is particularly vital in the diagnosis and prognosis of conditions like prostate cancer, where accurate identification of biomarkers such as carotenoids is essential. The success of this method in outperforming other statistical techniques underscores its potential as a valuable asset in medical spectral analysis and related fields.

\section{Conclusions and Limitations}

This paper, motivated by the challenge of detecting shapes in spectral data, introduces two novel test statistics,  $S_{m,n}$ and $P_{m,n}$, and proposes the DEEPEAST technique to  infer homogeneity of two samples. We have defined the concept of the same-attraction function, which encompasses our new statistics, and conducted comparative analyses with existing similar statistics. 
The strategic block permutation algorithm was developed to facilitate the comprehensive application of the DEEPEAST technique across various depth functions. Their asymptotic distributions were derived by applying Hoeffding decomposition and Cox-Reid expansion under reasonable  assumptions. Our extensive simulated power comparisons reveal that $P_{m,n}$ and $S_{m,n}$ exhibit superior performance in multivariate distributions. Furthermore, the application of the DEEPEAST technique in spectral sample comparisons conclusively shows that $P_{m,n}$ and $S_{m,n}$ outperform other test statistics in distinguishing differences between two samples of varying dimensions. 

This paper presents significant theoretical findings on the asymptotic distributions of $S_{m,n}$ and $P_{m,n}$. However, in practical applications, results may deviate from these theoretical distributions due to finite sample sizes. To address this, our proposed algorithm offers a method for obtaining approximate $p$-values. Additionally, to assess the performance across different depth functions, we selected three representative and widely used depth functions in our simulation study. The consistently strong performance of $S_{m,n}$ and $P_{m,n}$ across these depth functions highlights the robustness of these tests, making it highly unlikely that they will show inconsistency across all depth functions. It is also important to note that, although some information may be lost when projecting multivariate data into a one-dimensional space, our two novel test statistics are designed to retain a significant portion of that information. The simulation and real data analyses of both test statistics confirm the significant increase in information gain in practice, which is also supported by the theoretical basis. If the recovery of additional information is necessary for specific data analysis, appropriate dimensionality reduction techniques can be applied, see \cite{Small2011}.

Moreover, our research significantly contributes to two-sample testing, opportunities exist to extend this approach to multi-sample testing and apply it to various other fields. This paper creates an intriguing and challenging avenue for the theoretical research, promising to further enhance the use and applicability of our findings in broader statistical contexts.

\section*{Author Contributions}
Y.C., W.L., A.J., and X.S. designed research; Y.C., W.L., M.G., A.J., and X.S. performed research; A.J. and K.M. collected the prostate sample and RS; Y.C., W.L., A.J., and X.S. analyzed data; and Y.C., W.L., M.G., A.J., and X.S. wrote the paper.

\section*{Acknowledgement}
Dr. Shi’s work was   supported by the Natural Sciences and Engineering Research
Council of Canada under Grant RGPIN-2022-03264, the Interior Universities Research Coalition
and the BC Ministry of Health, the NSERC Alliance International Catalyst Grant ALLRP
590341-23, and the University of British Columbia Okanagan (UBC-O) Vice Principal Research in collaboration with UBC-O Irving K. Barber Faculty of Science.

\vspace*{-10pt}

\appendix

\renewcommand{\theequation}{\thesection\arabic{equation}}

\section*{Appendix}

\section{Details of Figure 1} \label{fig1-appendix}

\begin{figure}
\begin{adjustwidth}{-1em}{0em}
  \begin{subfigure}[b]{0.50\textwidth}
    \includegraphics[width=\textwidth]{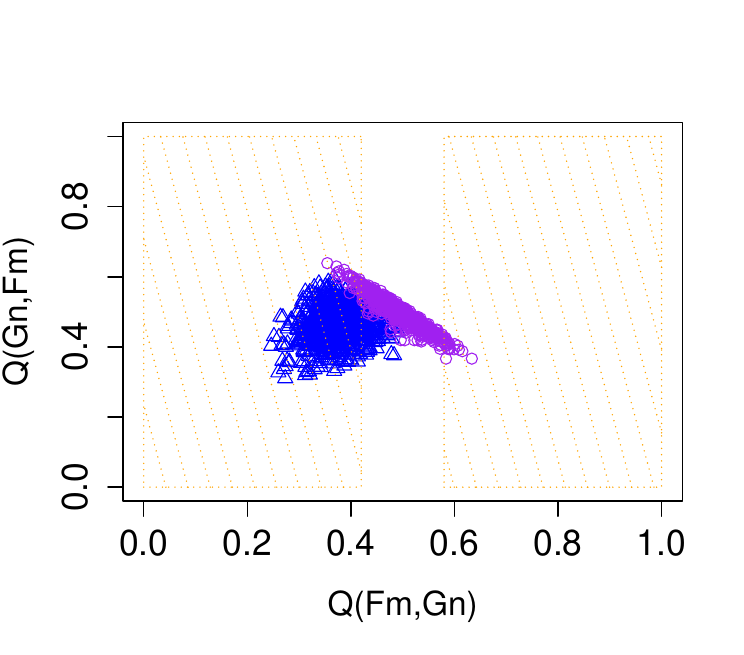}
    \caption{Rejection region of $Q(F_m,G_n)$ }
    \label{Q1}
  \end{subfigure}
  \begin{subfigure}[b]{0.50\textwidth}
    \includegraphics[width=\textwidth]{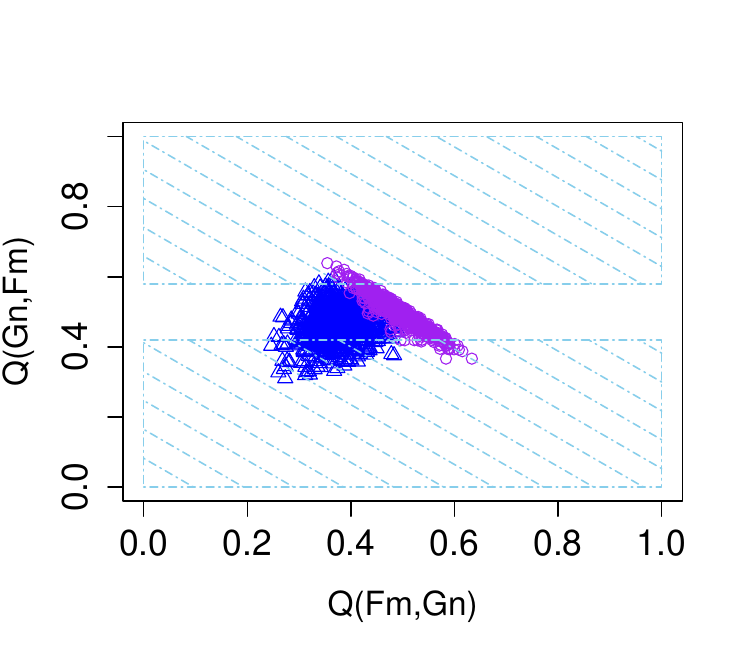}
    \caption{Rejection region of $Q(F_m,G_n)$}
    \label{Q2}
  \end{subfigure}
\end{adjustwidth}

\begin{adjustwidth}{-1em}{0em}
  \begin{subfigure}[b]{0.50\textwidth}
    \includegraphics[width=\textwidth]{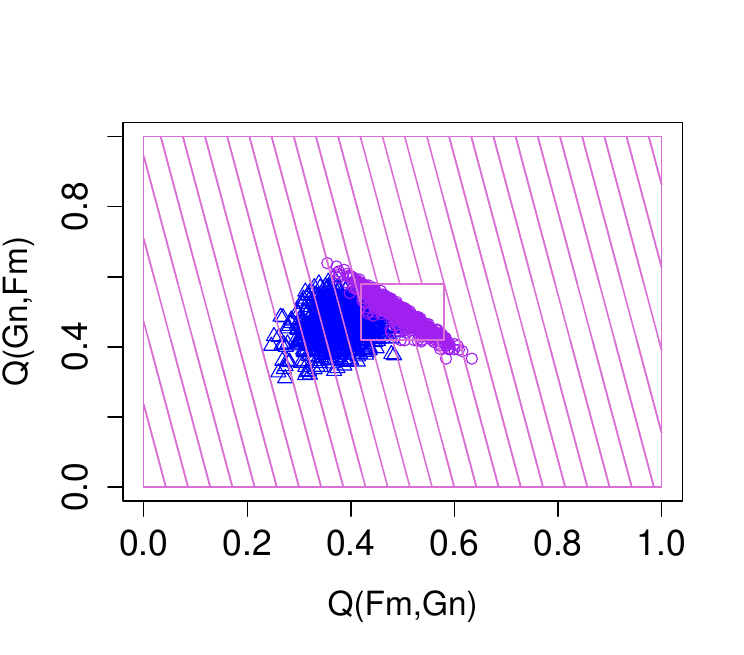}
    \caption{Rejection region of $M_{m,n}$}
    \label{M}
  \end{subfigure}
  \begin{subfigure}[b]{0.50\textwidth}
    \includegraphics[width=\textwidth]{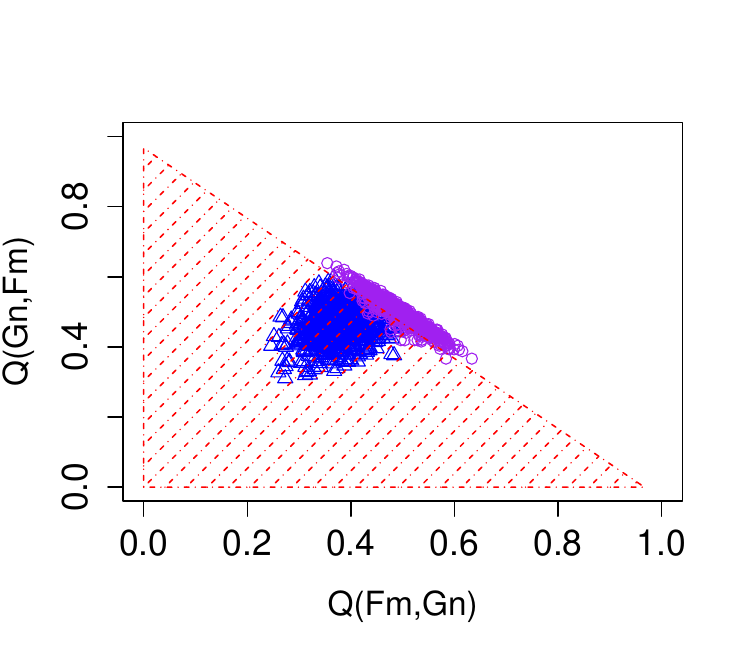}
    \caption{Rejection region of $S_{m,n}$}
    \label{S}
  \end{subfigure}
  
   \begin{subfigure}[b]{0.50\textwidth}
    \includegraphics[width=\textwidth]{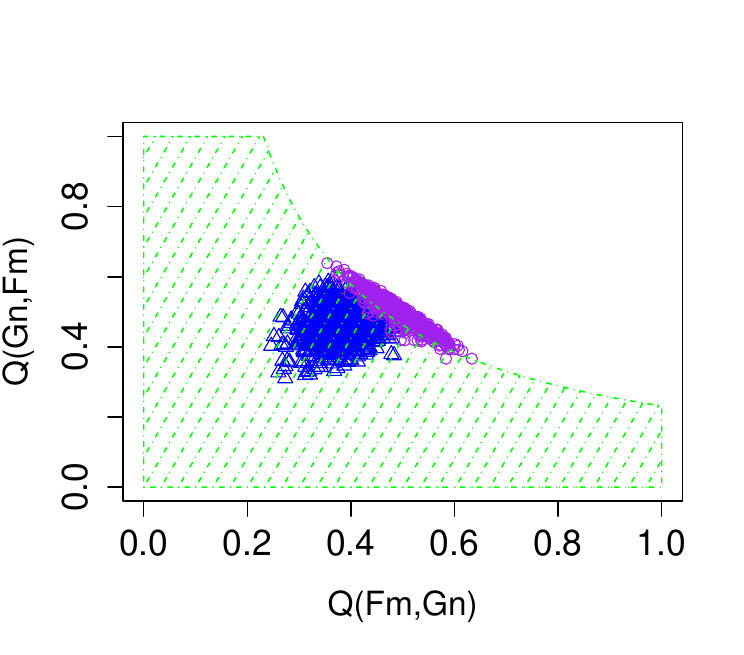}
    \caption{Rejection region of $P_{m,n}$}
    \label{P}
  \end{subfigure}
  
\end{adjustwidth}
\caption{Plots of $Q(F_m,G_n)$ and $Q(G_n, F_m)$, respectively for the univariate Euclidean depth under the null hypothesis $F=G=\mathcal{N}(0,1)$ (purple dots) and the alternative hypothesis $F=\mathcal{N}(0,1)$ and $G=\mathcal{N}(0.8,1.2)$ (blue triangles) corresponding to each shaded rejection region based on $Q(F_m,G_n)$, $Q(G_n, F_m)$, $M_{m,n}$, $S_{m,n}$, and $P_{m,n}$ from (a) to (e), respectively.}
\label{RR}
\end{figure}

\newpage

\section{Proof of Minimum Statistic}
\label{min-proof}
Based on the works of \cite{liu1993quality,zuo2006limiting}, and \cite{Shi2023}, we obtain the following expansion of the $Q$ statistic: $$Q( {G}_n, {F}_m)-1/2 = 1/2-Q( {F}_m, {G}_n)+o_p(n^{-1/2})+o_p(m^{-1/2}).$$ \cite{zuo2006limiting} showed that $$ \left[\frac{1}{12}(\frac{1}{m}+\frac{1}{n})\right]^{-\frac{1}{2}} (Q(F_m, G_n)-\frac{1}{2} ) \xrightarrow d \mathcal{N}(0,1).$$

Therefor, we have
\begin{align*}
&\min(Q(F_m, G_n), Q(G_n, F_m)) \\
= &\min (Q(F_m, G_n)-\frac{1}{2}, Q(G_n, F_m)-\frac{1}{2} )+\frac{1}{2}\\
=& -\left| Q(F_m, G_n)-\frac{1}{2} \right | +\frac{1}{2}, \text{under } H_0 
\end{align*}

Hence,

\begin{align*}
M_{m,n}^*  & = \left[\frac{1}{12}(\frac{1}{m}+\frac{1}{n})\right]^{-\frac{1}{2}} (\frac{1}{2}- \left| Q(F_m, G_n)-\frac{1}{2} \right | - \frac{1}{2}    ) \\
&=\left[\frac{1}{12}(\frac{1}{m}+\frac{1}{n})\right]^{-\frac{1}{2}}  \left| Q(F_m, G_n)-\frac{1}{2} \right |  \xrightarrow d \left | \mathcal{N}(0,1) \right |
\end{align*}

\section{Equivalence of minimum and maximum statistics}
\label{max=min}
Under null hypothesis, $M_{m,n}^*=\left[\frac{1}{12}(\frac{1}{m}+\frac{1}{n})\right]^{-\frac{1}{2}}  \left| Q(F_m, G_n)-\frac{1}{2} \right |$.

Then $(M_{m,n}^*)^2=\left[\frac{1}{12}(\frac{1}{m}+\frac{1}{n})\right]^{-1}  \left( Q(F_m, G_n)-\frac{1}{2} \right ) ^2$.

By \cite{zuo2006limiting}, $ \left[\frac{1}{12}(\frac{1}{m}+\frac{1}{n})\right]^{-\frac{1}{2}} (Q(F_m, G_n)-\frac{1}{2} ) \xrightarrow d \mathcal{N}(0,1),$ which means $ \left[\frac{1}{12}(\frac{1}{m}+\frac{1}{n})\right]^{-1}( (Q(F_m, G_n)-\frac{1}{2} ) )^2 \xrightarrow d \chi^2_1.$

Therefore, $(M_{m,n}^*)^2\xrightarrow d \chi^2_1$.

\section{Verification of Assumption 1-4} \label{validcation-assumption1-4}
Euclidean depth: The univariate Euclidean depth of any point $x\in R$ in a one-dimensional distribution $F$ is
defined as
$$
ED(x;F)=\frac{1}{1+(x-\mu_{F})^2},
$$
where $\mu_{F}$ is the mean of the distribution $F$.    
The empirical version of $ED(X;F)$ is $ED(X;F_{m})=1/(1+(X-\Bar{X})^2)$, where $\Bar{X}=\sum_{i=1}^{m}X_{i}/m$. Thus,
$$
|ED(X;F_{m})-ED(X;F)|^{\alpha}=\Big|\frac{(\Bar{X}-\mu_{F})(2X-\Bar{X}-\mu_{F})}{(1+(X-\Bar{X})^2)(1+(X-\mu_{F})^2)}\Big|^{\alpha}$$
Assume $\Bar{X}$ is  strongly consistent estimators of $\mu_{F}$ and $E(\Bar{X}-\mu_{F})^{\alpha}=O(m^{-\alpha/2})$. By H\"{o}lder's inequality, Assumption  \ref{ass 1} holds.  Assumption \ref{ass 2} holds if the depth is continuous or we require a rate. Similar to \cite{Gao24}, we convert the Euclidean depth to the Euclidean distance $d(x;F)=|x-\mu_{F}|$ as follows:
$$ 
F_{ED(X;F)}\Big(ED(Y;F)\Big)=1-F_{d(X;F)}\Big(d(Y;F)\Big).$$ 
Since $ED(X;F_{m})$ does not depend on the mean, without loss of generality, we further assume that $\mu_{F}=0$. We assume that the density function $f_{d(X;F)}(\cdot)$ is bounded and symmetric about the Y-axis from above. Let the condition $\Lambda_m$ be $\bar X$. Since $f_{d(X;F)}\Big(d(Y;F)\Big)$ is an even function of $Y$ and $\rho_{1}(Y;F_{m},F)|\Lambda_m$ is equal to zero or an odd function of $Y$, so
$$E_Y\Big[f_{d(X;F)}\Big(d(Y;F)\Big)\rho_{1}(Y;F_{m},F)\Big|\Lambda_{m}\Big]=0.$$
Thus, (\ref{r0}) and (\ref{r1}) hold. 

Since $\rho_{2}(Y;F_{m},F)|\Lambda_m$ is an even function of $Y$.  Thus, $\frac{\partial}{\partial(d(y_{j};F))}\Big[f_{d(x;F)}\Big(d(y_{j};F)\Big)\rho_{2}(y_{j};F_{m},F)\Big]$ is an odd function. Then, it has
\begin{eqnarray}
E\Big\{\frac{1}{n}\sum_{j=1}^{n}\frac{\partial}{\partial(d(y_{j};F))}\Big[f_{d(x;F)}\Big(d(y_{j};F)\Big)\rho_{2}(y_{j};F_{m},F)\Big]\Big\}=0.\nonumber
\end{eqnarray}
Furthermore, $\frac{\partial}{\partial(d(y_{j};F))}\Big[f_{d(x;F)}\Big(d(y_{j};F)\Big)\rho_{2}(y_{j};F_{m},F)\Big]$ conditionally independent for $j=1,\cdots,n$ and $\rho_{2}(Y;F_{m},F)=O_p(m^{-1})$, we use Cauchy's theorem to bound the partial derivatives in terms of integrals as well as Cauchy-Schwarz Inequality,
leading to 
\begin{eqnarray}
E\Big\{\frac{1}{n}\sum_{j=1}^{n}\frac{\partial}{\partial(d(y_{j};F))}\Big[f_{d(x;F)}\Big(d(y_{j};F)\Big)\rho_{2}(y_{j};F_{m},F)\Big]\Big\}^2=O(m^{-3}).\nonumber
\end{eqnarray}
Therefore, (\ref{r2}) and (\ref{rr2}) hold. Since $\rho_{3}(Y;F_{m},F)=O_p(m^{-3/2})$, we use Cauchy's theorem to bound the partial derivatives in terms of integrals, where the integrating function is $O_p(m^{-3/2})$, leading to 
$$
\sup\limits_{\xi}\Big\{\frac{\partial^2}{\partial^2d(Y;F)}\Big[f_{d(X;F)}\Big(d(Y;F)\Big)\rho_{3}(Y;F_{m},F)\Big]\Big|_{d(Y;F)=\xi}\Big\}= O_{p}(m^{-3/2})$$ 
Therefore, Assumption \ref{ass 3} holds. To check Assumption \ref{ass 4}, we consider the condition $\Lambda_{mn}$ to be $F_{m}$ and $G_{n}$.  For $i\neq k \in \{1,\cdots,m\}$ and $j \neq l\in \{1,\cdots,n\}$, there is
$$ E\Big\{E\Big[I(X_{i},Y_{j},F_{m}, G_{n})I(X_{k},Y_{l},F_{m},G_{n})\Big|\Lambda_{mn}\Big]\Big\}\nonumber\\
=E\Big\{E\Big[I(X_{i},Y_{j},F_{m},G_{n})\Big|\Lambda_{mn}\Big]E\Big[I(X_{k},Y_{l},F_{m},G_{n})\Big|\Lambda_{mn}\Big]\Big\}. $$
 Moreover, noting that $ED(X_{i};F_{m})$ and $ED(Y_{j};F_{m})$, $ED(X_{i};G_{n})$ and $ED(Y_{j};G_{n})$ are conditionally independent and have the same distribution given $\Lambda_{mn}$, we have $$E\Big[I(X_{i},Y_{j},F_{m},G_{n})\Big|\Lambda_{mn}\Big]=P\Big(\Big(ED(X_{i};F_{m})\leq ED(Y_{j};F_{m})\Big)\Big|\Lambda_{mn}\Big)-P\Big(ED(X_{i};G_{n})\leq ED(Y_{j};G_{n})\Big|\Lambda_{mn}\Big)=0.$$

Mahalanobis depth: For any point $x\in R^{d}$, the Mahalanobis depth of point $x$ is defined follows:
$$
MD(x;F)=\frac{1}{1+(x-\mu_{F})'\Sigma^{-1}_{F}(x-\mu_{F})},
$$
where $F$ is a $d$-dimensional distribution and $\mu_{F}$ and $\Sigma_{F}$ are the mean and covariance of $F$, respectively. 

The empirical version of $MD(X;F)$ is $MD(X;F_{m})=\frac{1}{1+(X-\Bar{X})'\Sigma_{F_{m}}^{-1}(X-\Bar{X})}$, where $\Bar{X}$ and $\Sigma_{F_{m}}$ are the sample mean and sample covariance, respectively.  \cite{zuo2006limiting} have already discussed Assumption \ref{ass 1} and \ref{ass 2} for the Mahalanobis depth. Next, we consdider Assumption \ref{ass 3} and \ref{ass 4}. Similarly, we also change the Mahalanobis depth to  the Mahalanobis distance $d(x;F)=\|\Sigma^{-\frac{1}{2}}_{F}(x-\mu_{F})\|_2$, and obtain
 $$ 
F_{MD(X;F)}\Big(MD(Y;F)\Big)=1-F_{d(X;F) }\Big(d(Y;F)\Big).$$
Meanwhile, we also assume that the density function $f_{d(X;F)}(\cdot)$ is bounded and symmetric about the Y-axis from above and that $\mu_{F}=0$. And, we consider the condition $\Lambda_m$ to be $\bar X$ and $\Sigma_{F_{m}}$.  Due to $\rho_{1}(Y;F_{m},F)|\Lambda_m$ is equal to zero or an odd function of $Y$, we have
\begin{eqnarray}
E_Y\Big[f_{d(X;F)}\Big(d(Y;F)\Big)\rho_{1}(Y;F_{m},F)\Big|\Lambda_{m}\Big]=0
.\nonumber
\end{eqnarray}
Again, since $\frac{\partial}{\partial(d(y_{j};F))}\Big[f_{d(x;F)}\Big(d(y_{j};F)\Big)\rho_{2}(y_{j};F_{m},F)\Big]$ is an odd function and conditionally independent for $j=1,\cdots,n$, $\rho_{2}(Y;F_{m},G_{n})=O_p(m^{-1})$ and $\rho_{3}(Y;F_{m},G_{n})=O_p(m^{-3/2})$, we also apply Cauchy's theorem to bound the partial derivatives in terms of integrals and Cauchy-Schwarz Inequality, so does the rest of the argument.

Halfspace depth: For any point $x\in R^{d}$, the Halfspace depth, also known for the Tukey depth, is defined as 
$$
HD(x;F)=\inf\{P(H_{x}): H_{x} \text{ is a closed half space containing } x \},
$$
where $P$ is the probability measure corresponding to $F$.     

If we replace the probability $P$ by the empirical probability, we obtain the sample version of $HD(x, F)$, denoted $HD(x, F_m).$ Since $f_{HD(X;F)}(HD(Y;F))$ is a constant, $\rho_{2}(Y;F_{m},G_{n})=O_p(m^{-1})$ and $\rho_{3}(Y;F_{m},G_{n})=O_p(m^{-3/2})$.
The remaining arguments are the same. 

Projection depth: For any point $x\in R^{d}$, the Projection depth is defined as 
$$
P D(x ; F)=1 /(1+O(x ; F)),
$$
where  
$
O(x; F)=\sup _{u \in S^{d-1}}\left|u^{\prime} x-\mu\left(F_u\right)\right| / \sigma\left(F_u\right)
$, $S^{d-1}=\{u:\|u\|=1\}, \mu(F)$ and $\sigma(F)$ are location and scale of $F$, respectively,  and $u^{\prime} X \sim F_u$ with $X \sim F$.  

The empirical version  of $PD(X;F)$ is $PD(X;F_{m})=\frac{1}{1+\sup _{u \in S^{d-1}}\left|u^{\prime} X-\mu\left(F_{mu}\right)\right| / \sigma\left(F_{mu}\right)}$, where $F_{mu}$ is the empirical distribution of $u^{\prime} X_1,\cdots,u^{\prime} X_m$. Assume that
$\mu(a X+b)=a \mu(X)+b$ and $\sigma(a X+b)=|a| \sigma(X)$ for any scalars $a, b \in  {R}^1$.
The rest of the verification is the same.
 
Spatial depth: 
For any point $x\in R^{d}$, the Spatial depth is defined as 
$$SD(x;F)=1-\Big\|E_X\frac{(x-X)}{\|x-X\|}\Big\|,\quad X\sim F.$$  

The empirical version  of $SD(X;F)$ is $SD(X;F_{m})=1-\Big\|\frac{1}{m}\sum\limits_{i=1}^{m}\frac{(X-X_{i})}{\|X-X_{i}\|}\Big\|$, which does not depend  on the mean of $F$.   
We also assume that $X$ is continuous and its density is symmetric around mean zero, $E\sup\limits_{x\in R^{d}}\Big\|\sum\limits_{i=1}^{m}\Big(\frac{(x-X_{i})}{\|x-X_{i}\|}-E\frac{(x-X_{i})}{\|x-X_{i}\|}\Big)\Big\|^{\alpha}=O(m^{\alpha/2})~ (\alpha>0)$, and the density of $SD(X;F)$ denoted as $f_{SD(X;F)}(\cdot)$ is bounded from above.  Again, (\ref{r2})-(\ref{rr3}) follows by $\rho_{2}(Y;F_{m},F)=O_p(m^{-1})$ and $\rho_{3}(Y;F_{m},F)=O_p(m^{-3/2})$.

\section{Verification of Assumption 5} \label{validation-assumption5}
To justify Assumption 5, without loss of generality we consider $\bm{\theta}(F)=E(X)=\mu_1\neq\bm{\theta}(G)=E(Y)=\mu_2$, where $X$ and $Y$ are adhering to normal distributions $F$ and $G$, respectively, each with a variance of 1. By Taylor expansions of $X$ and $Y$ around their expectation $E(X)=\mu_1$ and $E(Y)=\mu_2$ respectively, we have
\begin{align*}
   \frac{m+n}{mn} S_{m,n} &=  -\left[E\left\{1-F_{\chi^2}[(Y-\mu_1)^2]\right\}-E\left\{[F_{\chi^2}[(X-\mu_2)^2]\right\}\right] +o_p(1),  \\
       &=-1+2F_{\chi^2}[(\mu_2 -\mu_1)^2]+o_p(1), \\
      \frac{m+n}{mn} P_{m,n} &=   -\left[E\left\{1-F_{\chi^2}[(Y-\mu_1)^2]\right\} E\left\{1-F[(X-\mu_2)^2]\right\}-1/4\right]\\
 &=-\left\{1-F_{\chi^2}[(\mu_2-\mu_1)^2]\right\}^2+1/4 +o_p(1),
\end{align*}
where $F_{\chi^2}$ denotes the distribution function of $\chi^2_1$. It is evident that both $-1+2F_{\chi^2}[(\mu_2 -\mu_1)^2]$ and $-\left\{1-F_{\chi^2}[(\mu_2-\mu_1)^2]\right\}^2+1/4$  are monotonically increasing functions of $(\mu_2 -\mu_1)^2$. 
   
For multivariate distributions, we draw upon Theorem 6.1 in \citep{liu1993quality} to extend our analysis. The statistics $-[Q(F_m,G_n)+Q(G_n,F_m)-1]$ and $-[Q(F_m,G_n)Q(G_n,F_m)-1/4]$ can be represented as $1-Q(F,G)-Q(G,F)+o_p(1)$ and $1/4-Q(F,G)Q(G,F)+o_p(1)$, respectively. The focus then shifts to demonstrating that 
$$Q[(1-\beta)F+\beta G,(1-\beta)G+\beta F]+Q[(1-\beta)G+\beta F,(1-\beta)F+\beta G]>Q(F,G)+Q(G,F)$$ 
and 
$$Q[(1-\beta)F+\beta G,(1-\beta)G+\beta F]Q[(1-\beta)G+\beta F,(1-\beta)F+\beta G]>Q(F,G)Q(G,F)$$ for $0<\beta<1.$
As argued by \citep{liu1993quality},  $Q$ decreases  if there is a location shift or a scale increase, or both in terms of contamination: 
\begin{align*}
    Q(F,G)&<Q[F,(1-\beta)G+\beta F]\\
    &<Q[(1-)F+\{(1-\beta)G+\beta F\},(1-\beta)G+\beta F]\\
    &=Q[(1-\beta)F+\beta G,(1-\beta)G+\beta F],
\end{align*}
where $=\beta/(1-\beta)$. A similar argument holds for $G(G,F)$, leading to 
$$Q(G,F)<Q[G,(1-\beta)F+\beta G]<Q[(1-\beta)G+\beta F,(1-\beta)F+\beta G].$$

\section{Proof of Lemmas}  \label{lemma-proof}
Before proving the theorem, we first introduce the relevant lemmas and its proof.
\begin{lemma}[Theorem 1\citep{Cox}]\label{lem 1}
Let $Y=X_{0}+\epsilon X_{1}$ for some fixed $\epsilon>0$. If the joint density of $X_{0}$ and $X_{1}$ satisfie

(i) $\lim\limits _{x_0 \rightarrow-\infty} \frac{\partial^2}{\partial x_0^2} f\left(x_0, x_1\right)=0$,

(ii) $\sup\limits _{x_0}\left| \frac{\partial^2}{\partial x_0^2}f\left(x_0, x_1\right)\right|<K g\left(x_1\right)$, where $\int |x_1|^3 g\left(x_1\right) d x_1<\infty$,

(iii) $E\left(X_1^2 \mid X_0\right)<\infty$ a.e.,

then
$$
F_Y(y)=F_{X_0}(y)-\epsilon f_{X_0}(y) E\left(X_1 \mid X_0=y\right)+\frac{1}{2} \epsilon^2 \frac{\partial}{\partial y}\left\{f_{X_0}(y) E\left(X_1^2 \mid X_0=y\right)\right\}+O\left(\epsilon^3\right),
$$
where $F_{X}(\cdot)$ is the distribution function of $X$ and $f_{X}(\cdot)$ is the derivative of $F_{X}(\cdot)$.
\end{lemma}

\begin{lemma} \label{lem 2}
Under Assumption \ref{ass 4}, for $j\neq l$, we have
\begin{align}
E\Big(I(X_{i},Y_{j},F_{m},G_{n})E_{X}I(X,Y_{l},F_{m},G_{n})\Big)=E\Big(E_{X}I(X,Y_{j},F_{m},G_{n})E_{X}I(X,Y_{l},F_{m},G_{n})\Big)
=0;\label{r4}    
\end{align}
for $i\neq k$,
\begin{align}
E\Big(I(X_{i},Y_{j},F_{m},G_{n})E_{Y}I(X_{k},Y,F_{m},G_{n})\Big)=E\Big(E_{Y}I(X_{i},Y,F_{m},G_{n})E_{Y}I(X_{k},Y,F_{m},G_{n})\Big)=0;\label{r5} 
\end{align}

for $i\neq k,j\neq l$,
\begin{align}
E\Big(I(X_{i},Y_{j},F_{m},G_{n})I(X_{k},Y_{l},F_{m},G_{n})\Big)=0.\label{r6}    
\end{align}
for $i=k$ and $j\neq l$, 
\begin{eqnarray}
&&E\Big(I(X_{i},Y_{j},F_{m},G_{n})I(X_{k},Y_{l},F_{m},G_{n})\Big)\nonumber\\
&=&E\Big(I(X_{i},Y_{j},F_{m},G_{n})E_{Y_{l}}I(X_{k},Y_{l},F_{m},G_{n})\Big)\nonumber\\
&=&E\Big(E_{Y_{j}}I(X_{i},Y_{j},F_{m},G_{n})E_{Y_{l}}I(X_{k},Y_{l},F_{m},G_{n})\Big);\label{r7}
\end{eqnarray}
and, for $i\neq k$ and $j=l$,
\begin{eqnarray}
&&E\Big(I(X_{i},Y_{j},F_{m},G_{n})I(X_{k},Y_{l},F_{m},G_{n})\Big)\nonumber\\
&=&E\Big(I(X_{i},Y_{j},F_{m},G_{n})E_{X_{k}}I(X_{k},Y_{l},F_{m},G_{n})\Big)\nonumber\\
&=&E\Big(E_{X_{i}}I(X_{i},Y_{j},F_{m},G_{n})E_{X_{k}}I(X_{k},Y_{l},F_{m},G_{n})\Big).\label{r8}
\end{eqnarray}
\end{lemma}
The proof of this Lemma is similar to Lemma A7 of \cite{Gao24}, except that $G_{n}$  is added to the given conditions.
When 
$j\neq l$, based on the independence of $Y_{j}$ and $Y_{l}$ and Assumption \ref{ass 4}, we have
\begin{eqnarray}
&&E\Big(E_{X}I(X,Y_{j},F_{m},G_{n})E_{X}I(X,Y_{l},F_{m},G_{n})\Big)\nonumber\\
&=&E\Big\{E\Big[\Big(I(X_{i},Y_{j},F_{m},G_{n})E_{X}I(X,Y_{l},F_{m},G_{n})\Big)\Big|F_{m}, G_{n}, Y_j, Y_{l}\Big]\Big\}\nonumber\\
&=&E\Big(I(X_{i},Y_{j},F_{m},G_{n})E_XI(X,Y_{l},F_{m},G_{n})\Big)\nonumber\\
&=&E\Big\{E\Big[\Big(I(X_{i},Y_{j},F_{m},G_{n})E_XI(X,Y_{l},F_{m},G_{n})\Big)\Big|F_{m},G_{n}\Big]\Big\}\nonumber\\
&=&E\Big\{E_{X_i, Y_j}\Big[\Big(I(X_{i},Y_{j},F_{m},G_{n})\Big)\Big|F_{m},G_{n}\Big]E_{Y_l}\Big[E_{X}\Big(I(X,Y_{l},F_{m},G_{n})\Big)\Big|F_{m},G_{n}\Big]\Big\}\nonumber\\
&=&0.\nonumber
\end{eqnarray}
Therefore, 
(\ref{r4}) holds. Similarly, when $i\neq k$ and $i\neq k,j\neq l$, (\ref{r5}) and (\ref{r6}) also hold respectively.
When $i=k$ and $j\neq l$, it has
\begin{eqnarray}
&&E\Big(I(X_{i},Y_{j},F_{m},G_{n})I(X_{k},Y_{l},F_{m},G_{n})\Big)\nonumber\\
&=&E\Big\{E\Big[\Big(I(X_{i},Y_{j},F_{m},G_{n})I(X_{i},Y_{l},F_{m},G_{n})\Big)\Big|X_{i},Y_{j},F_{m},G_{n}\Big]\Big\}\nonumber\\
&=&E\Big\{I(X_{i},Y_{j},F_{m},G_{n})E_{Y_{l}}\Big[I(X_{i},Y_{l},F_{m},G_{n})\Big|X_{i},Y_{j},F_{m},G_{n}\Big]\Big\}\nonumber\\
&=&E\Big(I(X_{i},Y_{j},F_{m},G_{n})E_{Y_{l}}I(X_{k},Y_{l},F_{m},G_{n})\Big),\nonumber 
\end{eqnarray}
and
\begin{eqnarray}
&&E\Big(I(X_{i},Y_{j},F_{m},G_{n})I(X_{k},Y_{l},F_{m},G_{n})\Big)\nonumber\\
&=&E\Big\{E\Big[\Big(I(X_{i},Y_{j},F_{m},G_{n})I(X_{i},Y_{l},F_{m},G_{n})\Big)\Big|X_{i},F_{m},G_{n}\Big]\Big\}\nonumber\\
&=&E\Big\{E_{Y_{j}}\Big[I(X_{i},Y_{j},F_{m},G_{n})\Big|X_{i},F_{m},G_{n}\Big]E_{Y_{l}}\Big[I(X_{i},Y_{l},F_{m},G_{n})\Big|X_{i},F_{m},G_{n}\Big]\Big\}\nonumber\\
&=&E\Big(E_{Y_{j}}I(X_{i},Y_{j},F_{m},G_{n})E_{Y_{l}}I(X_{k},Y_{l},F_{m},G_{n})\Big)\nonumber 
\end{eqnarray}
Thus, 
(\ref{r7}) holds. Likewise, when $i\neq k$ and $j=l$, (\ref{r8}) also holds.\\

\begin{lemma}\label{lem 3}
Under the null hypothesis $F = G$, then under Assumption \ref{ass 3}, it has
\begin{eqnarray}
E\Big(I(x_{1},y_{1},F_{m},G_{n})\Big)^{2}=O(m^{-1}).\nonumber
\end{eqnarray}
\textbf{Proof of Lemma \ref{lem 3}.}  First, we have
\begin{eqnarray}
&&E\Big(I(x_{1},y_{1},F_{m},G_{n})\Big)^{2}\nonumber\\
&=&E\Big[ I(D(x_{1};F_{m})\leq D(y_{1};F_{m}))-I(D(x_{1};G_{n})\leq D(y_{1};G_{n})) \Big]^2\nonumber\\
&=&E\Big[P_{x_{1}}(D(x_{1};F_{m})\leq D(y_{1};F_{m}), D(x_{1};G_{n})> D(y_{1};G_{n}))\nonumber\\
&&+P_{x_{1}}(D(x_{1};F_{m})> D(y_{1};F_{m}), D(x_{1};G_{n})\leq D(y_{1};G_{n}))\Big]\nonumber\\
&=&E\Big[P_{x_{1}}\Big(D(x_{1};F)\leq D(y_{1};F)+D(y_{1};F_{m})-D(y_{1};F)-D(x_{1};F_{m})+D(x_{1};F),\nonumber\\
&& D(x_{1};F)> D(y_{1};F)+D(y_{1};G_{n})-D(y_{1};F)-D(x_{1};G_{n})+D(x_{1};F)\Big)
\Big]\nonumber\\
&&+E\Big[P_{x_{1}}\Big(D(x_{1};F)>D(y_{1};F)+D(y_{1};F_{m})-D(y_{1};F)-D(x_{1};F_{m})+D(x_{1};F),\nonumber\\
&& D(x_{1};F)\leq D(y_{1};F)+D(y_{1};G_{n})-D(y_{1};F)-D(x_{1};G_{n})+D(x_{1};F)\Big)
\Big]\nonumber\\
&=&T_{1}+T_{2}.\nonumber
\end{eqnarray}
By Lemma \ref{lem 1} and Assumptions  \ref{ass 1} and \ref{ass 3}, it has
\begin{eqnarray}
T_{1}&=&E\Big\{F_{D(x_{1};F)}\Big(D(y_{1};F)\Big)+f_{D(x_{1};F)}\Big(D(y_{1};F)\Big)\rho_{1}(y_{1};F_{m},F)\nonumber\\
&&+\frac{1}{2}\frac{\partial}{\partial(D(y_{1};F))}\Big[f_{D(x_{1};F)}\Big(D(y_{1};F)\Big)\rho_{2}(y_{1};F_{m},F)\Big]+O_{P}(m^{-3/2})\Big\}\nonumber\\
&&-E\Big\{F_{D(x_{1};F)}\Big(D(y_{1};F)\Big)+f_{D(x_{1};F)}\Big(D(y_{1};F)\Big)\rho_{1}(y_{1};G_{n},F)\nonumber\\
&&+\frac{1}{2}\frac{\partial}{\partial(D(y_{1};F))}\Big[f_{D(x_{1};F)}\Big(D(y_{1};F)\Big)\rho_{2}(y_{1};G_{n},F)\Big]+O_{P}(n^{-3/2})\Big\}\nonumber\\
&=&E\Big\{f_{D(x_{1};F)}\Big(D(y_{1};F)\Big)\Big(\rho_{1}(y_{1};F_{m},F)-\rho_{1}(y_{1};G_{n},F)\Big)\nonumber\\
&&+\frac{1}{2}\frac{\partial}{\partial(D(y_{1};F))}\Big[f_{D(x_{1};F)}\Big(D(y_{1};F)\Big)\Big(\rho_{2}(y_{1};F_{m},F)-\rho_{2}(y_{1};G_{n},F)\Big)\Big]\Big\}\nonumber\\
&&+O(n^{-3/2})+O(m^{-3/2})\nonumber\\
&=&O(m^{-1}).\label{b2}
\end{eqnarray}
The proof of $T_{2}$ is similar so it is omitted.
Then, we obtain the result of Lemma \ref{lem 3}.
\end{lemma}

\begin{lemma}\label{lem 4}
Under the null hypothesis $F = G$, then under Assumptions \ref{ass 3}-\ref{ass 4}, it has
\begin{eqnarray}
&&\int\int I(x,y,F_{m},G_{n})d(F_{m}-F)(x)d(G_{n}-G)(y)=O_{P}(m^{-3/2}). \label{f0}
\end{eqnarray}
The proof of this Lemma is similar to Lemma A8 of \cite{Gao24}, but here $G_{n}$ is used instead of $F$. The main difference lies in utilizing Cox and Reid's theorem up to the second-order expansion. The detailed proof is as follows:

\textbf{Proof of Lemma \ref{lem 4}.} By using integral discretization, it has
\begin{eqnarray}
&&\int\int I(x,y,F_{m},G_{n})d(F_{m}-F)(x)d(G_{n}-G)(y)\nonumber\\
&=&\frac{1}{mn}\sum_{i=1}^{m}\sum_{j=1}^{n}I(x_{i},y_{j},F_{m},G_{n})-\frac{1}{m}\sum_{i=1}^{m}E_{y}I(x_{i},y,F_{m},G_{n})\nonumber\\
&&-\frac{1}{n}\sum_{j=1}^{n}E_{x}I(x,y_{j},F_{m},G_{n})+E_{x,y}I(x,y,F_{m},G_{n}).\label{f1}
\end{eqnarray}
Then, given conditions such as $F_{m}$ and $G_{n}$, using the law of double expectation, we obtain
\begin{eqnarray}
&&E\int\int I(x,y,F_{m},G_{n})d(F_{m}-F)(x)d(G_{n}-G)(y)\nonumber\\
&=&E\Big[\frac{1}{mn}\sum_{i=1}^{m}\sum_{j=1}^{n}I(x_{i},y_{j},F_{m},G_{n})-\frac{1}{m}\sum_{i=1}^{m}E_{y}I(x_{i},y,F_{m},G_{n})\Big]\nonumber\\
&&-E\Big[\frac{1}{n}\sum_{j=1}^{n}E_{x}I(x,y_{j},F_{m},G_{n})-E_{xy}I(x,y,F_{m},G_{n})\Big]\nonumber\\
&=&E\Big\{E\Big[\Big(\frac{1}{nm}\sum_{i=1}^{m}\sum_{j=1}^{n}\Big(I(x_{i},y_{j},F_{m},G_{n})-E_{y}I(x_{i},y,F_{m},G_{n})\Big)\Big)\Big|F_{m},G_{n}\Big]\Big\}\nonumber\\
&&-E\Big\{E\Big[\Big(\frac{1}{n}\sum_{j=1}^{n}\Big(E_{x}I(x,y_{j},F_{m},G_{n})-E_{x,y}I(x,y,F_{m},G_{n})\Big)\Big)\Big|F_{m},G_{n}\Big]\Big\}\nonumber\\
&=&0,\label{f111}
\end{eqnarray}
and
\begin{eqnarray}
&&\text{Var}\Big(\int\int I(x,y,F_{m},G_{n})d(F_{m}-F)(x)d(G_{n}-G)(y)\Big)\nonumber\\
&=&\text{Cov}\Big(\frac{1}{mn}\sum_{i=1}^{m}\sum_{j=1}^{n}I(x_{i},y_{j},F_{m},G_{n}),
\frac{1}{mn}\sum_{k=1}^{m}\sum_{l=1}^{n}I(x_{k},y_{l},F_{m},G_{n})\Big)\nonumber\\
&&-2\text{Cov}\Big(\frac{1}{mn}\sum_{i=1}^{m}\sum_{j=1}^{n}I(x_{i},y_{j},F_{m},G_{n}),
\frac{1}{m}\sum_{k=1}^{m}E_{y}I(x_{k},y,F_{m},G_{n})\Big)\nonumber\\
&&-2\text{Cov}\Big(\frac{1}{mn}\sum_{i=1}^{m}\sum_{j=1}^{n}I(x_{i},y_{j},F_{m},G_{n}),\frac{1}{n}\sum_{l=1}^{n}E_{x}I(x,y_{l},F_{m},G_{n})
\Big)\nonumber\\
&&+\text{Cov}\Big(\frac{1}{m}\sum_{i=1}^{m}E_{y}I(x_{i},y,F_{m},G_{n}),
\frac{1}{m}\sum_{k=1}^{m}E_{y}I(x_{k},y,F_{m},G_{n})\Big)\nonumber\\
&&+\text{Cov}\Big(\frac{1}{n}\sum_{j=1}^{n}E_{x}I(x,y_{j},F_{m},G_{n}),
\frac{1}{n}\sum_{l=1}^{n}E_{x}I(x,y_{l},F_{m},G_{n})\Big)\nonumber\\
&&+\text{Cov}\Big(E_{x,y}I(x,y,F_{m},G_{n}), E_{x,y}I(x,y,F_{m},G_{n})\Big)\nonumber\\
&=:&A_{mn1}-2A_{mn2}-2A_{mn3}+A_{mn4}+A_{mn5}+A_{mn6}.\label{f2}
\end{eqnarray}
Firstly, we have
\begin{eqnarray}
A_{mn1}&=&\text{Cov}\Big(\frac{1}{mn}\sum_{i=1}^{m}\sum_{j=1}^{n}I(x_{i},y_{j},F_{m},G_{n}),
\frac{1}{mn}\sum_{k=1}^{m}\sum_{l=1}^{n}I(x_{k},y_{l},F_{m},G_{n})\Big)\nonumber\\
&=&\frac{1}{m^{2}n^{2}}\sum_{i=k}\sum_{j=l}\text{Cov}\Big(I(x_{i},y_{j},F_{m},G_{n}), I(x_{k},y_{l},F_{m},G_{n})\Big)\nonumber\\
&&+\frac{1}{m^{2}n^{2}}\sum_{i=k}\sum_{j\neq
l}\text{Cov}\Big(I(x_{i},y_{j},F_{m},G_{n}),I(x_{k},y_{l},F_{m},G_{n})\Big)\nonumber\\
&&+\frac{1}{m^{2}n^{2}}\sum_{i\neq k}\sum_{j=l}\text{Cov}\Big(I(x_{i},y_{j},F_{m},G_{n}),
I(x_{k},y_{l},F_{m},G_{n})\Big)\nonumber\\
&&+\frac{1}{m^{2}n^{2}}\sum_{i\neq k}\sum_{j\neq l}\text{Cov}\Big(I(x_{i},y_{j},F_{m},G_{n}),
I(x_{k},y_{l},F_{m},G_{n})\Big)\nonumber\\
&=&B_{mn1}+B_{mn2}+B_{mn3}+B_{mn4},\label{f3}
\end{eqnarray}
where
\begin{eqnarray}
B_{mn1}&=&\frac{1}{m^{2}n^{2}}\sum_{i=k}\sum_{j=l}\text{Cov}\Big(I(x_{i},y_{j},F_{m},G_{n}), I(x_{k},y_{l},F_{m},G_{n})\Big)\nonumber\\
&=&\frac{1}{mn}\text{Cov}\Big(I(x_{1},y_{1},F_{m},G_{n}), I(x_{1},y_{1},F_{m},G_{n})\Big)\nonumber\\
&=&\frac{1}{mn}\Big[E\Big(I(x_{1},y_{1},F_{m},G_{n})\Big)^{2}- \Big(EI(x_{1},y_{1},F_{m},G_{n})\Big)^{2}\Big].\nonumber
\end{eqnarray}
By Assumption \ref{ass 4} and Lemma \ref{lem 3}, we have
\begin{eqnarray}
B_{mn1}=O_{P}(m^{-2}n^{-1}),\label{30}
\end{eqnarray}
and by (\ref{r7}) in Lemma \ref{lem 2} and Assumption \ref{ass 4},
\begin{eqnarray}
B_{mn2}&=&\frac{1}{m^{2}n^{2}}\sum_{i=k}\sum_{j\neq
l}\text{Cov}\Big(I(x_{i},y_{j},F_{m},G_{n}),I(x_{k},y_{l},F_{m},G_{n})\Big)\nonumber\\
&=&\frac{1}{m^{2}n^{2}}\sum_{i=k}\sum_{j\neq
l}\Big[E\Big(I(x_{i},y_{j},F_{m},G_{n})I(x_{k},y_{l},F_{m},G_{n})\Big)\nonumber\\
&&-EI(x_{i},y_{j},F_{m},G_{n})EI(x_{k},y_{l},F_{m},G_{n})\Big]\nonumber\\
&=&\frac{1}{mn^{2}}\sum_{j\neq
l}\Big[E\Big(E_{y_{j}}I(x_{1},y_{j},F_{m},G_{n})E_{y_{l}}I(x_{1},y_{l},F_{m},G_{n})\Big)\nonumber\\
&&-EI(x_{1},y_{j},F_{m},G_{n})EI(x_{1},y_{l},F_{m},G_{n})\Big] \nonumber\\
&=&\frac{n-1}{mn}E\Big(E_{y_{1}}I(x_{1},y_{1},F_{m},G_{n})E_{y_{2}}I(x_{1},y_{2},F_{m},G_{n})\Big).\label{c3}
\end{eqnarray}

Similar to (\ref{c3}), by Assumption \ref{ass 4} and (\ref{r8}) in lemma \ref{lem 2}, we have
\begin{eqnarray}
B_{mn3}&=&\frac{1}{m^{2}n^{2}}\sum_{i\neq k}\sum_{j=l}\text{Cov}\Big(I(x_{i},y_{j},F_{m},G_{n}),
I(x_{k},y_{l},F_{m},G_{n})\Big)\nonumber\\
&=&\frac{1}{m^{2}n^{2}}\sum_{i\neq
k}\sum_{j=l}\Big[E\Big(I(x_{i},y_{j},F_{m},G_{n})I(x_{k},y_{l},F_{m},G_{n})\Big)\nonumber\\
&&-EI(x_{i},y_{j},F_{m},G_{n})EI(x_{k},y_{l},F_{m},G_{n})\Big]\nonumber\\
&=&\frac{1}{m^{2}n}\sum_{i\neq
k}\Big[E\Big(E_{x_{i}}I(x_{i},y_{1},F_{m},G_{n})E_{x_{k}}I(x_{k},y_{1},F_{m},G_{n})\Big)\nonumber\\
&&-EI(x_{i},y_{1},F_{m},G_{n})EI(x_{k},y_{1},F_{m},G_{n})\Big] \nonumber\\
&=&\frac{m-1}{mn}E\Big(E_{x_{1}}I(x_{1},y_{1},F_{m},G_{n})E_{x_{2}}I(x_{2},y_{1},F_{m},G_{n})\Big).
\label{d3}
\end{eqnarray}
By Assumption \ref{ass 4} and (\ref{r6}) of Lemma \ref{lem 2},
\begin{eqnarray}
B_{mn4}&=&\frac{1}{m^{2}n^{2}}\sum_{i\neq k}\sum_{j\neq l}\text{Cov}\Big(I(x_{i},y_{j},F_{m},G_{n}),
I(x_{k},y_{l},F_{m},G_{n})\Big)\nonumber\\
&=&\frac{1}{m^{2}n^{2}}\sum_{i\neq k}\sum_{j\neq l}\Big[EI(x_{i},y_{j},F_{m},G_{n})I(x_{k},y_{l},F_{m},G_{n})\nonumber\\
&&-EI(x_{i},y_{j},F_{m},G_{n})EI(x_{k},y_{l},F_{m},G_{n})\Big]=0.\label{e0}
\end{eqnarray}

Secondly, there is
\begin{eqnarray}
A_{mn2}&=&\text{Cov}\Big(\frac{1}{mn}\sum_{i=1}^{m}\sum_{j=1}^{n}I(x_{i},y_{j},F_{m},G_{n}),
\frac{1}{m}\sum_{k=1}^{m}E_{y}I(x_{k},y,F_{m},G_{n})\Big)\nonumber\\
&=&\frac{1}{m^{2}}\sum_{i=k}\text{Cov}\Big(I(x_{i},y_{1},F_{m},G_{n}), E_{y}I(x_{k},y,F_{m},G_{n})\Big)\nonumber\\
&&+\frac{1}{m^{2}}\sum_{i\neq k}\text{Cov}\Big(I(x_{i},y_{1},F_{m},G_{n}), E_{y}I(x_{k},y,F_{m},G_{n})\Big)\nonumber\\
&=&C_{mn1}+C_{mn2},\label{f4}
\end{eqnarray}
where
\begin{eqnarray}
C_{mn1}&=&\frac{1}{m^{2}}\sum_{i=k}\text{Cov}\Big(I(x_{i},y_{1},F_{m},G_{n}), E_{y}I(x_{k},y,F_{m},G_{n})\Big)\nonumber\\
&=&\frac{1}{m^{2}}\sum_{i=k}\Big[E\Big(I(x_{i},y_{1},F_{m},G_{n})E_{y}I(x_{k},y,F_{m},G_{n})\Big)\nonumber\\
&&-EI(x_{i},y_{1},F_{m},G_{n})E\Big(E_{y}I(x_{k},y,F_{m},G_{n})\Big)\Big]\nonumber\\
&=&\frac{1}{m}\Big[E\Big(E_{y_{1}}I(x_{1},y_{1},F_{m},G_{n})E_{y}I(x_{1},y,F_{m},G_{n})\Big)~(~by~(\ref{r7}) ~in~Lemma~\ref{lem 2}) \nonumber\\
&=&\frac{n}{n-1}B_{mn2}.\label{c6}
\end{eqnarray}
Under $H_{0}:F = G$, by Lemma \ref{lem 1}, Assumption  \ref{ass 1} and \ref{ass 3}, one has
\begin{eqnarray}
&&E\Big(E_{y_{1}}I(x_{1},y_{1},F_{m},G_{n})E_{y}I(x_{1},y,F_{m},G_{n})\Big)\nonumber\\
&=&E\Big\{ \Big[ f_{D(y_{1};F)}\Big(D(x_{1};F)\Big)\Big(\rho_{1}(x_{1};G_{n},F)-\rho_{1}(x_{1};F_{m},F)\Big)\nonumber\\
&&+\frac{1}{2}\frac{\partial}{\partial(D(x_{1};F))}\Big[f_{D(y_{1};F)}\Big(D(x_{1};F)\Big)\Big(\rho_{2}(x_{1};G_{n},F)-\rho_{2}(x_{1};F_{m},F)\Big)\Big]\nonumber\\
&&+O_{P}(n^{-3/2})+O_{P}(m^{-3/2})\Big]\times\nonumber\\
&&\Big[f_{D(y;F)}\Big(D(x_{1};F)\Big)\Big(\rho_{1}(x_{1};G_{n},F)-\rho_{1}(x_{1};F_{m},F)\Big)\nonumber\\
&&+\frac{1}{2}\frac{\partial}{\partial(D(x_{1};F))}\Big[f_{D(y;F)}\Big(D(x_{1};F)\Big)\Big(\rho_{2}(x_{1};G_{n},F)-\rho_{2}(x_{1};F_{m},F)\Big)\Big]\nonumber\\
&&+O_{P}(n^{-3/2})+O_{P}(m^{-3/2})\Big]\Big\}\nonumber\\
&=&E\Big\{f_{D(y_{1};F)}\Big(D(x_{1};F)\Big)\Big(\rho_{1}(x_{1};G_{n},F)-\rho_{1}(x_{1};F_{m},F)\Big)\nonumber\\
&&+\frac{\partial}{\partial(D(x_{1};F))}\Big[f_{D(y_{1};F)}\Big(D(x_{1};F)\Big)\Big(\rho_{2}(x_{1};G_{n},F)-\rho_{2}(x_{1};F_{m},F)\Big)\Big]\nonumber\\
&&+O_{P}(n^{-3/2})+O_{P}(m^{-3/2})\Big\}^2\nonumber\\
&=&O(m^{-1}),\label{c4}
\end{eqnarray}
and combining with Assumption \ref{ass 4}, we have
\begin{eqnarray}
&&E\Big(E_{y_{1}}I(x_{1},y_{1},F_{m},G_{n})E_{y}I(x_{1},y,F_{m},G_{n})\Big)\nonumber\\
&&-EI(x_{1},y_{1},F_{m},G_{n})EI(x_{1},y,F_{m},G_{n})=O(m^{-1}).\label{c5}
\end{eqnarray}
Thus, by (\ref{c3}), (\ref{c6}) and (\ref{c5}), it has
\begin{eqnarray}
B_{mn2}-C_{mn1}=O(m^{-3}). \label{c7}
\end{eqnarray}
By Assumption \ref{ass 4} and (\ref{r5}) in Lemma \ref{lem 2},
\begin{eqnarray}
C_{mn2}&=&\frac{1}{m^{2}}\sum_{i\neq k}\text{Cov}\Big(I(x_{i},y_{1},F_{m},G_{n}), E_{y}I(x_{k},y,F_{m},G_{n})\Big)\nonumber\\
&=&\frac{1}{m^{2}}\sum_{i\neq k}\Big[E\Big(I(x_{i},y_{1},F_{m},G_{n})E_{y}I(x_{k},y,F_{m},G_{n})\Big)\nonumber\\
&&-EI(x_{i},y_{1},F_{m},G_{n})E\Big(E_{y}I(x_{k},y,F_{m},G_{n})\Big)\Big]\nonumber\\
&=&0.\label{e1}
\end{eqnarray}

For $A_{mn3}$,
\begin{eqnarray}
A_{mn3}&=&\text{Cov}\Big(\frac{1}{mn}\sum_{i=1}^{m}\sum_{j=1}^{n}I(x_{i},y_{j},F_{m},G_{n}),\frac{1}{n}\sum_{l=1}^{n}E_{x}I(x,y_{l},F_{m},G_{n})
\Big)\nonumber\\
&=&\frac{1}{n^{2}}\sum_{j=l}\text{Cov}\Big(I(x_{1},y_{j},F_{m},G_{n}), E_{x}I(x,y_{l},F_{m},G_{n})\Big)\nonumber\\
&&+\frac{1}{n^{2}}\sum_{j\neq l}\text{Cov}\Big(I(x_{1},y_{j},F_{m},G_{n}), E_{x}I(x,y_{l},F_{m},G_{n})\Big)\nonumber\\
&=&D_{mn1}+D_{mn2},\label{f5}
\end{eqnarray}
where
\begin{eqnarray}
D_{mn1}&=&\frac{1}{n^{2}}\sum_{j=l}\text{Cov}\Big(I(x_{1},y_{j},F_{m},G_{n}), E_{x}I(x,y_{l},F_{m},G_{n})\Big)\nonumber\\
&=&\frac{1}{n^{2}}\sum_{j=l}\Big[E\Big(I(x_{1},y_{j},F_{m},G_{n})E_{x}I(x,y_{l},F_{m},G_{n})\Big)\nonumber\\
&&-EI(x_{1},y_{j},F_{m},G_{n})E\Big(E_{x}I(x,y_{l},F_{m},G_{n})\Big)\Big]\nonumber\\
&=&\frac{1}{n}\Big[E\Big(E_{x_{1}}I(x_{1},y_{1},F_{m},G_{n})E_{x}I(x,y_{1},F_{m},G_{n})\Big)\nonumber\\
&&-EI(x_{1},y_{j},F_{m},G_{n})EI(x,y_{j},F_{m},G_{n})\Big]~(~by~(\ref{r8}) ~in~Lemma~\ref{lem 2})  \nonumber\\
&=&\frac{m}{m-1}B_{mn3}.\label{d6}
\end{eqnarray}
By Lemma \ref{lem 1}, Assumption  \ref{ass 1} and \ref{ass 3}, we have
\begin{eqnarray}
&&E\Big(E_{x_{1}}I(x_{1},y_{1},F_{m},G_{n})E_{x_{2}}I(x_{2},y_{2},F_{m},G_{n})\Big)\nonumber\\
&=&E\Big\{\Big[f_{D(x_{1};F)}\Big(D(y_{1};F)\Big)\Big(\rho_{1}(y_{1};F_{m},F)-\rho_{1}(y_{1};G_{n},F)\Big)\nonumber\\
&&+\frac{1}{2}\frac{\partial}{\partial(D(y_{1};F))}\Big[f_{D(x_{1};F)}\Big(D(y_{1};F)\Big)\Big(\rho_{2}(y_{1};F_{m},F)-\rho_{2}(y_{1};G_{n},F)\Big)\Big]\nonumber\\
&&+O(n^{-3/2})+O(m^{-3/2})\Big]\times\nonumber\\
&&\Big[f_{D(x_{2};F)}\Big(D(y_{2};F)\Big)\Big(\rho_{1}(y_{2};F_{m},F)-\rho_{1}(y_{2};G_{n},F)\Big)\nonumber\\
&&+\frac{1}{2}\frac{\partial}{\partial(D(y_{2};F))}\Big[f_{D(x_{2};F)}\Big(D(y_{2};F)\Big)\Big(\rho_{2}(y_{2};F_{m},F)-\rho_{2}(y_{2};G_{n},F)\Big)\Big]\nonumber\\
&&+O(n^{-3/2})+O(m^{-3/2})\Big]\Big\} \nonumber\\
&=&E\Big\{f_{D(x_{1};F)}\Big(D(y_{1};F)\Big)\Big(\rho_{1}(y_{1};F_{m},F)-\rho_{1}(y_{1};G_{n},F)\Big)\nonumber\\
&&+\frac{1}{2}\frac{\partial}{\partial(D(y_{1};F))}\Big[f_{D(x_{1};F)}\Big(D(y_{1};F)\Big)\Big(\rho_{2}(y_{1};F_{m},F)-\rho_{2}(y_{1};G_{n},F)\Big)\Big]\nonumber\\
&&+O(n^{-3/2})+O(m^{-3/2})\Big\}^{2}\nonumber\\
&=&O(m^{-1}).\label{d4}
\end{eqnarray}
and combing this with Assumption 2, we have
\begin{eqnarray}
&&E\Big(E_{x_{1}}I(x_{1},y_{1},F_{m},G_{n})E_{x_{2}}I(x_{2},y_{1},F_{m},G_{n})\Big)\nonumber\\
&&-EI(x_{1},y_{1},F_{m},G_{n})EI(x_{2},y_{1},F_{m},G_{n})=O(m^{-1}).\label{d5}
\end{eqnarray}
Thus, by (\ref{d3}), (\ref{d6}) and (\ref{d5}) , it has
\begin{eqnarray}
B_{mn3}-D_{mn1}=O(m^{-3}). \label{d7}
\end{eqnarray}
and by Assumption \ref{ass 4} and (\ref{r4}) in 
Lemma \ref{lem 2},
\begin{eqnarray}
D_{mn2}&=&\frac{1}{n^{2}}\sum_{j\neq l}\text{Cov}\Big(I(x_{1},y_{j},F_{m},G_{n}), E_{x}I(x,y_{l},F_{m},G_{n})\Big)\nonumber\\
&=&\frac{1}{n^{2}}\sum_{j\neq l}\Big[E\Big(I(x_{1},y_{j},F_{m},G_{n})E_{x}I(x,y_{l},F_{m},G_{n})\Big)\nonumber\\
&&-EI(x_{1},y_{j},F_{m},G_{n})E\Big(E_{x}I(x,y_{l},F_{m},G_{n})\Big)\Big]\nonumber\\
&=&0. \text{ (by Lemma } \ref{lem 2}) \label{e2}
\end{eqnarray}
For $A_{mn4}$,
\begin{eqnarray}
A_{mn4}&=&\text{Cov}\Big(\frac{1}{m}\sum_{i=1}^{m}E_{y}I(x_{i},y,F_{m},G_{n}),
\frac{1}{m}\sum_{k=1}^{m}E_{y}I(x_{k},y,F_{m},G_{n})\Big)\nonumber\\
&=&\frac{1}{m^{2}}\sum_{i=k}\text{Cov}\Big(E_{y}I(x_{i},y,F_{m},G_{n}), E_{y}I(x_{k},y,F_{m},G_{n})\Big)\nonumber\\
&&+\frac{1}{m^{2}}\sum_{i\neq k}\text{Cov}\Big(E_{y}I(x_{i},y,F_{m},G_{n}), E_{y}I(x_{k},y,F_{m},G_{n})\Big)\nonumber\\
&=&E_{mn1}+E_{mn2},\label{f6}
\end{eqnarray}
where
\begin{eqnarray}
E_{mn1}&=&\frac{1}{m^{2}}\sum_{i=k}\text{Cov}\Big(E_{y}I(x_{i},y,F_{m},G_{n}), E_{y}I(x_{k},y,F_{m},G_{n})\Big)\nonumber\\
&=&\frac{1}{m^{2}}\sum_{i=k}\Big[E\Big(E_{y}I(x_{i},y,F_{m},G_{n})E_{y}I(x_{k},y,F_{m},G_{n})\Big)\nonumber\\
&&-E\Big(E_{y}I(x_{i},y,F_{m},G_{n})\Big)E\Big(E_{y}I(x_{k},y,F_{m},G_{n})\Big)\Big]\nonumber\\
&=&\frac{1}{m}E\Big(E_{y}I(x_{1},y,F_{m},G_{n})E_{y}I(x_{1},y,F_{m},G_{n})\Big)~(~by~Assumption~ \ref{ass 4})\nonumber\\
&=&C_{mn1},\label{c2}
\end{eqnarray}
and by (\ref{r5})  in Lemma \ref{lem 2}  and  Assumption  \ref{ass 4},
\begin{eqnarray}
E_{mn2}&=&\frac{1}{m^{2}}\sum_{i\neq k}\text{Cov}\Big(E_{y}I(x_{i},y,F_{m},G_{n}), E_{y}I(x_{k},y,F_{m},G_{n})\Big)\nonumber\\
&=&\frac{1}{m^{2}}\sum_{i\neq k}\Big[E\Big(E_{y}I(x_{i},y,F_{m},G_{n})E_{y}I(x_{k},y,F_{m},G_{n})\Big)\nonumber\\
&&-E\Big(E_{y}I(x_{i},y,F_{m},G_{n})\Big)E\Big(E_{y}I(x_{k},y,F_{m},G_{n})\Big)\Big]\nonumber\\
&=&0.\label{e3}
\end{eqnarray}

For $A_{mn5}$,
\begin{eqnarray}
A_{mn5}&=&\text{Cov}\Big(\frac{1}{n}\sum_{j=1}^{n}E_{x}I(x,y_{j},F_{m},G_{n}),
\frac{1}{n}\sum_{l=1}^{n}E_{x}I(x,y_{l},F_{m},G_{n})\Big)\nonumber\\
&=&\frac{1}{n^{2}}\sum_{j=l}\text{Cov}\Big(E_{x}I(x,y_{j},F_{m},G_{n}), E_{x}I(x,y_{l},F_{m},G_{n})\Big)\nonumber\\
&&+\frac{1}{n^{2}}\sum_{j\neq l}\text{Cov}\Big(E_{x}I(x,y_{j},F_{m},G_{n}), E_{x}I(x,y_{l},F_{m},G_{n})\Big)\nonumber\\
&=&F_{mn1}+F_{mn2}.\label{f7}
\end{eqnarray}
Similar to (\ref{c2}), by Assumption \ref{ass 4}, it has
\begin{eqnarray}
F_{mn1}&=&\frac{1}{n^{2}}\sum_{j=l}\text{Cov}\Big(E_{x}I(x,y_{j},F_{m},G_{n}), E_{x}I(x,y_{l},F_{m},G_{n})\Big)\nonumber\\
&=&\frac{1}{n^{2}}\sum_{j=l}\Big[E\Big(E_{x}I(x,y_{j},F_{m},G_{n})E_{x}I(x,y_{l},F_{m},G_{n})\Big)\nonumber\\
&&-E\Big(E_{x}I(x,y_{j},F_{m},G_{n})\Big)E\Big(E_{x}I(x,y_{l},F_{m},G_{n})\Big)\Big] \nonumber\\
&=&\frac{1}{n}E\Big(E_{x}I(x,y_{1},F_{m},G_{n})E_{x}I(x,y_{1},F_{m},G_{n})\Big)\nonumber\\
&=&D_{mn1},\label{d2}
\end{eqnarray}
and by (\ref{r4})  in  Lemma \ref{lem 2}  and  Assumption  \ref{ass 4},
\begin{eqnarray}
F_{mn2}&=&\frac{1}{n^{2}}\sum_{j\neq l}\text{Cov}\Big(E_{x}I(x,y_{j},F_{m},G_{n}), E_{x}I(x,y_{l},F_{m},G_{n})\Big)\nonumber\\
&=&\frac{1}{n^{2}}\sum_{j\neq l}\Big[E\Big(E_{x}I(x,y_{j},F_{m},G_{n})E_{x}I(x,y_{l},F_{m},G_{n})\Big)\nonumber\\
&&-E\Big(E_{x}I(x,y_{j},F_{m},G_{n})\Big)E\Big(E_{x}I(x,y_{l},F_{m},G_{n})\Big)\Big] \nonumber\\
&=& 0.\label{e4}
\end{eqnarray}
For $A_{mn6}$, by Assumption 2, we have
\begin{eqnarray}
A_{mn6}&=&\text{Cov}\Big(E_{x,y}I(x,y,F_{m},G_{n}), E_{x,y}I(x,y,F_{m},G_{n})\Big)\nonumber\\
&=&E\Big(E_{x,y}I(x,y,F_{m},G_{n})\Big)^2-\Big (E(E_{x,y}I(x,y,F_{m},G_{n}))\Big)^2=0.\label{l2}
\end{eqnarray}
Then, combining this with (\ref{f2})-(\ref{l2}), one has
\begin{eqnarray}
&&\text{Var}\Big(\int\int I(x,y,F_{m},G_{n})d(F_{m}-F)(x)d(G_{n}-G)(y)\Big)\nonumber\\
&=:&A_{mn1}-2A_{mn2}-2A_{mn3}+A_{mn4}+A_{mn5}+A_{mn6}\nonumber\\
&=&B_{mn1}+B_{mn2}+B_{mn3}+B_{mn4}-2C_{mn1}-2C_{mn2}-2D_{mn1}-2D_{mn2}\nonumber\\
&&+E_{mn1}+E_{mn2}+F_{mn1}+F_{mn2}+A_{mn6}\nonumber\\
&=&B_{mn1}+(B_{mn2}-C_{mn1})+(B_{mn3}-D_{mn1})+A_{mn6}\nonumber\\
&=&O(m^{-3}).\label{f8}
\end{eqnarray}
So, according to (\ref{f111}) and (\ref{f8}), by applying Markov’s inequality, we obtain the result of (\ref{f0}).

\end{lemma}

\begin{lemma}\label{lem 5}
\citep{zuo2006limiting}
Under Assumption \ref{ass 2}, one has
$$\int\int I(x,y,F_{m},F)dF(x)dG(y)=o(\Delta_{m}).$$
\end{lemma}

Lemma \ref{lem 5} is derived based on Assumption \ref{ass 2}. Specific details can be found in Lemma A5 of \cite{Gao24}. The proof is omitted.

\begin{lemma}\label{lem 6}
Under the null hypothesis $F = G$, if Assumptions  \ref{ass 1}-\ref{ass 4} and $\frac{m}{n}\rightarrow c$ hold, it has
\begin{eqnarray}
&&Q(F_{m},G_{n})-Q(F,G)\nonumber\\
&=&\frac{1}{n}\sum\limits_{j=1}^{n}\Big[F_{D(x;F)}\Big(D(y_{j};F)\Big)-\frac{1}{2}\Big]+\frac{1}{m}\sum\limits_{i=1}^{m}\Big[\frac{1}{2}-F_{D(y;F)}\Big(D(x_{i};F)\Big)\Big]\nonumber\\
&&+o(\Delta_{m})+O_{P}(m^{-1}),\nonumber
\end{eqnarray}
and
\begin{eqnarray}
&&Q(G_{n},F_{m})-Q(F,G)\nonumber\\
&=&\frac{1}{n}\sum\limits_{j=1}^{n}\Big[\frac{1}{2}-F_{D(x;F)}\Big(D(y_{j};F)\Big)\Big]+\frac{1}{m}\sum\limits_{i=1}^{m}\Big[F_{D(y;F)}\Big(D(x_{i};F)\Big)-\frac{1}{2}\Big]\nonumber\\
&&+o(\Delta_{m})+O_{P}(m^{-1}),\nonumber
\end{eqnarray}
where $Q(F,G)=\frac{1}{2}$.

\end{lemma}
\textbf{Proof of Lemma \ref{lem 6}.} The proof is analogous to the proof of Theorem 2 in \cite{Gao24}. So the proof is skipped here.

\section{Proof of Theorem 1} \label{proof-thm1}

Step 1: By Hoeffding decomposition
\begin{eqnarray}
&&Q(F_{m},G_{n})+Q(G_{n},F_{m})-1\nonumber\\
&=&\int\int I\Big(D(x;F_{m})\leq D(y;F_{m})\Big)dF_{m}(x)dG_{n}(y)\nonumber\\
&&+\int\int I\Big(D(y;G_{n})\leq D(x;G_{n})\Big)dF_{m}(x)dG_{n}(y)-1 \nonumber\\
&=&\int\int I\Big(D(x;F_{m})\leq D(y;F_{m})\Big)dF_{m}(x)dG_{n}(y)\nonumber\\
&&-\int\int I\Big(D(x;G_{n})\leq
D(y;G_{n})\Big)dF_{m}(x)dG_{n}(y)\nonumber\\
&=&\int\int I(x,y,F_{m},G_{n})dF_{m}(x)dG_{n}(y) \nonumber\\
&=&\int\int I(x,y,F_{m},G_{n})dF(x)dG_{n}(y)+\int\int I(x,y,F_{m},G_{n})dF_{m}(x)dG(y)+R_{mn} \text{(by Hoeffding decomposition)} \nonumber\\
&=&M_{mn1}+M_{mn2}+R_{mn}, \label{h0}
\end{eqnarray}
where
\begin{eqnarray}
R_{mn}&=&\int\int I(x,y,F_{m},G_{n})dF_{m}(x)dG_{n}(y)-\int\int I(x,y,F_{m},G_{n})dF(x)dG_{n}(y)\nonumber\\
&&-\int\int I(x,y,F_{m},G_{n})dF_{m}(x)dG(y).\label{h1}
\end{eqnarray}
Step 2: For the main terms $M_{mn1}$ and $M_{mn2} $

For $i=1,2,\ldots, m$, by Lemma \ref{lem 1},
\begin{eqnarray}
&&E_{y}\Big[I\Big(D(x_{i};F_{m})\leq D(y;F_{m})\Big)\Big]\nonumber\\
&=&P_{y}\Big(D(x_{i};F)\leq D(y;F)+D(y;F_{m})-D(y;F)-D(x_{i};F_{m})+D(x_{i};F)\Big)\nonumber\\
&=&1-F_{D(y;F)}\Big(D(x_{i};F)\Big)-f_{D(y;F)}\Big(D(x_{i};F)\Big)\rho_{1}(x_{i};F_{m},F)\nonumber\\
&&-\frac{1}{2}\frac{\partial}{\partial(D(x_{i};F))}\Big[f_{D(y;F)}\Big(D(x_{i};F)\Big)\rho_{2}(x_{i};F_{m},F)\Big]+O_{P}(m^{-3/2}).\label{h2}
\end{eqnarray}
 Meanwhile, it has
\begin{eqnarray}
&&E_{y}\Big[I\Big(D(x_{i};G_{n})\leq D(y;G_{n})\Big)\Big]\nonumber\\
&=&P_{y}\Big(D(x_{i};F)\leq D(y;F)+D(y;G_{n})-D(y;F)-D(x_{i};G_{n})+D(x_{i};F)\Big)\nonumber\\
&=&1-F_{D(y;F)}\Big(D(x_{i};F)\Big)-f_{D(y;F)}\Big(D(x_{i};F)\Big)\rho_{1}(x_{i};G_{n},F)\nonumber\\
&&-\frac{1}{2}\frac{\partial}{\partial(D(x_{i};F))}\Big[f_{D(y;F)}\Big(D(x_{i};F)\Big)\rho_{2}(x_{i};G_{n},F)\Big]+O_{P}(n^{-3/2}).\label{h3}
\end{eqnarray}
By (\ref{h2}) and (\ref{h3}), we obtain
\begin{eqnarray}
M_{mn1}&=&\int\int I(x,y,F_{m},G_{n})dF_{m}(x)dG(y)\nonumber\\
&=&\frac{1}{m}\sum_{i=1}^{m}E_{y}I(x_{i},y,F_{m},G_{n})\nonumber\\
&=&-\frac{1}{m}\sum_{i=1}^{m}f_{D(y;F)}\Big(D(x_{i};F)\Big)\Big(\rho_{1}(x_{i};F_{m},F)-\rho_{1}(x_{i};G_{n},F)\Big)\nonumber\\
&&-\frac{1}{2m}\sum_{i=1}^{m}\frac{\partial}{\partial(D(x_{i};F))}\Big[f_{D(y;F)}\Big(D(x_{i};F)\Big)\Big(\rho_{2}(x_{i};F_{m},F)-\rho_{2}(x_{i};G_{n},F)\Big)\nonumber\\
&&+O_{P}(n^{-3/2})+O_{P}(m^{-3/2}).\label{h6}
\end{eqnarray}
Similarly, for $j=1,2\ldots,n$, we have
\begin{eqnarray}
&&E_{x}\Big[I\Big(D(x;F_{m})\leq D(y_{i};F_{m})\Big)\Big]\nonumber\\
&=&P_{x}\Big(D(x;F)\leq D(y_{i};F)+D(y_{i};F_{m})-D(y_{i};F)-D(x;F_{m})+D(x;F)\Big)\nonumber\\
&=&F_{D(x;F)}\Big(D(y_{i};F)\Big)+f_{D(x;F)}\Big(D(y_{i};F)\Big)\rho_{1}(y_{j};F_{m},F)\nonumber\\
&&+\frac{1}{2}\frac{\partial}{\partial(D(y_{i};F))}\Big[f_{D(x;F)}\Big(D(y_{i};F)\Big)\rho_{2}(y_{i};F_{m},F)\Big]+O_{P}(m^{-3/2}),\label{h4}
\end{eqnarray}
and
\begin{eqnarray}
&&E_{x}\Big[I\Big(D(x;G_{n})\leq D(y_{j};G_{n})\Big)\Big]\nonumber\\
&=&P_{x}\Big(D(x;F)\leq D(y_{i};F)+D(y_{j};G_{n})-D(y_{j};F)-D(x;G_{n})+D(x;F)\Big)\nonumber\\
&=&F_{D(x;F)}\Big(D(y_{j};F)\Big)+f_{D(x;F)}\Big(D(y_{j};F)\Big)\rho_{1}(y_{j};F_{m},F)\nonumber\\
&&+\frac{1}{2}\frac{\partial}{\partial(D(y_{i};F))}\Big[f_{D(x;F)}\Big(D(y_{i};F)\Big)\rho_{2}(y_{i};G_{n},F)\Big]+O_{P}(n^{-3/2}),\label{h5}
\end{eqnarray}
By (\ref{h4}) and (\ref{h5}), one has
\begin{eqnarray}
M_{mn2}&=&\int\int I(x,y,F_{m},G_{n})dF(x)dG_{n}(y)\nonumber\\
&=&\frac{1}{n}\sum_{j=1}^{n}E_{x}I(x,y_{j},F_{m},G_{n})\nonumber\\
&=&\frac{1}{n}\sum_{j=1}^{n}f_{D(x;F)}\Big(D(y_{j};F)\Big)\Big(\rho_{1}(y_{j};F_{m},F)-\rho_{1}(y_{j};G_{n},F)\Big)\nonumber\\
&&+\frac{1}{2n}\sum_{j=1}^{n}\frac{\partial}{\partial(D(y_{j};F))}\Big[f_{D(x;F)}\Big(D(y_{j};F)\Big)\Big(\rho_{2}(y_{j};F_{m},F)-\rho_{2}(y_{j};G_{n},F)\Big)\Big]\nonumber\\
&&+O_{P}(n^{-3/2})+O_{P}(m^{-3/2}).\label{h7}
\end{eqnarray}
Step 3: For the residual term $R_{mn}$

By \eqref{h1}, we have
\begin{eqnarray}
R_{mn}
&=&\int\int I(x,y,F_{m},G_{n})d(F_{m}-F)(x)d(G_{n}-G)(y)\nonumber\\
&&-\int\int I(x,y,F_{m},G_{n})dF(x)dG(y).\nonumber
\end{eqnarray}
Therefore, combining Lemma \ref{lem 4} with Lemma \ref{lem 5}, we get
\begin{eqnarray}
R_{mn}=o(\Delta_{m})+O_{P}(m^{-3/2}).\label{h8}
\end{eqnarray}
By (\ref{r2}) and (\ref{rr2}) in the Assumption 1, (\ref{h6}), (\ref{h7}), and (\ref{h8}), there is
\begin{eqnarray}
S_{m, n}&=&-\frac{m n}{m+n}\Big[Q(F_m, G_n)+Q(G_n, F_m)-1\Big]\nonumber\\
&=&-\frac{m n}{m+n}\Big\{\frac{1}{m}\sum_{i=1}^{m}E_{y}I(x_{i},y,F_{m},G_{n})+\frac{1}{n}\sum_{j=1}^{n}E_{x}I(x,y_{j},F_{m},G_{n})+R_{mn}\Big\}\nonumber\\
&=&-\frac{m n}{m+n}\Big\{-\frac{1}{m}\sum_{i=1}^{m}f_{D(y;F)}\Big(D(x_{i};F)\Big)\Big(\rho_{1}(x_{i};F_{m},F)-\rho_{1}(x_{i};G_{n},F)\Big)\Big]\nonumber\\
&&+\frac{1}{n}\sum_{j=1}^{n}f_{D(x;F)}\Big(D(y_{j};F)\Big)\Big(\rho_{1}(y_{j};F_{m},F)-\rho_{1}(y_{j};G_{n},F)\Big)\Big]\nonumber\\
&&+O_{P}(m^{-3/2})+O_{P}(n^{-3/2})+R_{mn}\Big\}\nonumber\\
&=&-\frac{m n}{m+n}\Big\{-\frac{1}{m}\sum_{i=1}^{m}f_{D(y;F)}\Big(D(x_{i};F)\Big)\Big(\rho_{1}(x_{i};F_{m},F)-\rho_{1}(x_{i};G_{n},F)\Big)\nonumber\\
&&+\frac{1}{n}\sum_{j=1}^{n}f_{D(x;F)}\Big(D(y_{j};F)\Big)\Big(\rho_{1}(y_{j};F_{m},F)-\rho_{1}(y_{j};G_{n},F)\Big)\Big\}\nonumber\\
&&+o\Big(\frac{mn}{m+n}\Delta_{m}\Big)+O_{P}(m^{-1/2}).\label{h9}
\end{eqnarray}
Thus, Proof of (\ref{s1}) is complete. By (\ref{h9}) and Lemma \ref{lem 6}, it has
\begin{eqnarray}
P_{m, n}&=&-\frac{m n}{m+n}\Big[Q(F_m, G_n)\times Q(G_n, F_m)-\frac{1}{4}\Big]\nonumber\\
&=&-\frac{m n}{m+n}\Big[\Big(Q(F_m, G_n)-\frac{1}{2}+\frac{1}{2}\Big)\times \Big(Q(G_n, F_m)-\frac{1}{2}+\frac{1}{2}\Big)-\frac{1}{4}\Big]\nonumber\\
&=&-\frac{m n}{m+n}\Big[\Big(Q(F_m, G_n)-\frac{1}{2}\Big)\times \Big(Q(G_n, F_m)-\frac{1}{2}\Big)\nonumber\\
&&+\frac{1}{2}\Big(Q(F_m, G_n)+Q(G_n, F_m)-1\Big)\Big]\nonumber\\
&=&-\frac{m n}{m+n}\Big\{\Big[\frac{1}{n}\sum\limits_{j=1}^{n}\Big(F_{D(x;F)}\Big(D(y_{j};F)\Big)-\frac{1}{2}\Big)+\frac{1}{m}\sum\limits_{i=1}^{m}\Big(\frac{1}{2}-F_{D(y;F)}\Big(D(x_{i};F)\Big)\Big)\nonumber\\
&&o(\Delta_{m})+O_{P}(m^{-1})\Big]\times\Big[\frac{1}{n}\sum\limits_{j=1}^{n}\Big(\frac{1}{2}-F_{D(x;F)}\Big(D(y_{j};F)\Big)+\frac{1}{m}\sum\limits_{i=1}^{m}\Big(F_{D(y;F)}\Big(D(x_{i};F)\Big)-\frac{1}{2}\Big)\nonumber\\
&&+o(\Delta_{m})+O_{P}(m^{-1})\Big]\Big\}-\frac{mn}{2(m+n)}\Big\{-\frac{1}{m}\sum_{i=1}^{m}f_{D(y;F)}\Big(D(x_{i};F)\Big)\Big(\rho_{1}(x_{i};F_{m},F)-\rho_{1}(x_{i};G_{n},F)\Big)\Big]\nonumber\\
&&+\frac{1}{n}\sum_{j=1}^{n}f_{D(x;F)}\Big(D(y_{j};F)\Big)\Big(\rho_{1}(y_{j};F_{m},F)-\rho_{1}(y_{j};G_{n},F)\Big)+o(\Delta_{m})+O_{P}(m^{-3/2})\Big\}\nonumber\\
&=&-\frac{m n}{m+n}\Big\{\Big[\frac{1}{n}\sum\limits_{j=1}^{n}\Big(F_{D(x;F)}\Big(D(y_{j};F)\Big)-\frac{1}{2}\Big)+\frac{1}{m}\sum\limits_{i=1}^{m}\Big(\frac{1}{2}-F_{D(y;F)}\Big(D(x_{i};F)\Big)\Big)\Big]\times\nonumber\\
&&\Big[\frac{1}{n}\sum\limits_{j=1}^{n}\Big(\frac{1}{2}-F_{D(x;F)}\Big(D(y_{j};F)\Big)\Big)+\frac{1}{m}\sum\limits_{i=1}^{m}\Big(F_{D(y;F)}\Big(D(x_{i};F)\Big)-\frac{1}{2}\Big)\Big]\Big\}\nonumber\\
&&-\frac{mn}{2(m+n)}\Big\{-\frac{1}{m}\sum_{i=1}^{m}f_{D(y;F)}\Big(D(x_{i};F)\Big)\Big(\rho_{1}(x_{i};F_{m},F)-\rho_{1}(x_{i};G_{n},F)\Big)\nonumber\\
&&+\frac{1}{n}\sum_{j=1}^{n}f_{D(x;F)}\Big(D(y_{j};F)\Big)\Big(\rho_{1}(y_{j};F_{m},F)-\rho_{1}(y_{j};G_{n},F)\Big)\Big\}\nonumber\\
&&+o\Big(\frac{m n}{m+n}\Delta_{m}\Big)+O_{P}(m^{-1/2})
\nonumber\\
&=&\frac{m n}{m+n}\Big\{\Big[\frac{1}{n}\sum\limits_{j=1}^{n}\Big(F_{D(x;F)}\Big(D(y_{j};F)\Big)-\frac{1}{2}\Big)+\frac{1}{m}\sum\limits_{i=1}^{m}\Big(\frac{1}{2}-F_{D(y;F)}\Big(D(x_{i};F)\Big)\Big)\Big]^2\Big\}\nonumber\\
&&-\frac{mn}{2(m+n)}\Big\{-\frac{1}{m}\sum_{i=1}^{m}f_{D(y;F)}\Big(D(x_{i};F)\Big)\Big(\rho_{1}(x_{i};F_{m},F)-\rho_{1}(x_{i};G_{n},F)\Big)\nonumber\\
&&+\frac{1}{n}\sum_{j=1}^{n}f_{D(x;F)}\Big(D(y_{j};F)\Big)\Big(\rho_{1}(y_{j};F_{m},F)-\rho_{1}(y_{j};G_{n},F)\Big)\Big\}\nonumber\\
&&+o\Big(\frac{m n}{m+n}\Delta_{m}\Big)+O_{P}(m^{-1/2}).\label{h10}
\end{eqnarray}
Then, one can immediately obtain (\ref{s2}). The proof of Theorem \ref{the 1} is complete.

\section{Proof of Remark 3}
\label{euclidean-proof}

According to Theorem 1, for one-dimensional Euclidean depth, we can use Euclidean distance to replace the depth functions. Therefore,

\begin{eqnarray}
S_{m, n}
&=&-\frac{m n}{m+n}\Big\{ \frac{1}{m}\sum_{i=1}^{m}f_{d(y;F)}\Big(d(x_{i};F)\Big)\Big( E_y[d(x_i;F_m)-d(y_j;F_m)-d(x_i;G_n)+d(y_j;G_n)|d(x_i;F)=d(y_j;F)] \Big)\nonumber\\
&&-\frac{1}{n}\sum_{j=1}^{n}f_{d(x;F)}\Big(d(y_{j};F)\Big)\Big(
E_x[d(y_j;F_m) -d(x_i;F_m)-d(y_j;G_n)+d(x_i;G_n)|d(x_i;F)=d(y_j;F)] \Big)\Big\}\nonumber\\
&&+o\Big(\frac{m n}{m+n}\Delta_{m}\Big)+O_{P}(m^{-1/2}),
\label{q0}
\end{eqnarray}

where $f_{d(x;F)}(\cdot)$ is the density of Euclidean distance $d(x; F)$.

Then we have $d(x_i;F_m)=(x_i-\bar{x})^2, d(y_j;F_m)=(y_j-\bar{x})^2, d(x_i;G_n)=(x_i-\bar{y})^2, d(y_j;G_n)=(y_j-\bar{y})^2, d(x_i;F)=x_i^2, d(y_j;F)=y_j^2.$ Then, $f_{d(y;F)}(d(x_i;F))=f(x_i^2) $ and $f_{d(x;F)}(d(y_j;F))=f(y_j^2)$, where $f$ is density of $\chi^2_1$.

Therefore,

\begin{eqnarray}
S_{m, n}
&=&-\frac{m n}{m+n}\Big\{ \frac{1}{m}\sum_{i=1}^{m} f(x_i^2) 
\Big( E_y[(x_i-\bar{x})^2-(y_j-\bar{x})^2-(x_i-\bar{y})^2+(y_j-\bar{y})^2|x_i^2=y_j^2] \Big)\nonumber\\
&&-\frac{1}{n}\sum_{j=1}^{n} f(y_j^2) \Big(
E_x[(y_j-\bar{x})^2 -(x_i-\bar{x})^2-(y_j-\bar{y})^2+(x_i-\bar{y})^2|x_i^2=y_j^2 ] \Big)\Big\}\nonumber\\
&&+o\Big(\frac{m n}{m+n}\Delta_{m}\Big)+O_{P}(m^{-1/2}) \nonumber\\
&=& -\frac{m n}{m+n}\Big\{ \frac{1}{m}\sum_{i=1}^{m} f(x_i^2) 
\Big( -2x_i(\bar{x}-\bar{y}) \Big) -\frac{1}{n}\sum_{j=1}^{n} f(y_j^2) \Big( 2y_j(\bar{y}-\bar{x}) \Big)\Big\} \nonumber \\
&&+o\Big(\frac{m n}{m+n}\Delta_{m}\Big)+O_{P}(m^{-1/2}) \nonumber\\
&=& -\sqrt{\frac{mn}{m+n}}(\bar{y}-\bar{x}) \sqrt{\frac{mn}{m+n}} \Big\{ \frac{1}{m} \sum_{i=1}^m f(x_i^2) 2x_i - \frac{1}{n} \sum_{j=1}^n  f(y_j^2) 2y_j \Big\} \nonumber\\
&&+o\Big(\frac{m n}{m+n}\Delta_{m}\Big)+O_{P}(m^{-1/2})
\label{q1}
\end{eqnarray}

Since $\sqrt{\frac{mn}{m+n}}(\bar{y}-\bar{x}) \rightarrow \mathcal{N}(0,1)$ and $\sqrt{\frac{mn}{m+n}} [ \frac{1}{m} \sum_{i=1}^m f(x_i^2) 2x_i + \frac{1}{n} \sum_{j=1}^n  -f(y_j^2) 2y_j ] \rightarrow \mathcal{N}(0,\frac{2}{\sqrt{3}\pi})$, then $S_{m,n} \rightarrow -Z_1 Z_2$, where $Z_1 \sim \mathcal{N}(0,1)$ and $Z_2 \sim \mathcal{N}(0,\frac{2}{\sqrt{3}\pi})$, with $Cov(Z_1, Z_2)=-\frac{1}{\pi}$.


In a similar way, we can obtain the asymptotic distribution of $P_{m,n}$ under one-dimensional Euclidean depth.
Thus,
\begin{eqnarray}
P_{m, n}&=&-\frac{m n}{m+n}\Big\{\Big[-\frac{1}{n}\sum\limits_{j=1}^{n}\Big(F_{d(x;F)}\Big(d(y_{j};F)\Big)-\frac{1}{2}\Big) -\frac{1}{m}\sum\limits_{i=1}^{m}\Big(\frac{1}{2}-F_{d(y;F)}\Big(d(x_{i};F)\Big)\Big)\Big]  \nonumber\\
&& \times \Big[\frac{1}{n}\sum\limits_{j=1}^{n}\Big(F_{d(x;F)}\Big(d(y_{j};F)\Big)-\frac{1}{2}\Big) +\frac{1}{m}\sum\limits_{i=1}^{m}\Big(\frac{1}{2}-F_{d(y;F)}\Big(d(x_{i};F)\Big)\Big)\Big]\Big\}\nonumber\\
&&-\frac{mn}{2(m+n)}\Big\{\frac{1}{m}\sum_{i=1}^{m}f_{d(y;F)}\Big(d(x_{i};F)\Big)\Big( E_y[d(x_i;F_m)-d(y_j;F_m)-d(x_i;G_n)+d(y_j;G_n)|d(x_i;F)=d(y_j;F)] \Big)\nonumber\\
&&-\frac{1}{n}\sum_{j=1}^{n}f_{d(x;F)}\Big(d(y_{j};F)\Big)\Big(E_x[d(y_j;F_m) -d(x_i;F_m)-d(y_j;G_n)+d(x_i;G_n)|d(x_i;F)=d(y_j;F)]  \Big)\Big\}\nonumber\\
&&+o\Big(\frac{m n}{m+n}\Delta_{m}\Big)+O_{P}(m^{-1/2}), \label{2}
\end{eqnarray}
where $F_{D(x;F)}(\cdot)$ and $f_{D(x;F)}(\cdot)$ are the distribution function and density of $D(x; F)$, respectively.

Under one-dimensional Euclidean depth, $F_{d(y;F)} (d(x_i;F))=F(x_i^2)$ and $F_{d(x;F)}(d(y_j;F))=F(y_j^2)$, where $F$ is CDF of $\chi^2_1$, we have

\begin{eqnarray}
P_{m, n}
&=& -\frac{m n}{m+n}\Big\{\Big[-\frac{1}{n}\sum\limits_{j=1}^{n} \Big( F(y_j^2)-\frac{1}{2}\Big) -\frac{1}{m}\sum\limits_{i=1}^{m}\Big(\frac{1}{2}-F(x_i^2) \Big)\Big]  \nonumber\\
&& \times \Big[\frac{1}{n}\sum\limits_{j=1}^{n}\Big(F(y_j^2)-\frac{1}{2} \Big) +\frac{1}{m}\sum\limits_{i=1}^{m}\Big(\frac{1}{2}-F(x_i^2) \Big)\Big]\Big\}\nonumber\\
&&+\frac{1}{2}\left\{ -\sqrt{\frac{mn}{m+n}}(\bar{y}-\bar{x}) \sqrt{\frac{mn}{m+n}} \Big\{ \frac{1}{m} \sum_{i=1}^m f(x_i^2) 2x_i - \frac{1}{n} \sum_{j=1}^n  f(y_j^2) 2y_j \Big\} \right\} \nonumber\\
&&+o\Big(\frac{m n}{m+n}\Delta_{m}\Big)+O_{P}(m^{-1/2}) \\
&=& \frac{m n}{m+n} \Big[\frac{1}{n}\sum\limits_{j=1}^{n} \Big( F(y_j^2)-\frac{1}{2}\Big) +\frac{1}{m}\sum\limits_{i=1}^{m}\Big(\frac{1}{2}-F(x_i^2) \Big)\Big]^2  \nonumber\\
&&+\frac{1}{2}\left\{ -\sqrt{\frac{mn}{m+n}}(\bar{y}-\bar{x}) \sqrt{\frac{mn}{m+n}} \Big\{ \frac{1}{m} \sum_{i=1}^m f(x_i^2) 2x_i - \frac{1}{n} \sum_{j=1}^n  f(y_j^2) 2y_j \Big\} \right\} \nonumber\\
&&+o\Big(\frac{m n}{m+n}\Delta_{m}\Big)+O_{P}(m^{-1/2})
\end{eqnarray}

Since $ \sqrt{\frac{mn}{m+n}} [\frac{1}{m} \sum_{i=1}^m  \frac{1}{2}-F(x_i^2) +\frac{1}{n} \sum_{j=1}^n F(y_j^2)-\frac{1}{2} ] \rightarrow \mathcal{N}(0,\frac{1}{12})$,  $\sqrt{\frac{mn}{m+n}}(\bar{y}-\bar{x}) \rightarrow \mathcal{N}(0,1)$ and $\sqrt{\frac{mn}{m+n}} [ \frac{1}{m} \sum_{i=1}^m f(x_i^2) 2x_i + \frac{1}{n} \sum_{j=1}^n  -f(y_j^2) 2y_j ] \rightarrow \mathcal{N}(0,\frac{2}{\sqrt{3}\pi})$, then $P_{m,n} \rightarrow Z_3^2 -\frac{1}{2} Z_1 Z_2 $, where $Z_1 \sim \mathcal{N}(0,1)$, $Z_2 \sim \mathcal{N}(0,\frac{2}{\sqrt{3} \pi})$, and $Z_3 \sim \mathcal{N}(0,\frac{1}{12})$, with $Cov(Z_1, Z_2)=-\frac{1}{\pi}$. $Z_3$ is independent of $Z_1$ and $Z_2$.

\section{Extensions of Remark 3} \label{remark3-density}

\begin{figure}[H]
\begin{center}
\includegraphics[width=\textwidth]{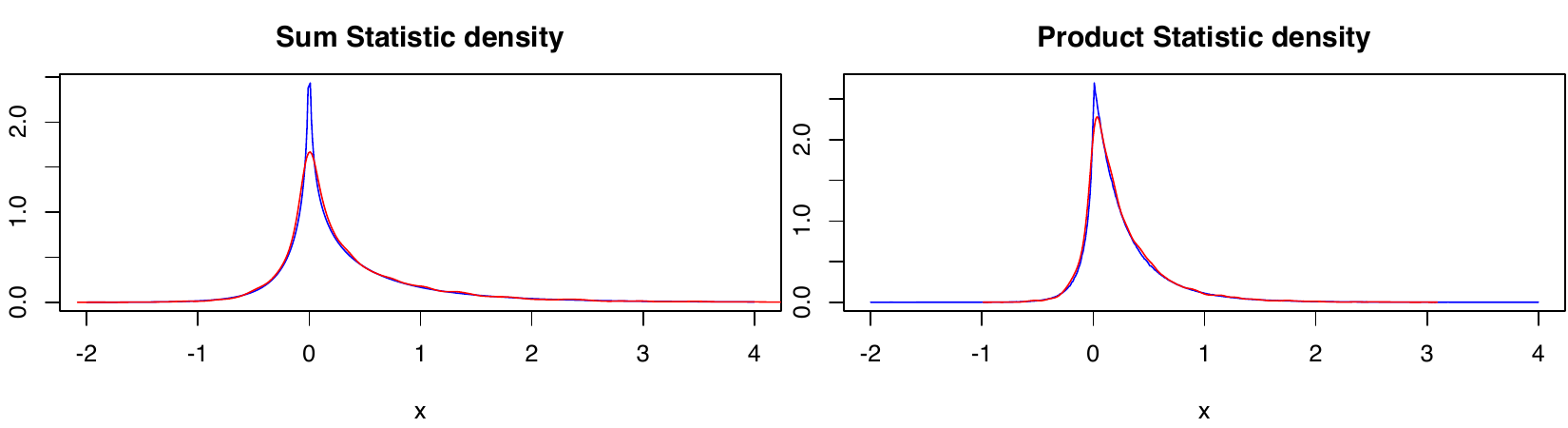}
\caption{The true (blue) probability density function vs. the estimated (red) density function for $f_S(x)$ on the left and $f_P(x)$ on the right respectively.}
\label{fig:density}
\end{center}
\end{figure}

According to the proof of Remark \ref{theorem1}, the approximated probability density functions of $S_{m,n}$ and $P_{m,n}$ can be denoted as $f_S(x)$ and $f_P(x)$, respectively: 
\begin{align*} 
f_S(x)&=\frac{1}{\pi \sqrt{\frac{2\pi-\sqrt{3}}{\sqrt{3} \pi^2 } }} e^{ \frac{\sqrt{3} x}{2-\frac{\sqrt{3}}{\pi} } }  \int_0^{\infty} \frac{1}{z_1}  e^{-\frac{1}{2-\frac{\sqrt{3}}{\pi}} (z_1^2+\frac{\sqrt{3}\pi x^2}{2 z^2_1} ) } \,dz_1,  \\
f_P(x)&=\int_{-\infty}^{\infty} \int_{-\infty}^{\infty} 12 f_{\chi^2}(12x+6z_1z_2) \frac{1}{2 \pi \sqrt{\frac{2\pi-\sqrt{3}}{\sqrt{3} \pi^2 } }} e^{-\frac{1}{2-\frac{\sqrt{3}}{\pi}} (z_1^2 +\sqrt{3} z_1 z_2 +\frac{\sqrt{3} \pi }{2} z_2^2) } \,dz_1 \,dz_2,
\end{align*}
where $f_{\chi^2}(12x+6z_1z_2)$ is the probability density of a $\chi_1^2$ at $12x+6z_1z_2$. Figure \ref{fig:density} compared $f_S(x)$ and $f_P(x)$ with simulated density for $m=n=1000$ and 10,000 repetitions, illustrating that both densities are right-skewed with sharp peaks. 

To further demonstrate the rate of convergence for distributions of $S_{m,n}$ and $P_{m,n}$, we conduct a simulation study. We generate samples where $m=n=100,\cdots,1000$ from $F=G=\mathcal{N}(0,1)$, each with a mean of 0 and a variance of 1. This simulation is repeated 10,000 times to compute the empirical $\alpha$ quantiles at levels $\alpha=0.2, 0.1, 0.05$ and $0.01$. 
   The empirical quantiles are then compared with the theoretical quantiles by assuming $m,n\rightarrow\infty$, based on the asymptotic distributions of $S_{m,n}$ and $P_{m,n}$ specified in \eqref{s3} and \eqref{s4}. This comparison is intended to quantitatively assess how well the finite sample distributions of $S_{m,n}$ and $P_{m,n}$ match their respective asymptotic behaviors.

As shown in Tables \ref{table:sum-type1-1} and \ref{table:prod-type1-1}, there is significant agreement between empirical and theoretical quantiles at different $\alpha$ levels. For all evaluated 
values, the empirical quantiles are very close to the theoretical quantiles, except for $\alpha=0.01$ which requires a larger sample size. This observation holds true even with relatively small sample sizes, thus demonstrating the fast convergence rate of the asymptotic distributions of $S_{m,n}$ and $P_{m,n}$ in approximating the behavior of their finite sample counterparts.

\begin{table}
\centering
\begin{tabular}{ |p{0.5cm}|p{1cm}|p{1cm}|p{1cm}|p{1cm}|p{1cm}|p{1cm} |p{1cm} |p{1cm}|p{1cm}|} 
\hline 
$\alpha$  & 100 & 200 & 300 & 400 & 500 & 600   & 800   & 1000 & $\infty$\\
\hline 
 0.2 & 0.6500 & 0.6525 & 0.6567 & 0.6200 & 0.6352 &0.6425   &0.6650   &0.6561 & 0.6417 \\ 
 \hline
 0.1 & 1.1150 & 1.1128 & 1.1483 & 1.1051 & 1.1021 & 1.1393   & 1.1600   & 1.1390 &1.1312 \\
 \hline
 0.05 & 1.6150 & 1.6100 & 1.6601 & 1.6350 &1.6171 & 1.6959   & 1.7025  & 1.6523 & 1.6566 \\
 \hline
0.01 & 2.8052 & 2.7803 & 2.8767 & 2.9901 & 2.8530 & 2.9526   & 2.9463   & 2.9671 &2.9608\\
\hline 
\end{tabular}
\caption{Empirical quantiles vs. theoretical quantiles of $S_{m,n}$ for different $\alpha$ with $m=n=100, \cdots, 1000$}
\label{table:sum-type1-1}
\end{table}

 \begin{table}
\centering
\begin{tabular}{ |p{0.5cm}|p{1cm}|p{1cm}|p{1cm}|p{1cm}|p{1cm}|p{1cm} |p{1cm} |p{1cm}|p{1cm}|} 
\hline 
$\alpha$  & 100 & 200 & 300 & 400 & 500 & 600   & 800 &   1000 & $\infty$ \\
\hline 
 0.2 & 0.4342 & 0.4372 & 0.4471 & 0.4331 & 0.4324 & 0.4400   & 0.4528  & 0.4457 & 0.4379\\ 
 \hline
 0.1 & 0.6656 & 0.6729 & 0.6869 & 0.6714 & 0.6701 & 0.6775   & 0.7014   & 0.6840 & 0.6818\\
 \hline
 0.05 & 0.9008 & 0.9187 & 0.9402 & 0.9232 & 0.9337 & 0.9532   & 0.9606   & 0.9452 & 0.9384\\
 \hline
0.01 & 1.4527 & 1.5048 & 1.5473 & 1.5930 & 1.5522 & 1.5883   & 1.5673   & 1.5739 & 1.5706 \\
\hline 
\end{tabular}
\caption{Empirical quantiles vs. theoretical quantiles of $P_{m,n}$ for different $\alpha$ with $m=n=100, \cdots, 1000$}
\label{table:prod-type1-1}
\end{table}

\section{Proof of Theorem 2}
\label{thm2-proof}
As each block from the same distribution are the same, we can express  $||\bm{\theta}(F)-\bm{\theta}(G)||$ as $||\sum_{i=1}^{b_1}\bm{\theta}(F_i)/b_1-\sum_{j=1}^{b_2}\bm{\theta}(G_{b_1+j})/b_2||$, where $F_i$ and $G_{b_1+j}$ are distributions of $i-$th block and $b_1+j-$th block in combined samples $x_1, \dots, x_m, y_1, \dots, y_n$, respectively.  For the permutation $(\pi(1), \dots, \pi(N))$, we have $||\sum_{i=1}^{b_1}\bm{\theta}(F_{\pi(i)})/b_1-\sum_{j=1}^{b_2}\bm{\theta}(G_{\pi(b_1+j)})/b_2||$, which can be expressed as 
$$||\frac{(b_1-c)\bm{\theta}(F)+c\bm{\theta}(G)}{b_1}-\frac{(b_2-c)\bm{\theta}(G)+c\bm{\theta}(F)}{b_2}||=||\bm{\theta}(F)-\bm{\theta}(G)|||1-\frac{c}{b_1}-\frac{c}{b_2}|,$$
where $0\leq c\leq\min(b_1,b_2).$

We note that $$||\sum_{i=1}^{b_1}\bm{\theta}(F_i)/b_1-\sum_{j=1}^{b_2}\bm{\theta}(G_{b_1+j})/b_2||=\sum_{i=1}^{b_1}\bm{\theta}(F_{\pi(i)})/b_1-\sum_{j=1}^{b_2}\bm{\theta}(G_{\pi(b_1+j)})/b_2||$$ with probability less than $\frac{2 m! n!}{(m+n)!}$, i.e., the  permutations  occur only within each sample with probability $\frac{m! n!}{(m+n)!}$ for $c=0$ or two samples are exchanged when $m=n$ with probability $\frac{m! m!}{(2m)!}$ for $c=b_1=b_2$. Except these case, for $0< c<\min(b_1,b_2).$ we have 
$$||\sum_{i=1}^{b_1}\bm{\theta}(F_i)/b_1-\sum_{j=1}^{b_2}\bm{\theta}(G_{b_1+j})/b_2||>||\sum_{i=1}^{b_1}\bm{\theta}(F_{\pi(i)})/b_1-\sum_{j=1}^{b_2}\bm{\theta}(G_{\pi(b_1+j)})/b_2||.$$

That's because $-1<1-\frac{\min(b_1,b_2)-1}{b_1}-\frac{\min(b_1,b_2)-1}{b_2}\leq 1-\frac{c}{b_1}-\frac{c}{b_2}\leq 1-\frac{1}{b_1}-\frac{1}{b_2}<1$.

As $s\rightarrow\infty$, under Assumption A, the permuted new $S_{m,n}$ and $P_{m,n}$ are less than their original ones in probability. In addition, the $p$-value$_S$ and $p$-value$_P$ converge to the probabilities of permuted new $S_{m,n}$ and $P_{m,n}$   greater than their original ones as $\mathcal{C} \rightarrow\infty$, which is zero and less than $\alpha$. So, Theorem \ref{theorem2} follows.

\end{document}